\documentclass[manuscript,screen]{acmart}
\usepackage{caption}
\usepackage{subcaption}
\usepackage{array}
\usepackage{multirow, booktabs}
\AtBeginDocument{%
  \providecommand\BibTeX{{%
    \normalfont B\kern-0.5em{\scshape i\kern-0.25em b}\kern-0.8em\TeX}}}

\setcopyright{none}

\begin{document}

\title{Decoy Effect In Search Interaction}
\subtitle{Understanding User Behavior and Measuring System Vulnerability}

\author{Nuo Chen}
\email{pleviumtan@outlook.com}
\orcid{0000-0001-8600-8203}
\affiliation{%
  \institution{The Hong Kong Polytechnic University}
  \state{HK SAR}
  \country{China}
}
\affiliation{%
  \institution{Waseda University}
  \state{Tokyo}
  \country{Japan}
}

\author{Jiqun Liu}
\orcid{0000-0003-3643-2182}
\affiliation{%
  \institution{The University of Oklahoma}
  \state{OK}
  \country{USA}}
\email{jiqunliu@ou.edu}

\author{Hanpei Fang}
\orcid{0009-0009-5117-5621}
\affiliation{%
  \institution{Waseda University}
  \state{Tokyo}
  \country{Japan}
}

\author{Yuankai Luo}
\orcid{0000-0003-3844-7214}
\affiliation{%
  \institution{The Hong Kong Polytechnic University}
  \state{HK SAR}
  \country{China}
}

\author{Tetsuya Sakai}
\orcid{0000-0002-6720-963X}
\email{tetsuyasakai@acm.org}
\affiliation{%
  \institution{Waseda University}
  \city{Tokyo}
  \country{Japan}}

\author{Xiao-Ming Wu}
\email{xiao-ming.wu@polyu.edu.hk}
\orcid{0000-0002-3130-0554}
\affiliation{%
  \institution{The Hong Kong Polytechnic University}
\state{HK SAR}
  \country{China}
}
\renewcommand{\shortauthors}{Chen, et al.}

\begin{abstract}
  This study addresses~(1)~the influence of the decoy effect, a cognitive bias where the presence of an inferior item alters preferences between two options, on users' search interactions and~(2)~the measurement of information retrieval systems' \textit{vulnerability} to the decoy effect\footnote{This article is a follow-up study of the NTCIR EVIA paper of Chen \textit{et al.}~\citep{chendecoy}. The Experiment 1 in this article is mainly based on the work of the EVIA paper, but the Experiment 1 in this article is conducted on one more new dataset (THUIR2018~\citep{Liu2018www}) in comparison to the EVIA paper. Experiment 2 and Experiment 3 in this article are entirely new and have not been reported before.}. From the perspective of user behavior, this study investigates the influence of the decoy effect in information retrieval (IR) by examining how decoy results affect users' interaction on search engine result pages (SERPs), particularly in terms of click-through likehood, browsing dwell time, and perceived document usefulness. 
   We conducted an experiment based upon regression analysis on user interaction logs from three user study datasets which in total encompass 24 topics, 841 unique search sessions, and 2,685 queries. The findings indicate that decoys significantly increase the likelihood of document clicks and perceived usefulness. To investigate whether the influence of the decoy varies across different levels of task difficulty and user knowledge, we ran an additional experiment on one of the three datasets, which encompasses 6 topics, 166 search sessions and 652 queries. 
   The results indicate that when the task is less challenging, users are more likely to click on a document with a decoy. Additionally, they spend more time on the target document and assign it a higher usefulness score. Furthermore, users with lower knowledge levels about the topic tend to give higher usefulness ratings to the target document. 
  
    Regarding IR system evaluation, this study provides empirical insights into measuring the vulnerability of text retrieval models to potential decoy effect. An evaluation metric, namely DEcoy Judgement and Assessment VUlnerability (DEJA-VU), is proposed to evaluate the possibility of a retrieval model ranking results in a way that could trigger decoy biases. The experiments on the TREC 19 DL passage retrieval task and the TREC 20 DL passage retrieval task demonstrate that ColBERT and SPLADE show higher relevance-oriented retrieval effectiveness while also displaying lower vulnerability to decoy effect. 
    
    Overall, this work advances the understanding of decoy effect, a well-established concept in cognitive psychology and behavioral economics, in a novel application field (\textit{i.e.,} Information Retrieval). It contributes to modeling users' search behavior in the context of cognitive biases, as well as assessment of the vulnerability of systems and ranking algorithms to the decoy effect.
    
    
\end{abstract}

\ccsdesc[500]{Information systems~Users and interactive retrieval}

\keywords{cognitive bias, web search, search interaction, evaluation measure}


\maketitle

\section{Introduction}

Understanding the cognitive processes, behavioral patterns, and decision-making mechanisms of users during interactions with search systems is a fundamental research focus in interactive Information Retrieval (IR). In recent years, the exploration of \textit{cognitive biases} and their impact on the information seeking 
behaviors and outcomes has garnered increasing attention~\citep{liu2023behavioral,Azzopardi2021}. The cognitive bias is a systematic pattern of deviations in thinking which may lead to irrational judgements and problematic decision-making~\citep{Tversky1974, Tversky1992,Ariely2011-ed}. Contrary to the (over)simplified assumptions of \textit{globally rational} users~(\textit{i.e.,} users will rationally and comprehensively weigh the benefits and costs incurred during the search process before making decisions such as clicking or re-querying.) that form the basis of various existing user models and evaluation metrics, users are frequently affected by a range of systematic cognitive biases, emotions, mental shortcuts and heuristics~\citep{Agosto2002,Lopatovska2014,liu2023,liu2023behavioral, draws2023}. As a result, predictions made by models based on the assumption of global rationality could show significant divergence from the actual decisions and retrospective assessments of users~\citep{Liu2020,liu2023behavioral, zhang2020}.

\begin{figure}[t]
    \centering
    \includegraphics[width=.4\linewidth]{
    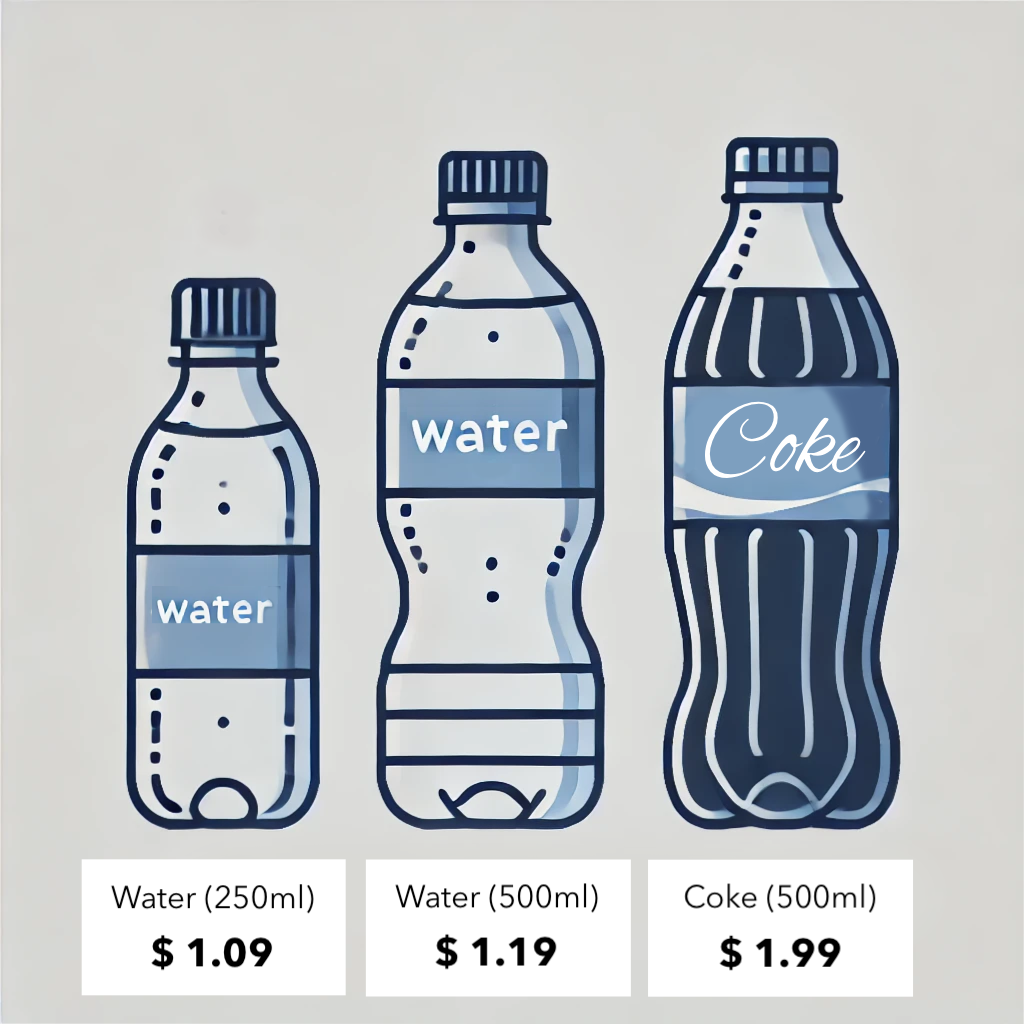
    }
    \caption{\label{fig:water}An example of the decoy effect. A customer might waver between the 500ml water and the coke. Yet, with a bottle of 250ml water presenting beside the 500ml water, the customer might lean towards the 500ml water. The image is generated by DALL·E-3 and manually edited by the authors.}
\end{figure}

The \textbf{decoy effect}, which is one kind of cognitive biases, describes a situation in which individuals alter their preference between two initial choices when introduced to a third~(\textit{i.e.,} the decoy), which is asymmetrically inferior to one of the initial choices~\citep{Huber1982}. Figure~\ref{fig:water} illustrates an example of the decoy effect in shopping decision-making. In a shop, a customer who intends to buy a beverage might waver between a 500ml bottle of water~(for $\$1.19$) and a bottle of coke with a similar size (for $\$1.99$). While the water is more affordable, the coke may offer a superior taste, making the decision challenging. The final choice may rest on the consumer's relative utility assessment of these options. Yet, with a  250ml  bottle of water for $\$1.09$ presenting beside the 500ml water, the customer might lean towards the 500ml water, as they perceive a substantial relative gain from the comparison of the 500ml water and the 250ml water:~spending an additional $\$0.10$ to purchase a  500ml bottle of water, compared to the 250ml one, evidently presents a highly economical deal. In the above example, the 250ml water serves as the \textit{decoy} to the \textit{target} 500ml water. 

\begin{figure}[t]
    \centering
    \includegraphics[width=.99\linewidth]{
    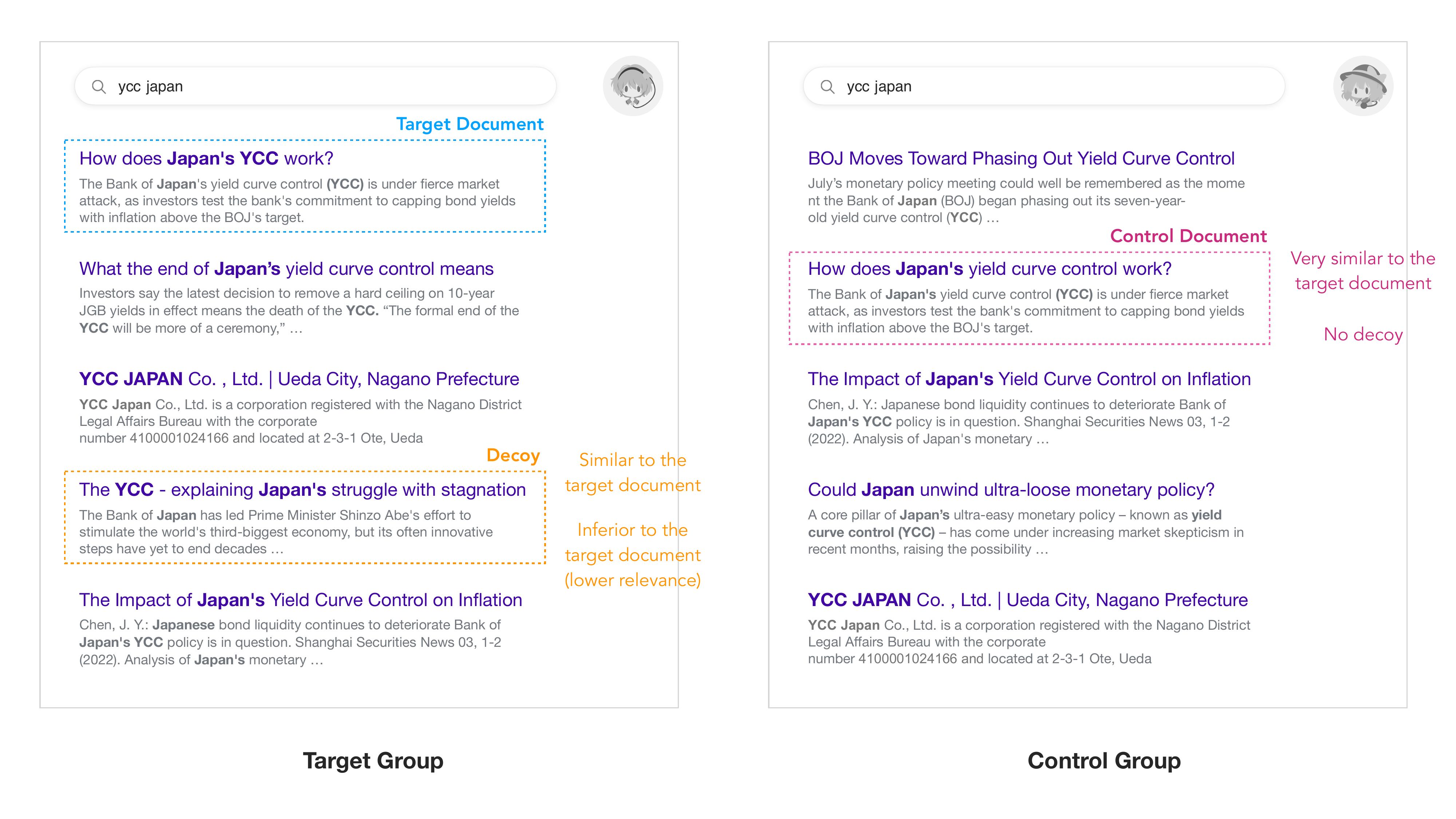
    }
    \caption{
       This illustration depicts two SERPs. In the SERP on the left, the document outlined by the orange dashed line functions as a decoy for the document outlined by the light blue dashed line, forming a ``decoy pair.'' In Experiment 1 and Experiment 2, documents such as the one outlined by the light blue dashed line, which have at least one decoy present, are designated as ``target documents.'' Conversely, in the SERP on the right, the document outlined by the rose-colored dashed line is highly similar to the document outlined by the light blue dashed line but lacks a decoy. Such documents are classified as ``control documents'' in both experiments, serving as a control group for the target documents in the absence of a decoy.
    \label{fig:decoy_search_example}}
\end{figure}

The investigation of decoy effects is of significant practical importance, as evidenced through both empirical studies and real-world applications, such as marketing and e-commerce
~\citep{wu2020decoy,wu2020profit, Teppan2010,teppan2015, mo2022decoy}. In the field of IR, gaining insights into the decoy effect can enable researchers to gain a better understanding of the preferences and judgments of real-life users (as opposed to simulated agents) towards information objects, leading to more accurate predictions of their behavior. 
However, the decoy effect has not received significant attention from the IR community up to the present. Eickhoff~\citep{Eickhoff2018} examined the impact of a decoy document on  thresholds and strategies in crowdsourcing relevance judgments, showing that assessors could increase the relevance rating of target document when it is shown with the decoy document. Nevertheless, Eickhoff~\citep{Eickhoff2018} only focused on crowdsourcing assessors operating within the annotation interface and few researches currently addresses how the decoy effect influences user interactions on Search Engine Result Pages (SERPs).

The SERP on the left in Figure~\ref{fig:decoy_search_example} can be viewed as an example of the decoy effect in search interactions. Suppose the document $t$, encircled by the light blue dashed line, is similar to the document $d$, which is encircled by the orange dashed line, and $d$ is substantially inferior to $t$ in quality (in the context of this study, quality is equated with relevance). If $t$ is dissimilar to several nearby documents, then $t$ can be considered the target document (similar to the 500ml bottle of mineral water in Figure~\ref{fig:water}), while $d$ serves as the decoy (similar to the 250ml bottle of mineral water in Figure~\ref{fig:water}). The other nearby documents, \textit{i.e.,} documents that are not enclosed by the dashed lines, are similar to the coke depicted in Figure~\ref{fig:water}.

To address the research gap mentioned above, 
we seek to understand how decoy effect at the document level influences users’ interaction behaviors on SERPs, such as clicks, browsing dwell time, and usefulness judgments. Specifically, our work seeks to answer following \textbf{Research Questions} (RQs):~
\begin{itemize}
    \item \textbf{RQ1}:~How, and to what extent, the presence of a decoy influences the likelihood of a document being clicked, the browsing duration on it, and its perceived usefulness?
    \item  \textbf{RQ2}:~Does the extent to which the decoy effect influences user behavior vary with task difficulty? If so, how does the decoy effect influence user behavior differently across varying levels of task difficulty? 
    \item  \textbf{RQ3}:~Does the impact of the decoy effect on user behavior relate to the user's level of prior knowledge about the search topic? If so, how does the decoy effect influence user behavior differently across varying levels of the user's prior knowledge scale?
    \item  \textbf{RQ4}:~How can we measure the vulnerability of rankers to decoy effect? 
\end{itemize}

To empirically demonstrate the impact of the decoy effect on user-document interaction behavior, we conduct user log mining and analysis on three publicly available user behavior datasets (refer to Section~\ref{ch:4}), identifying a total of $982 + 922 + 413 = 2,317$ decoy pairs, with a total of $318 + 376 + 98 = 792$ distinct documents containing at least one decoy, across the three datasets. We utilize Ordinary Least Squares (OLS) regression for analysis to observe the influence of the presence of decoys on the likelihood of document clicks, browsing duration time, and perceived usefulness score (refer to Section~\ref{ch:5}). In addition, we investigate whether there are differences in the impact of the decoy effect on user behavior under different task difficulties and user prior knowledge levels (refer to Section~\ref{ch:6}). Finally, we address how to evaluate the system's vulnerability to the decoy effect. In a broader sense, the \textit{vulnerability} of IR systems refers to the risk of irrational user behavior and biased decisions that may arise when the user is interacting with the system. In this paper, we define vulnerability as the number of decoy pairs returned in search results by an IR system at a given level of effectiveness. we analyzed several text retrieval models, calculating the number of decoy pairs, nDCG scores~\citep{Jarvelin-2002}, and recall scores on 97 topics from TREC 19 and 20 DL passage retrieval tracks~\cite{craswell2020overview,craswell2021overview}. The findings suggest that vulnerability cannot be solely measured by decoy pairs. We propose a heuristic metric, namely \textbf{DEcoy Judgement and Assessment VUlnerability (DEJA-VU)}, for assessing this vulnerability~(refer to Section~\ref{ch:7}).
 
The experimental result demonstrates that:~(1)~While keeping other conditions constant, when a decoy is present, in comparison to when it is absent, there is an increase in the likehood of a document being clicked and its perceived usefulness. (2)~While keeping other conditions constant, when the search task is more difficult, users are less likely to click on documents having a decoy compared to situations where the search task is less difficult. They also spend less time on documents having a decoy and provide lower usefulness scores for such documents. (3)~While keeping other conditions constant, when users have a lower level of prior knowledge about the retrieval topic, they tend to assign higher usefulness scores to the document having a decoy. (4)~
According the score of DEJA-VU, TCT-ColBERT and SPLADE++ show higher effectiveness and lower vulnerability to the decoy effect, especially with a smaller cutoff.

The main contributions of our paper are threefold:

\begin{itemize}
    \item This study extends the understanding of decoy effect, a well-established concept in cognitive psychology and behavioral economics, in a novel application field (\textit{i.e.,} interactive IR). By analyzing logs collected from real users, this study provides empirical evidence on how the presence of a decoy influences user interactions, such as click patterns, browsing durations, and perceived usefulness of documents, on SERPs. As far as we know, this is the first study to address the influence of the decoy effect on users' open domain information seeking behavior.

    \item  This study uniquely contributes to the understanding of how the decoy effect shapes user behavior under different search task difficulties and the user's prior knowledge levels. The finding enhances previous work on how tasks and search tasks influence users' information seeking and search behavior~(\textit{e.g.,}~\citep{Liu2010,liu2010b, Campbell1988,capra2017, Kelly2015}) from the perspective of cognitive bias.

    \item We propose a metric for evaluating the vulnerability of text retrieval systems to the decoy effect. This metric can guide developers towards improving existing text retrieval systems by enhancing effectiveness while mitigating the vulnerability to the decoy effect. 
    By adding new bias dimensions to current user modeling frameworks, it enhances the human-centered evaluation of search systems, particularly in how they can accommodate and mitigate cognitive biases in user interactions. 
\end{itemize}


The structure of the entire text is as follows: Section~\ref{ch:2} introduces previous work in the areas of cognitive biases, behavioral economics, and interactive search; Section~\ref{ch:rq} presents our research questions; Section~\ref{ch:4} provides an overview of the user study datasets used in the first two experiments and how we processed the datasets; Section~\ref{ch:5} describes the methodology and experimental results of our first experiment; Section~\ref{ch:6} outlines the methodology and experimental results of our second experiment; Section~\ref{ch:7} details the methodology and experimental results of our third experiment. Section~\ref{ch:8} summarizes our findings and discusses some open questions.

\section{Related Work}
\label{ch:2}
This section introduces the fundamental concepts and the interdisciplinary approaches underpinning our study.

\subsection{Cognitive Biases in Interactive Information Seeking and Retrieval}
\label{ch:2_1}
Insights from cognitive psychology and behavioral economics suggest that, \textit{cognitive biases} arise from one’s limited cognitive ability when there are not enough resources to properly collect and process available information~\citep{kruglanski1983,Evans1990-EVABIH}. Due to cognitive biases, one’s decisions under uncertainty can systematically deviate from what is expected given rational decision-making models~\citep{Tversky1974, Tversky1991, Tversky1992,Ariely2011-ed}.

Characterizing the cognitive processes, behavioral patterns, and decision-making mechanisms that users exhibit whilst engaging with search systems constitutes a core research endeavor within the domain of interactive Information Retrieval~(IR). Contrary to the implicit assumptions prevalent in the majority of exisiting user models, users do not exhibit global rationality (\textit{e.g.} always pursuing optimal utility, having full access to information for decision-making and unlimited cognitive resources for analyzing gains and costs). Instead, they are \textit{boundedly rational}, which means that they are frequently affected by a range of systematic cognitive biases, emotions, mental shortcuts and heuristics~\citep{Agosto2002,Lopatovska2014,liu2023,liu2023behavioral}, which usually lead to significant divergence between the predictions made by models based on the assumption of global rationality and the actual decisions and retrospective assessments of users~\citep{Liu2020,liu2023behavioral, zhang2020}. Previous studies in the field of interactive IR have  demonstrated that, due to the inherent nature of cognitive biases, certain individuals are more susceptible or more likely to be influenced by biased judgments arising from interaction contexts~(\textit{e.g.,} cognitive load, domain knowledge)~\citep{liao2013,boon2023} and individual characteristics~\citep{PENNYCOOK2019}.

In search and recommendation contexts, interactions between individuals and systems could lead to the incorporation of behavioral signals, influenced by \textit{cognitive biases}, into datasets used for training machine learning algorithms, thereby potentially magnifying existing system biases~\citep{Azzopardi2021, liu2023behavioral}. Cognitive biases might also result in significant deviations in users' behaviors and judgements from optimal or desired outcomes. Consequently, this could give rise to unfair decisions and outcomes between users who are more susceptible to certain biases and contextual triggers and those who are not~\citep{liu2023}.

Therefore, with an increasing number of individuals turning to search and recommendation systems to access and utilize information for life decisions, the influence of cognitive biases on the information retrieval process is drawing heightened attention from IR researchers. Thus far, a lot of studies have explored the influence of cognitive biases such as the anchoring effect~\citep{Shokouhi2015}, the priming effect~\citep{Scholer2013, chandar2012}, the ordering effect~\citep{BANSBACK2014}, the confirmation bias~\citep{draws2023}, and the reference dependence effect~\citep{Liu2020, wang2023investigating} on document examination, relevance judgment, and evaluation of whole-session search satisfaction. 

\subsection{Decoy Effect}
In this paper, we specifically shed light on one of the cognitive effects, the \textit{decoy effect}. The \textit{decoy effect}, which is one kind of cognitive biases, describes a situation in which individuals alter their preference between two initial choices when introduced to a third~(\textit{i.e.,} the decoy), which is asymmetrically inferior to one of the initial choice~\citep{Huber1982}. Figure~\ref{fig:water} illustrates an example of the decoy effect in shopping decision-making. In a shop, a customer who intends to buy a beverage might hesitate between a 500ml bottle of water (priced at $\$1.19$) and a similarly sized bottle of soda (priced at $\$1.49$). While the water is more affordable, the soda may offer a superior taste, making the decision challenging. The final choice may rest on the consumer's relative utility assessment of these options. However, when a 250ml bottle of water (priced at $\$1.09$) is displayed alongside the 500ml water, the customer may incline towards the 500ml water, discerning a notable relative advantage from the comparison of the 500ml water and the 250ml water:~spending an additional $\$0.10$ to purchase a  500ml bottle of water, compared to the 250ml one, evidently presents a highly economical deal. In the above example, the 250ml water serves as the \textit{decoy} to the \textit{target} 500ml water.

In the fields of marketing and e-commerce, there have been some studies exploring the impact of the decoy effect~\citep{wu2020decoy,wu2020profit, Teppan2010,teppan2015, mo2022decoy}. However, in the field of information retrieval, it is not clear how the decoy effect influences user interactions with and evaluations on Search Engine Result Pages. The work most closely related to ours in theme is that of Eickhoff~\cite{Eickhoff2018}, which shows that when a relevant item is presented alongside two non-relevant items, with one non-relevant item being distinctly inferior (i.e., the decoy), assessors tend to rate the superior non-relevant document as more relevant. Nevertheless,~\citet{Eickhoff2018} only focused on crowdsourcing assessors operating within the annotation interface and our study addresses how the decoy effect could influence user interactions on SERPs, which is a broader scenario. 

\subsection{Ranking Models in Information Retrieval}
Traditional IR ranking models rely on exact lexical matching, such as Boolean retrieval, BM25~\citep{robertson94, robertson09}, and statistical language models~\citep{laferty2001}. These retrieval models, also known as~\textit{Bag of Words}~(BOW) models, are based on sparse vector representation and process queries by organizing documents into inverted indices, wherein each unique term is associated with an inverted list that stores information regarding the documents in which it appears. However, the token-based sparse representation of text cannot fully capture the semantic nuances of each term within the entire textual context. These retrieval models thus suffer from the problem of vocabulary mismatch or semantic mismatch~(\textit{i. e.,} relevant documents may not contain terms that appear in the query). 

One approach to deal with the vocabulary mismatch is to use \textit{dense vectors}, which represent the text in a continuous vector space with predefined dimensions and the dimension is not dependent on the length of the text. The advantage of this approach is that text with similar semantics is typically represented by vectors that are close to each other in the vector space. Ranking models based on dense vectors is referred to as\textit{ dense retrieval} models. Dense retrieval models include Dense Passage Retriever (DPR)~\citep{karpukhin-etal-2020-dense}, Contriever~\citep{izacard2021contriever}, Approximate nearest neighbor Negative Contrastive Learning (ANCE)~\citep{xiong2020approximate}, ColBERT~\citep{khattab20}, Sentence-BERT (SBERT)~\citep{reimers-gurevych-2019-sentence}, and so forth. Many dense retrieval models utilize BERT~\citep{devlin-etal-2019-bert} for encoding queries and passages (\textit{e.g.,}~\citep{karpukhin-etal-2020-dense,xiong2020approximate, khattab20, reimers-gurevych-2019-sentence}) and utilize techniques such as contrastive learning~(\textit{e.g.,} ~\citep{karpukhin-etal-2020-dense,xiong2020approximate}) or or Siamese Network~(\textit{e.g.,}~\citep{reimers-gurevych-2019-sentence}) during the training process, achieving better semantic matching effectiveness compared to BM25 algorithm on benchmarks such as MS MARCO\footnote{https://microsoft.github.io/msmarco/}. 



On the other hand, \textit{sparse representations} based on pre-training language models~(PLMs) have also garnered increasing interest because they inherently inherit desirable properties of lexical models in their design. For example,  \textbf{CO}ntextualized \textbf{I}nverted \textbf{L}ist (COIL)~\citep{gao-etal-2021-coil} learns dense term-level representations to perform contextualized lexical matching; uniCOIL~\citep{lin2021brief} further simplifies the approach by learning a single weight per term. \textbf{SP}arse \textbf{L}exical \textbf{A}n\textbf{D} \textbf{E}xpansion (SPLADE)~\citep{Formal_2021} directly learns high-dimensional sparse representations that are capable of jointly performing expansion and re-weighting with the help of the PLM's masked language modeling head and sparse regularization. These sparse retrieval models also achieved better semantic matching ability compared to BM25 algorithm on benchmarks such as MS MARCO.   

In this study, we do not focus on improving the specific structure of ranking algorithms. Instead, we are are interested in whether different ranking models return document retrieval results with varying degrees of vulnerability to the decoy effect, and how to measure such vulnerability~(refer to Section~\ref{ch:7}). 

\subsection{The Evaluation of Information Retrieval Systems}
The Evaluation of information retrieval systems is one of the main research interests in the information retrieval community. Existing evaluation methods can be broadly divided into two classes, \textit{user-based}~(or online) evaluation and \textit{test collection-based}~(or offline) evaluation~\cite{Voorhees-2002}. 

Online methods rely on users' implicit or explicit online signals (such as A/B tests, click models~\citep{Chuklin-2013}, time-based models~\citep{Sumucker-2012-TBG, Sakai-2013}, machine learning based methods~\citep{wang2014, jiang2015, mehrotra2017, chen2023sigir}, \textit{etc.}) as feedback to measure the effectiveness of IR systems, which goes beyond the main scope of this article.

Offline evaluation is often built upon different assumptions and simulations regarding the process of a user interacting with a search system~\cite{Sanderson-2010, liu2022toward}. A range of evaluation metrics involving explicit or implicit user behaviour models have been proposed and examined on test collections, including Discounted Cumulative Gain~(DCG)~\cite{Jarvelin-2002} and its variants,  Rank-Biased Precision~(RBP)~\cite{Moffat-2008}, Expected Reciprocal Rank~(ERR)~\cite{Chapelle-2009}, INST~\citep{moffat2017},~\textit{etc}. As described in Section~\ref{ch:2_1}, most offline evaluation metrics treat users as globally rational decision makers when simulating interactions with search engines, but this assumption has been increasingly challenged recently. In recent years, with the growing knowledge about users’ cognitive biases, some works in the field of IR system evaluation began to introduce cognitive biases into the construction and meta-evaluation of evaluation metrics~\citep{zhang2020,chen2022,chen2023}. However, there is few work that incorporates the decoy effect into the calculation of IR system evaluation metrics. To address this gap, in this study, we propose an offline evaluation metric to measure the potential vulnerability to the decoy effect of IR systems~(refer to Section~\ref{ch:7_4}). Unlike most offline evaluation metrics mentioned earlier, the DEJA-VU score is not based on how users accumulate gains from relevant documents during the interaction with SERP. Instead, the DEJA-VU score considers the system's ability to return highly relevant documents~(the higher the better) while also accounting for the presence of decoy pairs in the results~(the lower the better). This is because the factors affecting the number of decoy pairs are complex, so we anticipate that the metric scores will prefer systems that return more highly relevant documents while returning fewer decoy pairs.

\section{Research Questions}
\label{ch:rq}
To address the research gap mentioned above, in this study, we seek to understand how the decoy
effect at the document level influences users’ interaction behaviors on SERPs, such as clicks, browsing dwell time, and usefulness perceptions. Specifically, our work sought to answer following \textbf{Research Question} (RQ):
\begin{itemize}
    \item \textbf{RQ1}:~How, and to what extent, the presence of a decoy influences the likelihood of a document being clicked, the browsing duration on it, and its perceived usefulness?
    \item  \textbf{RQ2}:~To what extent does impact of decoy items on user behavior vary across tasks of varying difficulty?
    \item  \textbf{RQ3}:~ How is the behavioral impact of decoy item associated with users' topic knowledge? 
    \item  \textbf{RQ4}:~How should we measure the vulnerability of information retrieval systems to decoy effect? 
\end{itemize}

\begin{figure}[t]
    \centering
    \includegraphics[width=.85\linewidth]{
    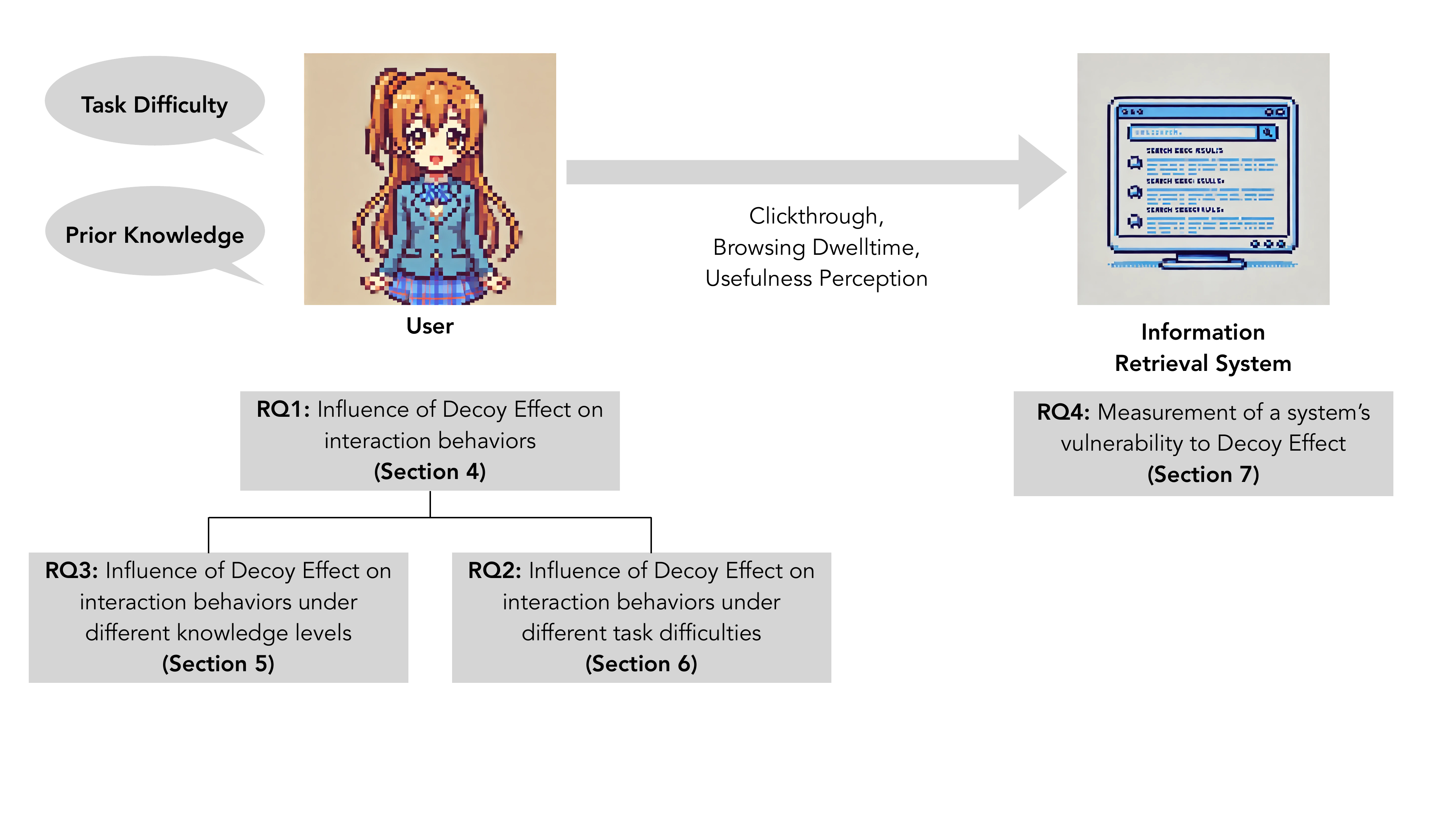
    }
    \caption{\label{fig:decoy_rq} The relationships among the research questions in this study and the corresponding sections they relate to.} 
\end{figure}

The four questions presented are interconnected, forming the multi-layered research framework of this study. \textbf{RQ1} lays the groundwork by providing a fundamental understanding of the influence of the decoy effect on users' interaction and perception. \textbf{RQ2} and \textbf{RQ3} delve deeper into the complexity of the influence of the decoy effect from two distinct dimensions: task difficulty and users' prior knowledge, respectively. \textbf{RQ4} applies the insights from the first three RQs to practical system design, seeking a balance between improving effectiveness and mitigating the vulnerability to the decoy effect. Figure~\ref{fig:decoy_rq} illustrates how our research questions were formulated from the perspective of user-system interaction, as well as the interrelationships between them. 

Furthermore, these four research questions aim to identify and understand human biases within information retrieval systems, and to measure such biases, for instance, the decoy effect in this study, to guide developers in formulating algorithms to mitigate these biases in practice. This approach is critical for enhancing the reliability and user-centered nature of information retrieval systems, ensuring that they not only return relevant results but also do so in a manner that is cognizant of and resilient to inherent human biases.

\section{User Behavior Datasets and Data Processing}
\label{ch:4}
\subsection{Summary of the User Behavior Datasets}
\label{ch:dataset}
\begin{figure}[t]
    \centering
    \includegraphics[width=.75\linewidth]{
    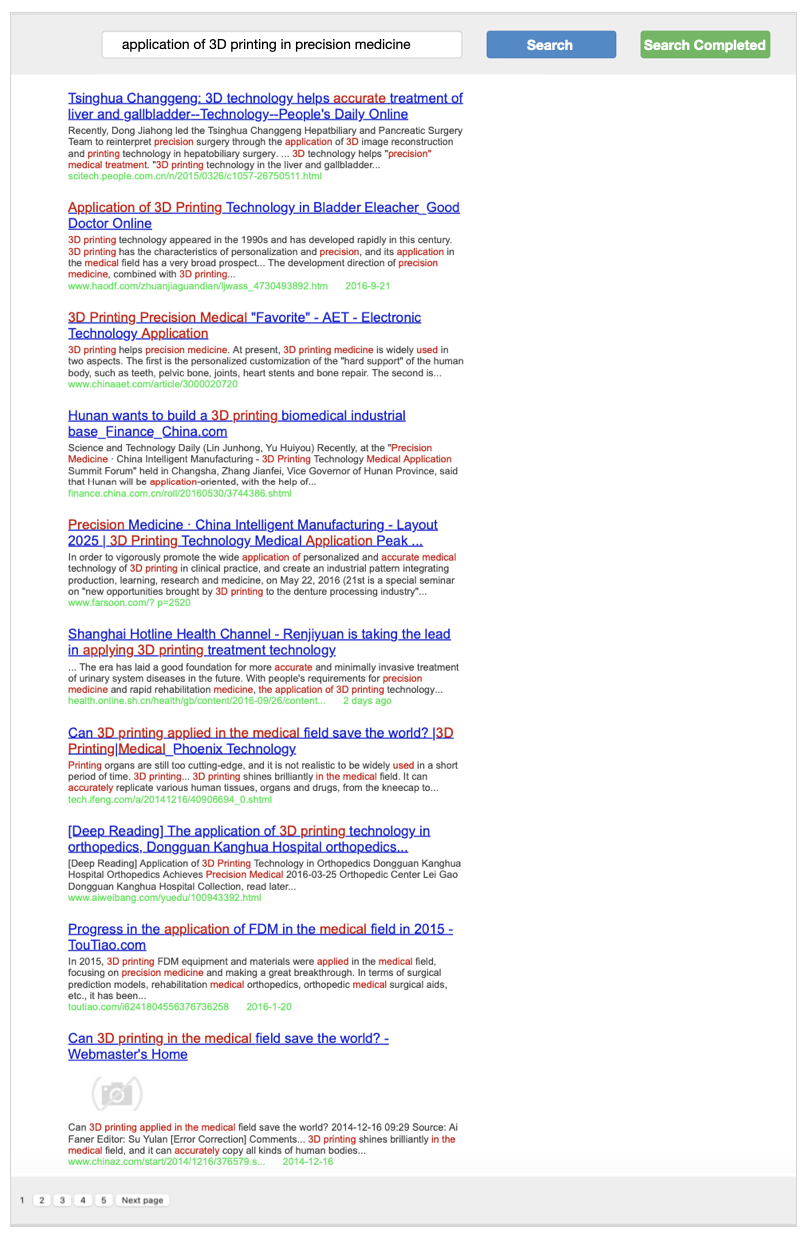
    }
    \caption{\label{fig:serp}
    An example of the SERP interfaces used for collecting user behavior in the user study datasets. Sourced from the THUIR2018 dataset~\cite{Liu2018www}. The original webpage was in Chinese, and the image shows the interface after being machine-translated.}
\end{figure}

In this section, we introduce the user study datasets employed in our first two experiments. All these datasets are publicly available and the data are collected under a controlled laboratory setting. During the process of data collection, participants were instructed to perform several complex search tasks utilizing commercial search engines in Chinese, and the interaction behavior is trailed by a plugin on the browser. Figure~\ref{fig:serp} displays an example of the SERP returned by a commercial search engine in the user studies.

The THUIR2016 dataset~\citep{mao2016} encompasses a total of $9$ topics, $225$ search sessions and $933$ queries, along with the title and snippets on the SERP of each query. It also contains $4$-level (from $1$ (lowest)  to $4$ (highest)) user self-rating usefulness scores for the items they clicked and $5$-level (from $0$ (lowest)  to $4$ (highest)) graded relevance labels  collected from external assessors. 

The THU-KDD dataset~\citep{Liu2019} encompasses a total of $9$ topics, $450$ search sessions and around $1100$ queries, along with the title and snippets on the SERP of each query. It also contains $4$-level (from $0$ (lowest)  to $3$ (highest)) user self-rating usefulness scores  for the items they clicked and $4$-level (from $1$ (lowest)  to $4$ (highest)) graded relevance labels collected from external assessors. In our experiment, we remapped the usefulness scores to a scale from $1$ (lowest) to $4$ (highest) to be consistent with other datasets.

The THUIR2018 dataset~\citep{Liu2018www} encompasses a total of $6$ topics, $166$ search sessions and around $652$ queries, along with the html file of the SERP of each query. It also contains $4$-level (from $1$ (lowest)  to $4$ (highest))  user self-rating usefulness scores for the items they clicked and $4$-level graded relevance labels (from $1$ (lowest)  to $4$ (highest)) collected from external assessors. Additionally, participants were requested to provide an evaluation of the task difficulty and knowledge in the task both before starting tasks and upon completing tasks, utilizing a scale ranging from $1$ (low) to $5$ (high).

Table~\ref{tab:dataset_summary} provides a summary of the three datasets. The detailed information about the datasets can be found in Appendix~\ref{ch:appendix_dataset}.

\begin{table}[htbp]
\centering
\begin{tabular}{p{2.5cm}ccccccc}
\toprule
\textbf{Dataset} & \textbf{Language} &\textbf{\#topics} & \textbf{\#sessions} & \textbf{\#queries} & \textbf{Usefulness Lv.} & \textbf{Rel. Lv.} & \textbf{\#participants}  \\ 
\midrule
THUIR2016~\citep{mao2016} & Chinese & 9 & 225 & 933 & 4 & 5 & 25 \\ 
THU-KDD~\citep{Liu2019} & Chinese  & 9 & 450 & 1, 092 & 4 & 4 & 50\\ 
THUIR2018~\citep{Liu2018www} & Chinese & 6 & 166 & 652 & 4 & 4 & 28 \\
\bottomrule
\end{tabular}
\caption{Summary of datasets used in this study. \textit{Rel.} stands for ``relevance'' and \textit{Lv.} stands for ``level''.}
\label{tab:dataset_summary}
\end{table}

\subsection{Data Processing Flow}
\label{ch:data_processing}

To identify potential decoy instances in user logs, we first propose a definition of a decoy pair. A pair of documents, composed of the \textit{target document} and the \textit{decoy document} $(t, d)$, constitutes a \textit{\textbf{decoy pair}} if and only if the following conditions are met: (1) $t$ and $d$ share certain degree of similarity in content, albeit not identical, i.e., $S_{\min} \leq \text{similarity}(t, d) \leq S_{\max}$, where $S_{\min}$ and $S_{\max}$ respectively represent the minimum and maximum similarity thresholds; (2) $d$ is inferior in quality to $t$, i.e., $\text{quality}(t) > \text{quality}(d)$; (3) the position $t$ and $d$ within a SERP is close enough, i.e., $|rank(t)-rank(d)| \leq \Delta_{\mathrm{rank}}$. In this study, we use the relevance score as a measure of the \textit{quality} of a document. In practice, depending on the scenario, other evaluation methods can also be selected to measure the construct of \textit{quality}, such as the readability and informativeness of the document. In news-related scenarios, other metrics like groundedness and freshness can also be considered. 

Following the definition above, in this experiment, we calculated the cosine similarity between each pair of documents under the same topic and designated $S_{\min}$ as the 99th percentile of document similarity in each dataset. For $S_{\max}$, we set it as $0.95$. In THUIR2016 dataset $S_{\min}$ stands at $0.626$; in THU-KDD dataset $S_{\min}$ stands at $0.594$, and in THUIR2018 dataset .$S_{\min}$ stands at $0.574$. For the second condition, we employ the relevance scores given by external assessors as the measurement of document quality, mandating that $\text{relevance}(t) - \text{relevance}(d) \geq 2$ to ensure that the decoy is substantially inferior to the target. For the third condition, we require the absolute value of the difference of the rank between $t$ and $d$ is smaller than or equal to 5 ($\Delta_{\mathrm{rank}} = 5$). Regarding the window length of 5, our rationale is as follows: In fullscreen mode on commonly used modern screen sizes, approximately 5 results can be displayed on the same screen simultaneously. We processed the top 10 documents in each SERP on all datasets adhering to the aforementioned three conditions, and we identified $982$ records of decoy pairs involving $318$ distinct target documents in the THUIR2016 dataset; $922$ records of decoy pairs involving $376$ distinct target documents in the THU-KDD dataset; and $413$ records of decoy pairs involving $98$ distinct target documents in the THUIR2018 dataset. In the following discourse, we denote the set consisting of all target documents in a corpus as $\mathcal{T}$.

To investigate whether user interactions with documents are disparate when no decoy is present compared to situations with a decoy, we assign some documents not in $\mathcal{T}$ to the control group~(\textit{i.e.,} \textit{control documents}), adhering to the following condition: A document $c$ which is not in the set of target documents~(\textit{i.e.,} $c \notin \mathcal{T}$) is considered a control document if and only if it matches a target document $t \in \mathcal{T}$ such that ~$\text{similarity}(c,t) \geq S_{\mathbf{control}}$ and ~$|\text{relevance}(c) - \text{relevance}(t)| <= 2$. We denote the set of all such $c$ as $\mathcal{C}$, and the set of all $t$ that can match with at least one $c$ as $\mathcal{T}^\prime$, where $ \mathcal{T}^\prime \subset \mathcal{T}$. As mentioned above, we calculated the cosine similarity between each pair of documents under the same topic and designated $S_{\mathbf{control}}$ to the $99.5$th percentile of document similarity in each dataset. In THUIR2016 dataset $S_{\mathbf{control}}$ stands at $0.709$; in THU-KDD dataset $S_{\mathbf{control}}$ stands at $0.676$; and in THUIR2018 dataset, $S_{\mathbf{control}}$ stands at $0.621$. According to the aforementioned condition, we have identified $741$ qualifying control documents in the THUIR2016 dataset, $1790$ in the THU-KDD dataset and $219$ in the THU-KDD dataset. 

\begin{figure}[t]
    \centering
    \includegraphics[width=.99\linewidth]{
    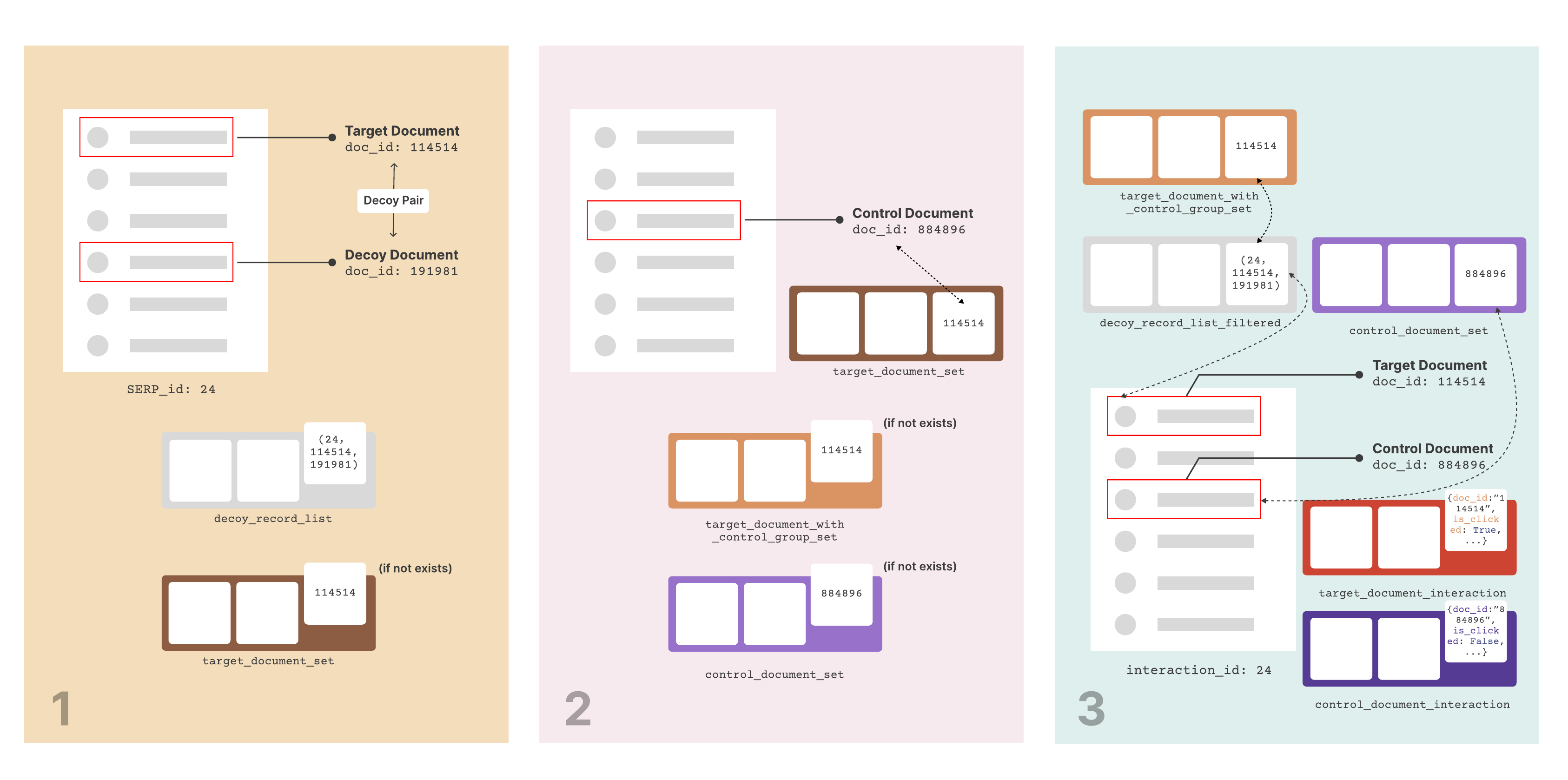
    }
    \caption{\label{fig:data_process} Data processing flow on THUIR2016, THU-KDD and THUIR2018. (1)~First, we filter out decoy pairs from the search results and identify the SERP numbers where these decoy pairs are located. (2)~Next, we determine documents that are similar to the target documents in the decoy pairs but do not have a decoy, serving as the control group. (3)~Finally, we extract user interactions with both the target documents and control documents for further analysis.} 
\end{figure}

We then extracted interaction records of all control documents in the three datasets, obtaining $1384$ records from THUIR2016, $2770$ records from THU-KDD and $548$ records from THUIR2018. Subsequently, from the records of decoy pairs in the three datasets~(with $982$ records from THUIR2016, $922$ from THU-KDD and $413$ from THUIR2018 respectively), we filter out all records where $t \in \mathcal{T}^\prime$, obtaining $768$, $839$ and $219$ records respectively. Note that, for decoy pairs from the same SERP interaction $i$, there could be situations where the same target document corresponds to multiple decoy documents. In our filtering process, we ensure that for a given SERP interaction $i$ and a given target document $t$, only one record is eventually extracted. We concatenate the interaction records of target documents and control documents, ultimately obtaining three document interaction record lists of lengths $2123$ for THUIR2016, $3598$ for THU-KDD and $767$ for THUIR2018 respectively. The three lists of interactions will be employed for the subsequent data analysis. In the subsequent analysis, we process the interaction signals as follows: for documents that have not been clicked, their usefulness score is assigned a value of 0, and their browsing duration is also set to 0. Figure~\ref{fig:data_process} provides a brief outline of our data processing workflow.

\section{Experiment~1:~The Impact of Decoy on Interaction Behaviors}
\label{ch:5}
In order to address \textbf{RQ1}, in this section, we introduce the experiment conducted to investigate whether the presence of a decoy influences the probability of a document being clicked, its duration of browsing, and its percevied usefulness when compared to the conditions where no decoy is present. In the following text, we refer to documents that can be associated with a decoy in the context of SERP as \textit{\textbf{target documents}}, and documents that cannot be associated with a decoy are referred to as \textit{\textbf{control documents}}. For the detailed definition, see Section~\ref{ch:data_processing}.

\subsection{Preliminary Analysis}
\begin{figure*}[htbp]
\centering
\begin{subfigure}[t]{0.32\textwidth}
    \centering
    \includegraphics[width=4.8cm]{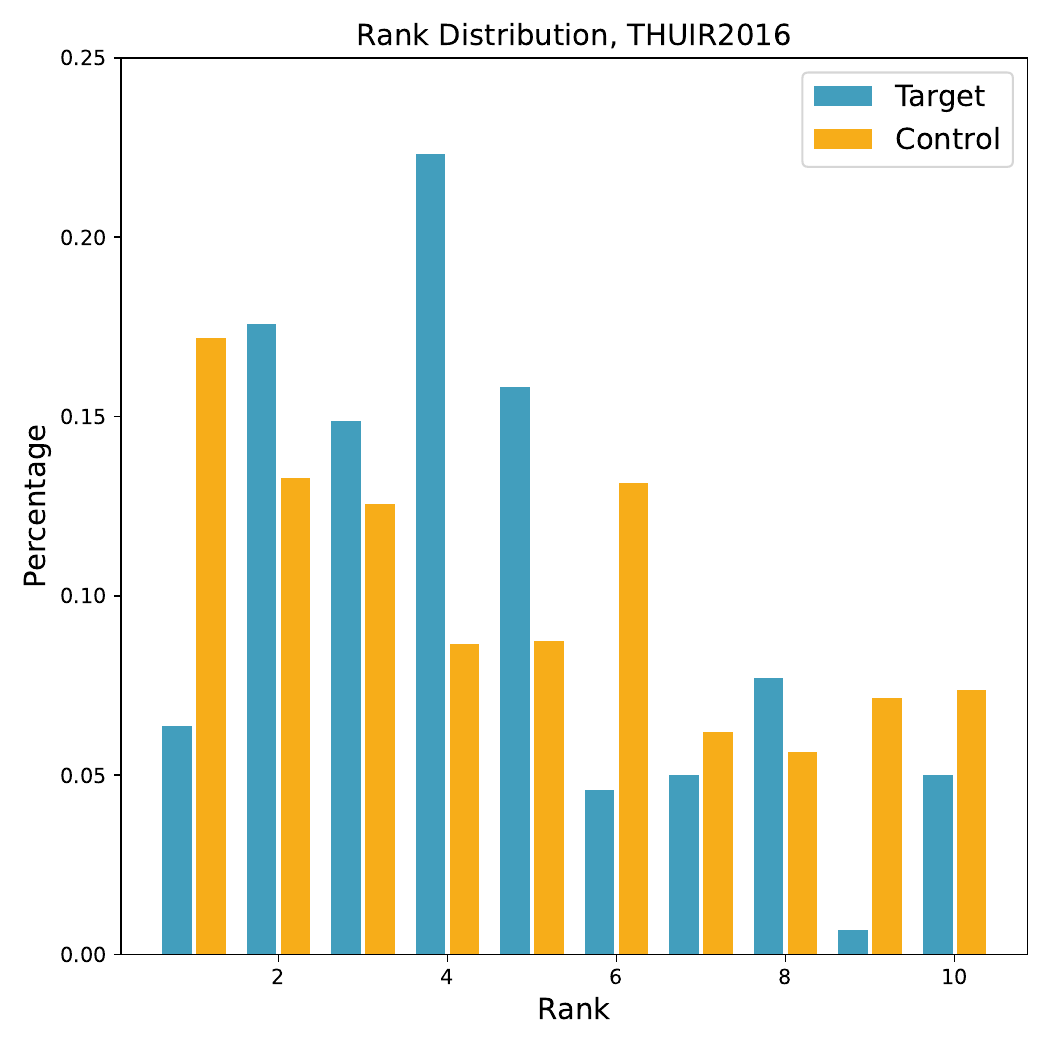}
    \caption*{}
\end{subfigure}
\begin{subfigure}[t]{0.32\textwidth}
    \centering
    \includegraphics[width=4.8cm]{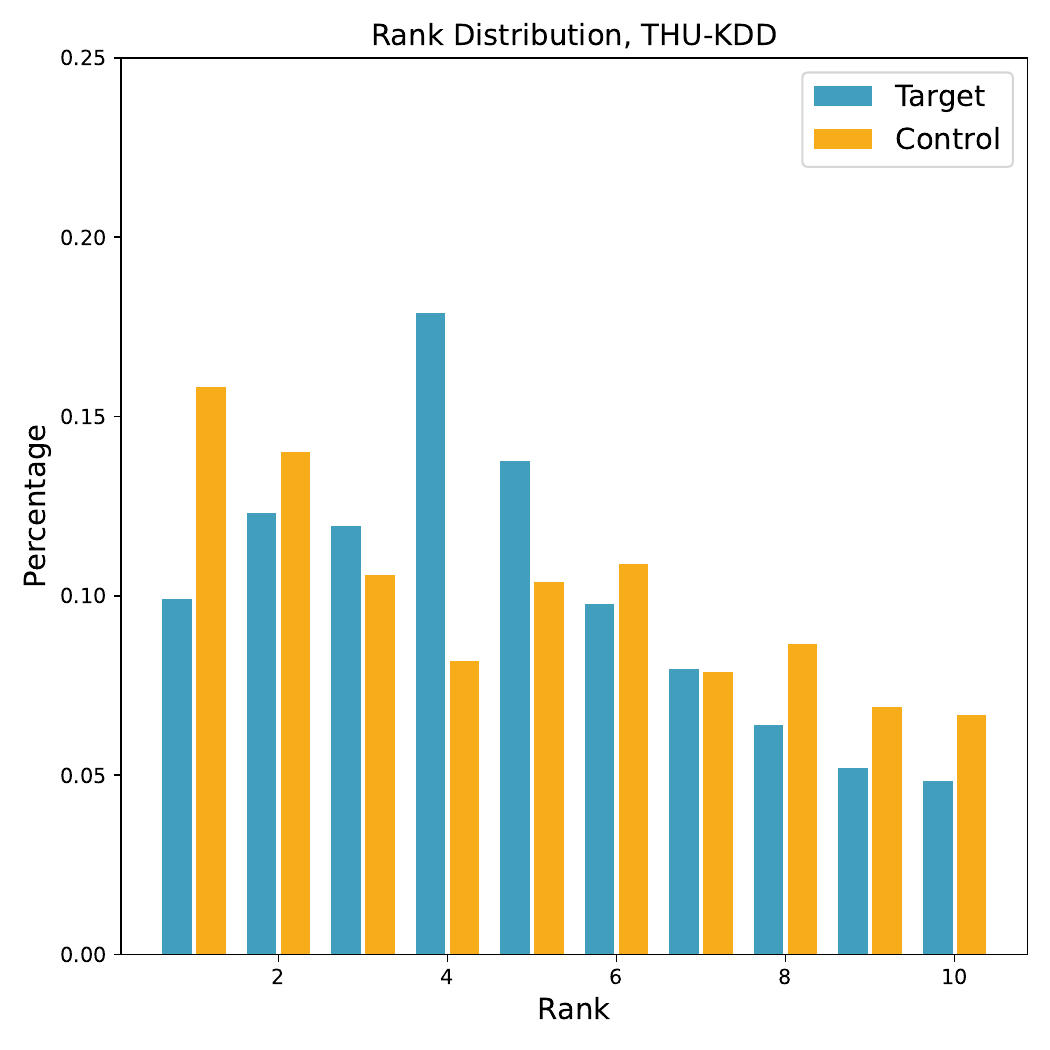}
    \caption*{}
\end{subfigure}
\begin{subfigure}[t]{0.32\textwidth}
    \centering
    \includegraphics[width=4.8cm]{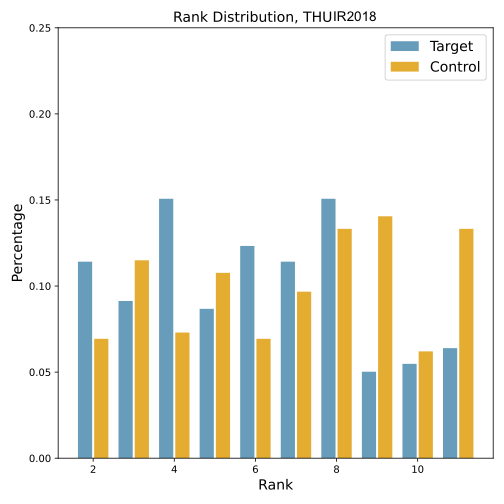}
    \caption*{}
\end{subfigure}
\caption{The distribution of rank for both the target group and control group across THUIR2016~(left), THU-KDD~(middle), and THUIR2018~(right).}
\label{fig:dist}
\end{figure*}

\begin{table}[h!]
\centering
\begin{tabular}{lccccccccc}
\toprule
 & \multicolumn{3}{c}{THUIR2016} & \multicolumn{3}{c}{THU-KDD} & \multicolumn{3}{c}{THUIR2018} \\
 & Target & Control & Sig. Lv. & Target & Control & Sig. Lv. & Target & Control & Sig. Lv. \\
\midrule
Clickthrough prob. & 0.139 & 0.139 & - & 0.227 & 0.207 & - & 0.306 & 0.175 & *** \\
Browsing duration  & 5.323 & 4.677 & - & 8.181 & 6.744 & - & 79.403 & 28.200 & *** \\
Usefulness score   & 0.361 & 0.322 & - & 0.867 & 0.780 & - & 0.598 & 0.237 &  *** \\
\midrule
\# Observations & 982 & 741 &   &  922 & 1790 &  &  413 & 219 &  \\

\bottomrule
\end{tabular}
\caption{The clickthrough probability as well as the mean value of browsing duration and usefulness score across THUIR2016, THU-KDD and THUIR2018. \textit{Sig. Lv.} stands for ``Significance Level''. ***  indicates $ p < 0.001$ under t-test.}
\label{tab:mean_metrics}
\end{table}

Table~\ref{tab:mean_metrics} reports the clickthrough proportion as well as the mean value of browsing duration and usefulness score across THUIR2016, THU-KDD and THUIR2018. From Table~\ref{tab:mean_metrics} one can observe that, apart from the THUIR2018 dataset, the differences between the target group and control group in terms of click probability, browsing duration, and usefulness are not statistically significant under t-test. Nevertheless, we observed inconsistencies in the distribution of document rankings between the target group and control group, which could potentially affect the results presented in 
and Table\ref{tab:mean_metrics}.  Figure~\ref{fig:dist} shows the distribution of rank for both the target group and control group across the THUIR2016, THU-KDD, and THUIR2018. From Figure~\ref{fig:dist} one can observe that, across both datasets, the distribution of target documents and control documents over ranks substantially diverges. Taking into account that the rank position might also influence the click-through rate, browsing duration, and usefulness evaluation of a document, the results presented in 
and Table\ref{tab:mean_metrics} could be subject to rank position bias~\citep{craswell08,Wu2012}. Hence, it is necessary to factor out any latent effects stemming from position biases on our results.

\subsection{Regression Analysis}
To control the impact caused by \textit{rank position}, we employ regression analysis to investigate the relationships between the presence of a decoy and the probability of clicks, browsing duration, and usefulness scores. We constructed three regression models, taking whether the document is clicked (\texttt{is\_clicked}), browsing duration  (\texttt{duration}), and usefulness score (\texttt{usefulness}) as dependent variables respectively, and the presence of a decoy (\texttt{has\_decoy}), the document's rank (\texttt{rank}), task ID (\texttt{task\_id}), and user ID (\texttt{user\_id}) as independent variables. For \texttt{is\_clicked}, we employ Logistic regression, and for \texttt{duration} and \texttt{usefulness}, we resort to Ordinary Least Squares (OLS) regression. Note that since \texttt{usefulness} is actually an ordinal variable in the three datasets, performing logistic regression is another feasible approach. But if we conceptualize \textit{usefulness} as akin to the monetary value a user is willing to pay for a document~\citep{moffat2022batch}, then we can treat usefulness as a continuous variable (although, due to data collection constraints, the datasets only has values of 1.0, 2.0, 3.0, and 4.0), and thus use OLS regression. Equation~\ref{eq:reg} presents the structure of the regression equation we employed. 

\begin{equation}
\label{eq:reg}
    y = \beta + \alpha x_\mathrm{has\_decoy} + \sum_{r=2}^{R} w_r  x_{r} +  \sum_{t=2}^{T} w_t  x_{t} +  \sum_{s=2}^{S} w_s  x_{s}
\end{equation}

In Equation~\ref{eq:reg}, $\beta$ represents the intercept, while $x_\mathrm{has\_decoy}$ denotes \texttt{has\_decoy} and $\alpha$ denotes the regression coefficient for \texttt{has\_decoy}. The term $R$ signifies the deepest document rank position, and in our experiments, we have $R = 10$. The variable $x_{r}$ is a binary indicator, taking the value of 1 if the current document's rank position is $r$, and 0 otherwise; $w_{r}$ is the regression coefficient associated with $x_{r}$. $T$ stands for the number of tasks. In the datasets THUIR2016, THU-KDD, and THUIR2018, $T$ is respectively 9, 9, and 6. The variable $x_{t}$ is another binary indicator, which is set to 1 if the current document originates from the $t$-th task, and 0 otherwise; $w_{t}$ is the regression coefficient for $x_{t}$. $S$ stands for the number of participants. For the datasets THUIR2016, THU-KDD, and THUIR2018, the values of $S$ are 25, 50, and 28, respectively. The variable $x_{s}$ is a binary indicator that takes the value of 1 if the behavioral signals for the current document were collected from the $s$-th participant, and 0 otherwise; and $w_{s}$ is its associated regression coefficient. In the regression equation, the values for $r$, $t$, and $s$ start from 2 to circumvent the issue of multicollinearity. Finally, $y$ represents the dependent variable in the regression equation. For the click behavior, given that we employed Logistic regression, the relation is given by $y = 1 /(1 + \exp^{-\texttt{is\_clicked}})$; for browsing duration and usefulness assessment, $y$ is equated to \texttt{duration} and \texttt{usefulness} respectively.

Note that unlike common practices in computer science, in econometrics, regression models are predominantly employed for interpretation rather than for prediction. In a multiple regression model, each coefficient tells people the impact on the dependent variable of a one-unit change in that independent variable, holding all other independent variables constant~\citep{SMITH2012}. In this study, we focus on elucidating how, and to what extent, the presence of a \textit{decoy} influences the likelihood of a document being clicked, the browsing duration on it, and its perceived usefulness. Hence, we do not partition the dataset into training and test subsets; instead, we perform regression on the entirety of the data. Including \texttt{rank}, \texttt{task\_id}, and \texttt{user\_id} as independent variables serves to use them as \textit{control variables} to mitigate the potential influences from rank position, task type, and individual characteristics on the outcomes, thus better elucidating how variations in \texttt{has\_decoy} would affect the values of \texttt{is\_clicked}, \texttt{duration}, and \texttt{usefulness}. 

\begin{table}[htbp]
\centering
\begin{tabular}{cccc}
\toprule
  & THUIR2016 & THU-KDD  & THUIR2018 \\
\midrule
 \texttt{is\_clicked} & 0.363* & 0.217* & 0.879***\\
 \texttt{duration}  &  1.916 & 1.913 & 51.521*** \\
 \texttt{usefulness}  & 0.136 ** & 0.156 ** & 0.358*** \\
 \midrule
 \# Observations & 2123 & 3598 & 767\\
\bottomrule
\end{tabular}
\caption{The regression coefficient ($\alpha$) of the independent variable \texttt{has\_decoy} with the dependent variables \texttt{is\_clicked}, \texttt{duration}, and \texttt{usefulness} on THUIR2016, THU-KDD and THUIR2018. *, ** and *** respectively indicate  \( p < 0.05 \), \(  p < 0.01 \), and \( p < 0.001 \).}
\label{table:reg}
\end{table}

Table~\ref{table:reg} shows the regression coefficient ($\alpha$) of the independent variable \texttt{has\_decoy} ($h$) with the dependent variables \texttt{is\_clicked}, \texttt{duration} and \texttt{usefulness}. As previously stated,   our focus in this research is to elucidate in what manner and to what extent the presence of a decoy impacts whether a document is clicked, the browsing duration, and the usefulness scores, and \texttt{rank}, \texttt{task\_id}, and \texttt{user\_id} are included merely to control for the effects brought by rank position, task, and individual characteristic respectively. Therefore, to maintain brevity in the main text, we omit the reporting of the constant as well as the regression coefficients of \texttt{rank}, \texttt{task\_id}, and \texttt{user\_id} in the above tables. The complete regression results are reported in Appendix~\ref{ch:appendix_regression1}.

From Table~\ref{table:reg}, one can observe that: across all datasets, the presence of a \textit{decoy} could exert a positive influence on the likelihood of being clicked (coefficient = $0.363$, $0.217$ and  $0.879$ respectively) and on the usefulness score (coefficient = $0.136$, $0.156$  and $0.358$ respectively), all with a statistical significance at the level of $p < 0.05$. The existence of a \textit{decoy} also seems to exert a positive impact on duration (coefficient = $1.916$, $1.913$ and $51.521$ respectively), but this result is only statistically significant in the THUIR2018 dataset. This result indicates that, given the document rank, type of task, and individual characteristics, when a \textit{decoy} is present, in comparison to the case when the decoy is absent, the likelihood of a document~(the \textbf{target}) being clicked would increase; the usefulness score of the target perceived by the user would be elevated. 


\section{Experiment~2:~The Impact of Task Difficulty and User Knoweldge on Decoy Effect}
\label{ch:6}
\begin{figure*}[htbp]
\centering
\begin{subfigure}[t]{0.75\textwidth}
    \centering
    \includegraphics[width=10cm]{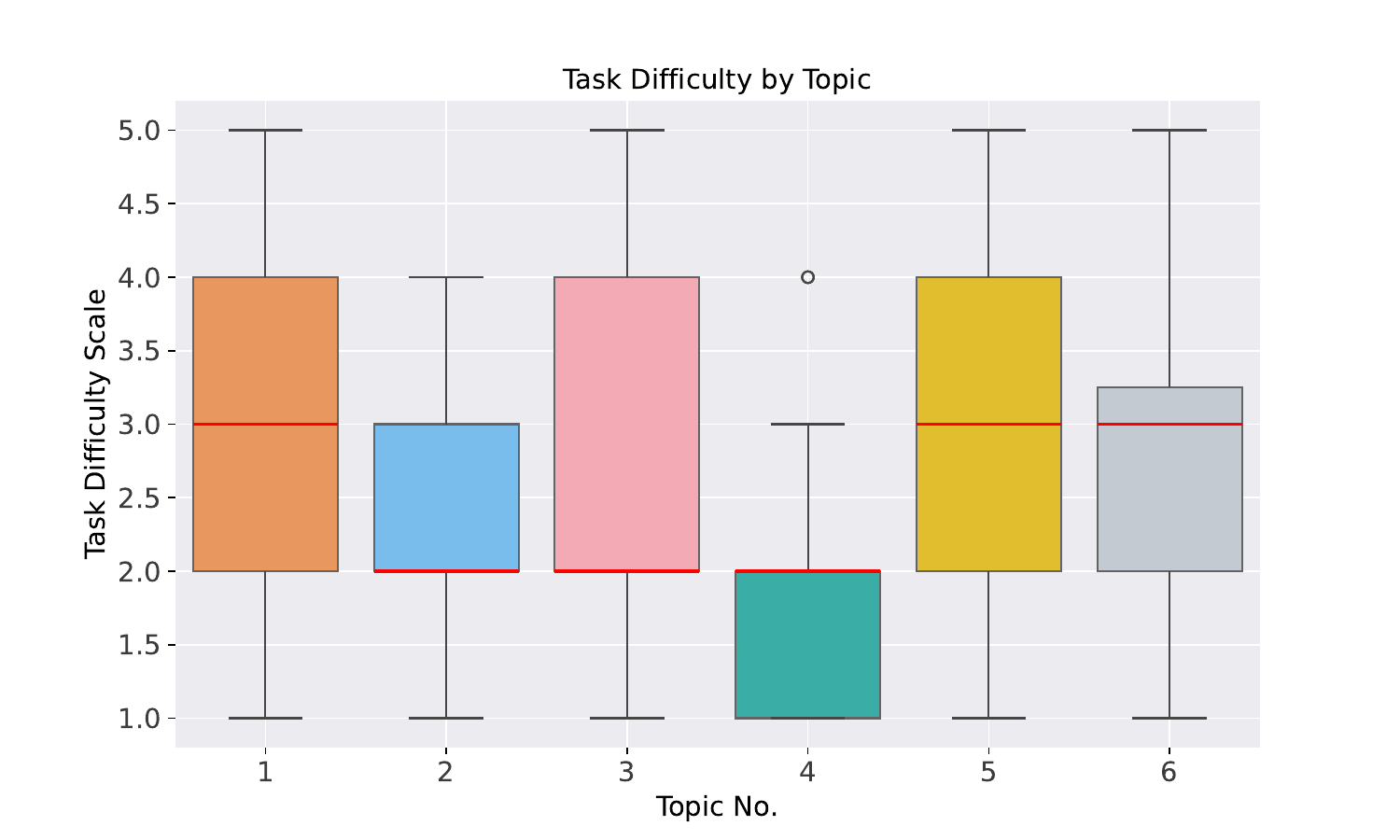}
    \caption*{}
\end{subfigure}
\begin{subfigure}[t]{0.75\textwidth}
    \centering
    \includegraphics[width=10cm]{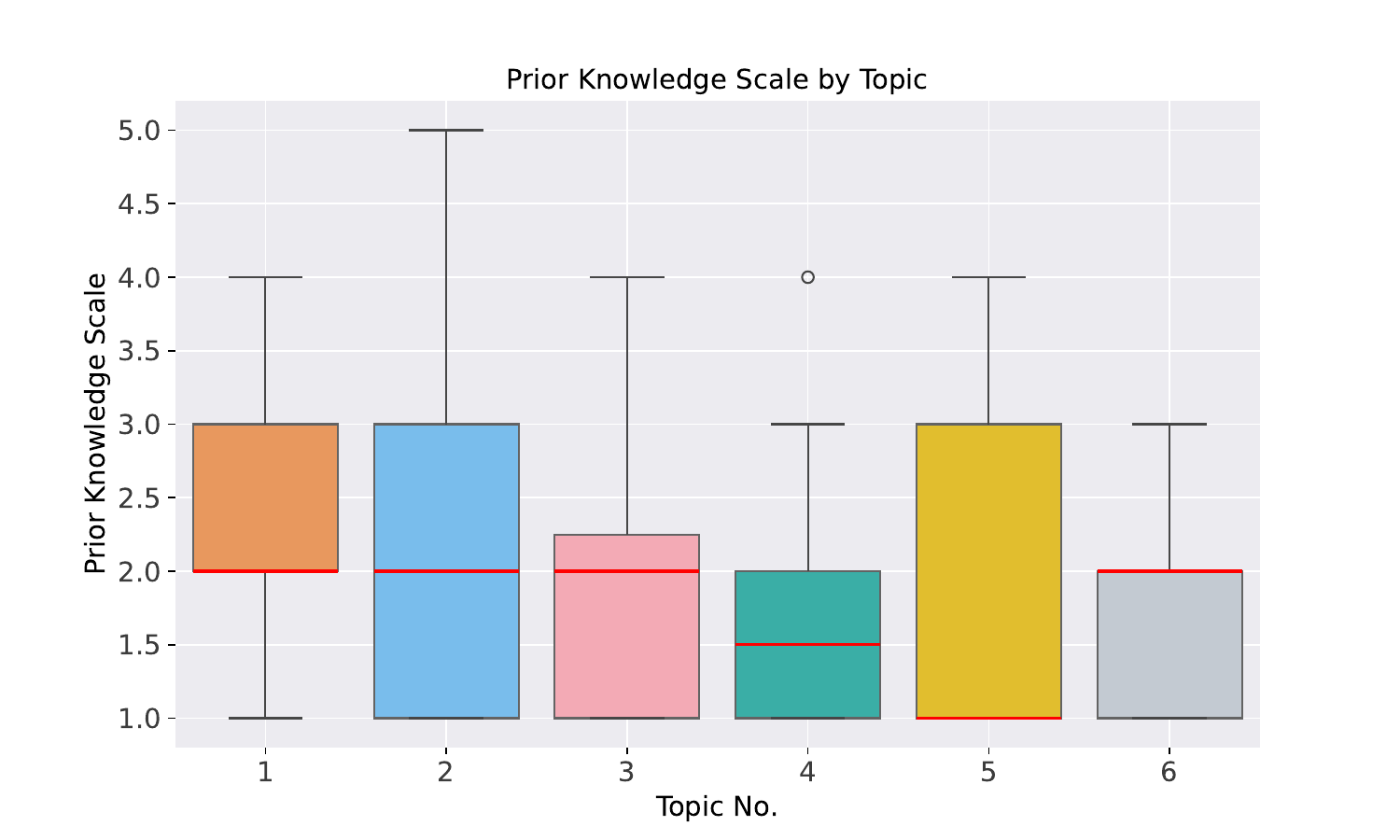}
    \caption*{}
\end{subfigure}
\caption{The distribution of task difficulty and prior knowledge scale reported by users across different topics in the THUIR2018 dataset. The red line represents the median.}
\label{fig:diff_kn_by_topic}
\end{figure*}

Through Experiment 1, we found that, when a document has a decoy, users are more likely to click on the document and tend to give it a higher usefulness score. This indicates the potential impact of cognitive biases on user interaction behavior.  Additionaly, previous studies in the field of interactive IR have demonstrated that, the nature of a user's task could impact on their information-seeking behaviors~\citep{Kim2005, Liu2010}. In more specific terms, specific task natures such as task type, task complexity and difficulty, as well as the stage of the task, have been identified as influential factors affecting users' assessments of usefulness and relevance~\citep{bystrom1995task,bystrom2002information, white2006,liu2010b}. On the other hand, previous empirical studies in interactive IR have shown that users' prior knowledge of the topic may influence their search behavior, such as the dwell time during searches~\citep{Liu2010,liu2010b,mao2018tois}, or decisions on what results to click or skip~\citep{Sanchiz2017, Monchaux2015}. Previous work has also shown that, due to the inherent nature of cognitive biases, certain individuals are more susceptible or more likely to be influenced by biased judgments arising from interaction contexts~(\textit{e.g.,} cognitive load, domain knowledge)~\citep{liao2013,boon2023} and individual characteristics~\citep{PENNYCOOK2019}. Consequently, building upon Experiment 1 and previous work, we propose two additional research questions:
\begin{itemize}
    \item Does the extent to which the decoy effect influences user behavior vary according to the nature of the task, such as its difficulty? If so, how does the decoy effect influence user behavior differently across varying levels of task difficulty~(\textbf{RQ2})? 
    \item Does the impact of the decoy effect on user behavior relate to the user’s existing level of knowledge about the search topic? If so, does the influence of the decoy effect on user behavior vary at different levels of existing knowledge~(\textbf{RQ3})? 
\end{itemize}

To address \textbf{RQ2} and \textbf{RQ3}, we conducted an additional analysis on the THUIR2018 dataset. As described in Section~\ref{ch:dataset}, the THUIR2018 dataset includes user-reported levels of task difficulty and prior knowledge about the task, both before starting and after completing the task. Figure~\ref{fig:diff_kn_by_topic} shows the distribution of task difficulty and prior knowledge scale scores across different topics and Table~\ref{tab:task_difficulty} presents the mean values of user-reported post-task difficulties for each task and the mean values of user-reported prior-task knowledge on each task. We designate the three tasks with the highest average post-task difficulty as ``high difficulty'' and  the three tasks with the lowest average prior-task knowledge as ``low knowledge'. According to this criterion, a total of $315$ records out of $767$ records in the THUIR2018 dataset are from tasks of high difficulty, accounting for approximately $41.07\%$ of the total; a total of $306$ records out of $767$ records in the THUIR2018 dataset are from tasks of low knowledge, accounting for approximately $39.9\%$ of the total. For the above binary classification method, we provide the following rationale. First, dividing the data into binary categories helps to simplify the analytical model, making the results more intuitive and easier to interpret while reducing the complexity of the model. This approach also avoids potentially unnecessary complex comparisons between multiple categories.
Second, binary classification can enhance the statistical significance of the variables. Multiple categories may result in insufficient sample sizes for each category, thereby weakening the power of statistical tests. 

Subsequently, based on Equation~\ref{eq:reg}, we introduced the following new variables:~ \texttt{Is\_High\_Difficulty~(IHD)}, \texttt{Has\_Decoy\_is\_High\_Difficulty~(HDHD)}, \texttt{Is\_Low\_Knowledge~(ILK)}, \texttt{Has\_Decoy\_is\_Low Knowledge~(HDLK)}.

These variables are all binary variable. \texttt{IHD} is set to $1$ if the document is from a task designated as ``high difficulty'', otherwise it is set to $0$. \texttt{HDHD} is an interaction term of \texttt{has decoy} and \texttt{IHD}. If a document has a decoy AND is from a task designated as ``high difficulty'', then \texttt{HDHD} is set to 1; otherwise, it is set to 0. \texttt{ILK} is set to $1$ if the document is from a task designated as ``low knowledge'', otherwise it is set to $0$. \texttt{HDLK} is an interaction term of \texttt{has\_decoy} and \texttt{ILK}. If a document has a decoy AND is from a task designated as ``low knowledge'', then \texttt{HDLK} is set to 1; otherwise, it is set to 0. 
The new regression model is presented as follows.
\begin{equation}
\label{eq:reg2}
\begin{aligned}
    y = & \ \beta + \alpha_\mathrm{has\_decoy} x_\mathrm{has\_decoy} + \alpha_\mathrm{HDHD}  x_\mathrm{HDHD} 
        + \alpha_\mathrm{ILK} x_\mathrm{ILK} + \alpha_\mathrm{HDLK} x_\mathrm{HDLK} + \sum_{r=2}^{R} w_r  x_{r} + \sum_{t=2}^{T} w_t x_{t} 
        + \sum_{s=2}^{S} w_s  x_{s}
\end{aligned}
\end{equation}

In Equation~\ref{eq:reg2}, 
$x_\mathrm{has\_decoy}$ represents \texttt{has\_decoy} and $\alpha_\mathrm{has\_decoy}$ represents the regression coefficient for it; $x_\mathrm{HDHD}$ represents \texttt{HDHD} and $\alpha_\mathrm{HDHD}$  represents the regression coefficient for it; $x_\mathrm{ILK}$ represents\texttt{ILK} and $\alpha_\mathrm{ILK}$ represents the regression coefficient for it; $x_\mathrm{HDLK}$ represents \texttt{HDLK} and $\alpha_\mathrm{HDLK}$  represents the regression coefficient for it. The definition of other symbols is consistent with Equation~\ref{eq:reg}. Note that the variable \texttt{IHD} was not included in the regression model to avoid the issue of multicollinearity, as \texttt{IHD} can be obtained through a linear combination of $x_t$.

Table~\ref{table:reg2} shows the regression coefficients $\alpha_1$ and $\alpha_2$ with the dependent variables \texttt{is\_clicked}, \texttt{duration} and \texttt{usefulness}. The complete regression results are reported in Appendix~\ref{ch:appendix_regression2}. From Table~\ref{table:reg2} it can be observed that:~(1)~Under the condition that other variables are held constant, when the task difficulty is relatively low (\texttt{is\_high\_difficulty} = 0), compared to the absence of a decoy (\texttt{has\_decoy} = 0), the presence of a decoy (\texttt{has\_decoy\_high\_difficulty} = 0, \texttt{has\_decoy} = 1) results in a substantial increase in the likelihood of the target document being clicked (coefficient = $1.304$, $p < 0.001$), the time spent on the target document (coefficient = $89.96$, $p < 0.001$), and the usefulness rating (coefficient = $0.59$, $p < 0.001$).~(2)~Conversely, under the condition that other variables are held constant, when the task difficulty is relatively high (\texttt{is\_high\_difficulty} = 1), compared to the absence of a decoy (\texttt{has\_decoy} = 0), the presence of a decoy (\texttt{has\_decoy\_high\_difficulty} = 1, \texttt{has\_decoy} = 1) only results in a $12.0\%$ increase in the likelihood of the target document being clicked, a $6.5$ seconds increase in the time spent on the target document, and a $0.085$ increase in the usefulness rating for the target document. In other words, considering the presence of a decoy, compared to scenarios with low task difficulty, when the task difficulty is relatively high, the likelihood of the target document being clicked has a substantial decrease (coefficient = $118.4\%$, $p < 0.001$); the duration spent on the target document (coefficient = $83.5$, $p < 0.001$) and the usefulness rating for the target document also substantially decreases (coefficient = $0.505$, $p < 0.001$).~(3)~Under the condition of keeping other factors constant, when users have lower prior knowledge of the search topic, compared to the scenario without decoys, users tend to assign higher usefulness scores to a document when it contains decoys (coefficient = $0.275$, $p < 0.05$).

For \textbf{RQ2} and \textbf{RQ3}, our result suggests that the impact of the decoy effect on user behavior varies under different task difficulties and different prior knowledge levels. In summary, when the task difficulty is lower or the user's prior knowledge level is lower, the decoy effect tends to have a greater impact on user interaction behaviors, such as clicks, browsing duration, and usefulness evaluations. Compared to more challenging tasks, the decoy effect exerts a greater influence in simpler search tasks, manifesting as users being more likely to click on target documents, spending more time on target documents, and potentially assigning higher usefulness ratings to target documents. The impact of the decoy effect on user behavior is relatively small when the search task is of high difficulty. When users have a lower level of prior knowledge, the decoy effect tends to have a greater impact on their behavior, manifesting as the assignment of higher usefulness scores to a document that includes decoys.

A possible explanation for this observation can be drawn from the field of psychology and cognitive science. Based on the \textit{dual process} theory~\citep{Kahneman11,jonathan2013, Wason1974-WASDPI}, there are two conceptual systems within the human brain: System 1, which is automatic, fast, and intuitive; and System 2, which is conscious, slow, and analytical. When System 1 dominates thinking, it can lead to faster decision-making, albeit potentially error-prone. In contrast, when System 2 thinking is engaged, it is typically more reliable but requires more cognitive effort. Some researchers (\textit{e.g.,}~\citet{jonathan2013}) argue that System 1 is assumed to produce default responses unless these are overridden by distinctive higher-order reasoning processes associated with System 2. Literature from the field of pyschology also suggests that prior knowledge can be associated with different information selection processes~\citep{SCHWIND20122280}, which may also affect users' behavior when interacting with search systems.

Thus, a possible explanation for what we observed in Experiment 2 is that: (1)~When individuals are tasked with more challenging search tasks, users may need to expend more cognitive effort to memorize, understand and analyze information from search results~\cite{Campbell1988, capra2017, Kelly2015}, which requires them to engage more intensively with System 2 cognitive processes. Therefore, they tend to utilize System 2, subsequently mitigating the impact of the decoy effect; (2)~When users have a lower level of prior knowledge (i.e., they are less familiar with the search topic), users may tend to take cognitive shortcuts and rely more on System 1 in scenarios where they need to process a large amount of information quickly to make decisions. As prior knowledge can be associated with different information selection processes~\citep{SCHWIND20122280}, users with low prior knowledge mayo insufficiently select or evaluate the collected information, thereby making them more susceptible to the decoy effect. It is worth noting that, in addition to the failure to adequately collect and evaluate information, there may be other factors influencing the strength of the decoy effect, such as the user's cognitive abilities. Due to the limitations of the dataset, this aspect cannot be tested in our experiment. Future research could examine how different types of cognitive abilities influence the extent to which users are affected by the decoy effect during interactions with web pages.

\begin{table}[htbp]
\centering
\begin{tabular}{c c c c c c c}
\toprule
Task ID & 1 & 2 & 3 & 4 & 5 & 6 \\
\midrule
Average Post-task Difficulty & \textbf{3.0} & 2.52 & 2.68 & 2.04 & \textbf{3.04} & \textbf{2.89} \\
Average Prior Knowledge & {2.33} & {2.15} & 2.11 & \textit{1.64} & \textit{1.75} & \textit{1.75} \\
\bottomrule
\end{tabular}
\caption{Average post-task difficulty for each task on the THUIR2018 dataset. Bold font indicates a task that is classified as ``high difficulty'' and italic text indicating tasks classified as ``low knowledge''.}
\label{tab:task_difficulty}
\end{table}
\begin{table}[htbp]
\centering
\begin{tabular}{cccc}
\toprule
 & $\alpha_\mathbf{HD}$ & $\alpha_\mathbf{HDHD}$ & $\alpha_\mathbf{HDLK}$ \\
\midrule
 \texttt{is\_clicked} & 1.064** & -1.156*  & 0.698\\
 \texttt{duration}  & 78.890*** & -84.899***  & 31.535\\
 \texttt{usefulness}  & 0.494 *** & -0.519 ***  & 0.275* \\
  \midrule
\# Observations & \multicolumn{3}{c}{767}\\
\bottomrule
\end{tabular}
\caption{The regression coefficients of the independent variable \texttt{has\_decoy}~($\alpha_\mathbf{HD}$), \texttt{has\_decoy\_high\_difficulty}~($\alpha_\mathbf{HDHD}$), and \texttt{has\_decoy\_low\_knowledge}~($\alpha_\mathbf{HDLK}$) with the dependent variables \texttt{is\_clicked}, \texttt{duration}, and \texttt{usefulness} on THUIR2018. *, ** and *** respectively indicate  \( p < 0.05 \), \(  p < 0.01 \), and \( p < 0.001 \).}
\label{table:reg2}
\end{table}

\section{Experiment~3:~Vulneratbility of Retrieval Models to Decoy Effect in Ranking}
\label{ch:7}
With the widespread adoption of search systems, people have increasingly come to rely on information retrieved from search systems for making decisions, including important life decisions such as medical, political and financial choices. However, as previously discussed, users are often influenced by various cognitive biases~(\textit{e. g.,} the decoy effect) when interacting with search systems, leading to decisions that deviate from optimal outcomes. In critical contexts such as medical diagnosis, criminal judgments or information consumption, cognitive biases can even result in dangerous decisions and have negative societal consequences. The results returned by search systems have the potential to magnify and exacerbate users' cognitive biases~\citep{baya2018, cho2004}. Building on the findings of Experiments 1 and 2, which primarily explore the influence of the decoy effect on user behaviors, it becomes crucial to shift our focus towards the search systems themselves. Understanding the extent to which the results returned by search systems may lead users into the influence of the decoy effect, and designing a reasonable metric to measure such vulnerability of search systems, becomes particularly important, as previous studies have almost neglected these tasks.

Therefore, we come up with the following research question~(\textbf{RQ4}): How to measure the vulnerability of information retrieval systems to the decoy effect? In commonly used sparse and dense retrieval models, which models can achieve higher effectiveness while having lower vulnerability to the decoy effect?

In this section, we utilize the Microsoft Machine Reading Comprehension (MS MARCO)~\citep{bajaj2018ms} dataset to investigate the effectiveness of various retrievers and their vulnerability to the decoy effect at different cutoff depths $k$. We investigate these aspects across 97 topics of the passage retrieval task from the Text REtrieval Conference~(TREC) 2019 Deep Learning~(DL)  Track collection~\citep{craswell2020overview} and the TREC 2020 DL Track collection~\citep{craswell2021overview}. To measure the effectiveness of a retrieval model, we use nDCG@$k$ and Recall@$k$ as the evaluation metrics. To measure a retriever's vulnerability to the decoy effect, we first analyze the relationship between effectiveness and the number of decoy pairs, followed by introducing a heuristic metric for measuring the vulnerability of retrieval models to decoy effect in result presentation.

\subsection{Summary of The Dataset and Collections}
MS MARCO~\citep{bajaj2018ms} comprises an extensive dataset collection tailored for deep learning applications in the realm of information retrieval. In our experiment, we use MS MARCO Version 1. The dataset consists of $1,010,916$ anonymized questions obtained from Bing's search query logs, $8,841,823$ passages extracted from $3,563,535$ web documents retrieved by Microsoft Bing. 

The TREC DL Track aims to investigate information retrieval in the context of large-scale training data, with the objective to make large-scale datasets publicly available for deep learning-based information retrieval methods and to provide a standardized grounding for comparing various information retrieval approaches.  
In our experiment, we selected $43$ and $54$ topics respectively from the passage retrieval tasks of the TREC 19 DL and TREC 20 DL tracks, all of which are accompanied by relevance (qrel) annotations. Passages were evaluated using a four-point relevance scale: Not Relevant (0), Related (1), Highly Relevant (2), and Perfect (3). It is important to note that the ``Related'' rating, despite its name, indicates that while a passage pertains to the same general topic, it fails to directly answer the question. 

\subsection{Experimental Setting}
\label{subsec:exp3setting}
\begin{figure}[t]
    \centering
    \includegraphics[width=.95\linewidth]{
    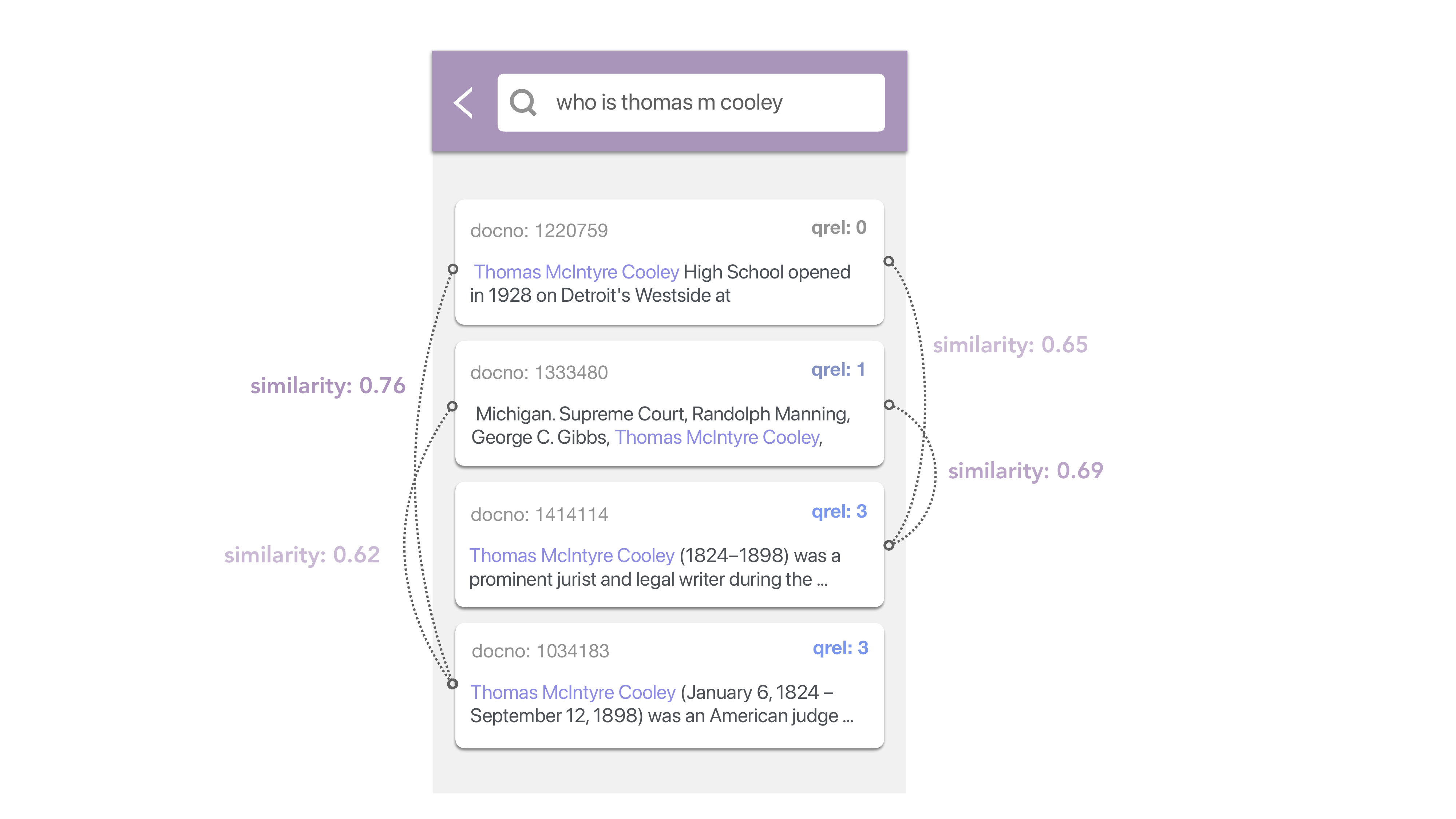
    }
    \caption{\label{fig:decoy_pair_example}A case where a target document can correspond to multiple potential decoy documents. For example, both 1220759 and 1333480 can be considered as decoys for 1414114, and they can also be considered as decoys for 1034183. Therefore, there can be up to 4 potential decoy pairs. But when calculating, we only consider one decoy pair for each target document under each topic. The document with the highest similarity to the target document is considered a decoy. Therefore, in this example, there are only two decoy pairs, (1034183, 1220759) and (1414114, 1333480), respectively, will be identified.}
\end{figure}

Our experiment adopted several retrievers based on sparse or dense vector representations for comparison: (1)~BM25~\citep{robertson94}, (2)~SPLADE++~\citep{formal22}, (3)~uniCOIL~\citep{lin2021brief}, (4)~ANCE~\citep{xiong2020approximate}, (5)~TCT-ColBERT~\citep{lin-etal-2021-batch}, (6)~SBERT~\citep{reimers-gurevych-2019-sentence}. The main rationale for selecting the above retrievers is that they are either commonly used text retrieval models (\textit{e.g.,} BM25, ANCE and SBERT) or variants of commonly used text retrieval models (\textit{e.g.,} SPLADE~\cite{Formal_2021}, COIL~\cite{gao-etal-2021-coil}, ColBERT~\cite{lin-etal-2021-batch}). Another reason for selecting the above retrievers is that \texttt{pyserini}~\citep{Lin_etal_SIGIR2021_Pyserini} toolkit offers pre-built indexes of them on MS MARCO v1 with predefined parameters or weights, which facilitates the comparison of our experimental result with those of previous work~(\textit{e.g.,} \cite{lin2021brief, lin-etal-2021-batch}). Since the focus of our work is on evaluating the results returned by these retrievers rather than improving the rankers themselves, we offer a brief overview of these models in the following part. For detailed information about the models, please refer to the original papers. 

\begin{itemize}
    \item \textbf{BM25}~\citep{robertson94}: BM25 is a probabilistic information retrieval model based on sparse vector representation. It computes the document scores by considering the \textit{term frequency}~(TF) and \textit{inverse document frequency}~(IDF) of each term. Specifically, BM25 adjusts the importance of each term during the computation process, controlling the impact of term frequency and document length through parameters such as $k1$, $k2$, and $b$.
    \item \textbf{SPLADE++}~\citep{formal22}: The SPLADE model~\citep{Formal_2021} is a retrieval approach that employs high-dimensional sparse vector representation. It integrates the Masked Language Modeling (MLM) head of a pre-trained language model (PLM) with sparse regularization, enabling joint term expansion and re-weighting. In our experiments, we utilized the enhanced version of SPLADE, named SPLADE++, which incorporates modifications to the pooling mechanism and introduces models trained with distillation.
    \item \textbf{uniCOIL}~\citep{lin2021brief}: Contextualized Lexical Retriever~(COIL)~\citep{gao-etal-2021-coil} is a retrieval approach based on sparse vector representation. The scoring of COIL is based on vector similarities between query-document overlapping term contextualized representations. UniCOIL~\citep{lin2021brief} is a variant of COIL that reduces the token dimension of COIL to 1 and is directly compatible with inverted indexes.
    \item \textbf{ANCE}~\citep{xiong2020approximate}: ANCE is a method applicable for training dense retrieval model. It leverages asynchronous updates of Approximate Nearest Neighbor (ANN)~\citep{johnson2017billionscale} indices to globally select hard negatives from the entire corpus for training, which is concurrently updated with the learning process to select more representative negative training instances.
    \item \textbf{TCT-ColBERT}~\citep{lin-etal-2021-batch}: ColBERT~\citep{khattab20} is a multi-vector dense retriever which uses the delayed interaction approach with a dual-encoder design. It encodes queries and documents separately with BERT~\citep{devlin-etal-2019-bert} and efficiently calculating their similarity. This allows it to leverage the outstanding natural language understanding ability of deep language models while speeding up query processing. Tightly-Coupled Teacher ColBERT~(TCT-ColBERT)~\citep{lin-etal-2021-batch} employs knowledge distillation to accelerate the ColBERT retriever and is claimed to approach the performance of ColBERT while significantly improving speed by several orders of magnitude. 
    \item \textbf{SBERT}~\citep{reimers-gurevych-2019-sentence}: Sentence-BERT (SBERT) is a modification of the BERT network. It utilizes siamese and triplet network structures to generate semantically meaningful sentence embeddings in fixed-sized vectors for input sentences, enabling the sentence to be compared using cosine similarity. The retriever based on SBERT maintains the accuracy of BERT in determining text similarity while having an improved computing efficiency. 
\end{itemize}

In our experiments, we utilize the pre-built indexes provided by the \texttt{pyserini} toolkit to evaluate the above-mentioned retrievers~(without extra expansions) on a total of $86$ topics, consisting of $43$ from the TREC 2019 DL and $54$ from the TREC 2020 DL passage retrieval tasks\footnote{Regarding the parameter settings for each retriever, please refer to: https://castorini.github.io/pyserini/2cr/msmarco-v1-passage.html}. For each topic, the retrievers return a ranked list of the top 1000 passages. 

Similar to the setup in Section~\ref{ch:data_processing}, we define that a pair of documents~(passages)\footnote{Due to convention, we still use the term ``document" here, but to be more precise, the text passages used in this experiment should be referred to as ``passages". Unless specifically distinguished, within this section, we consider ``document" and ``passage" as synonyms.}, composed of the target document and the decoy document $(t, d)$, constitutes a decoy pair if and only if the following conditions are met:~(1)~The similarity between target document $t$ and decoy document $d$ measured by cosine similarity is greater than or equal to $0.6$ and less than $0.95$;~(2)~The qrel score for $t$ is greater than or equal to 2, and the qrel score for $d$ is less than or equal to 1.~(3)~The absolute value of the rank distance between $t$ and $d$ is less than or equal to 5. Note that here condition (2) differs slightly from Section~\ref{ch:data_processing}. The rationale for it is as follows:~since TREC 19 DL and TREC 20 DL provide more detailed relevance annotation standards for each qrel level, it can be considered that there is a substantial difference in quality between passages with qrel scores of 0 or 1 and passages with qrel scores of 2 or 3. Furthermore, as shown in Figure~\ref{fig:decoy_pair_example}, when calculating decoy pairs for each topic, there may be cases where one target document corresponds to multiple potential decoy documents. In our calculation, we consider only the
document with the highest similarity to the target document as the decoy. Therefore, there is only one decoy pair for each target document under each topic.

Following the methodology mentioned above, we calculate (1)~the average number of decoy pairs, (2)~the recall score and (3)~the nDCG score for each retriever's output on each topic, with the cutoff~($k$) ranging from 10 to 1000 incrementally by 10.

\subsection{Experimental Result and Analysis}
\begin{figure*}[htbp]
\centering
\begin{subfigure}[t]{0.32\textwidth}
    \centering
    \includegraphics[width=5cm]{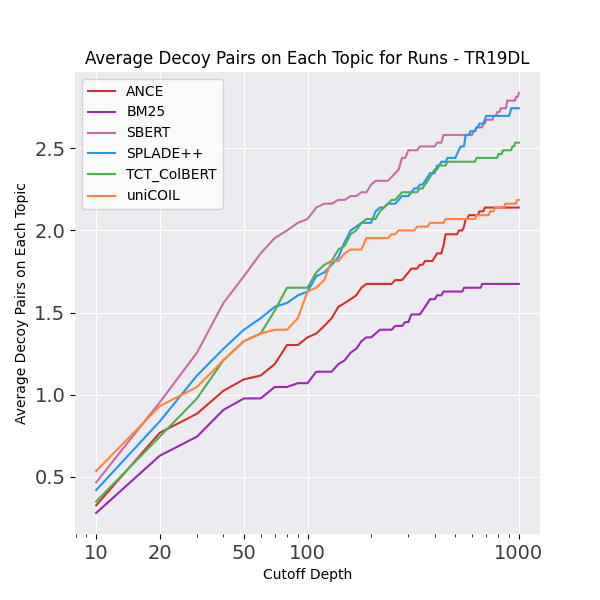}
    \caption*{}
\end{subfigure}
\begin{subfigure}[t]{0.32\textwidth}
    \centering
    \includegraphics[width=5cm]{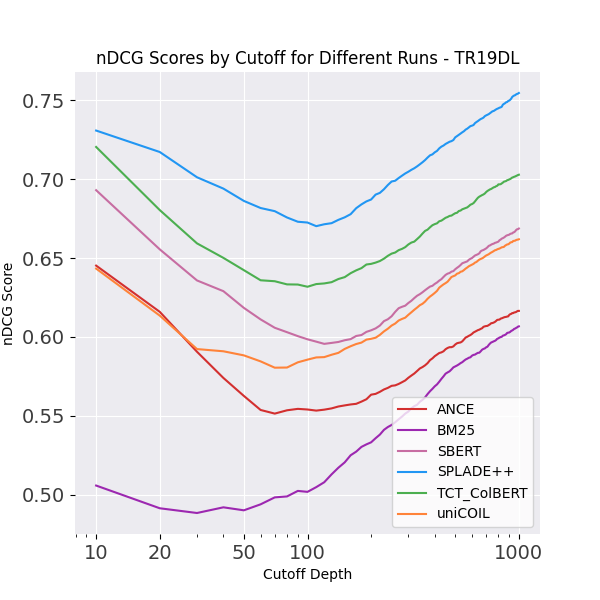}
    \caption*{}
\end{subfigure}
\begin{subfigure}[t]{0.32\textwidth}
    \centering
    \includegraphics[width=5cm]{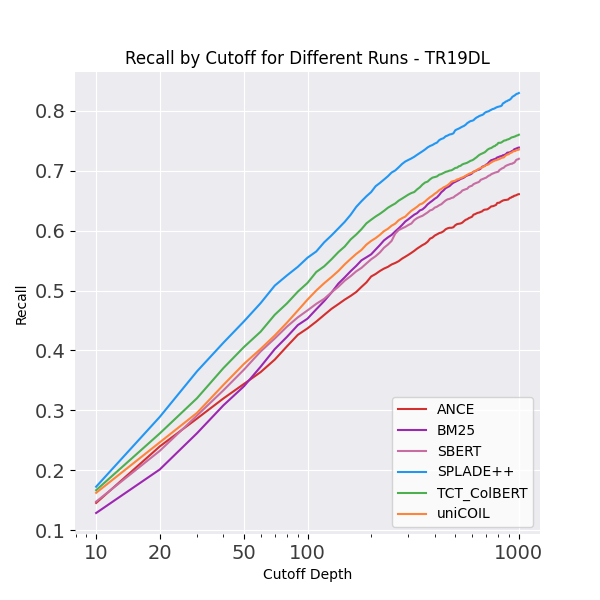}
    \caption*{}
\end{subfigure}

\begin{subfigure}[t]{0.32\textwidth}
    \centering
    \includegraphics[width=5cm]{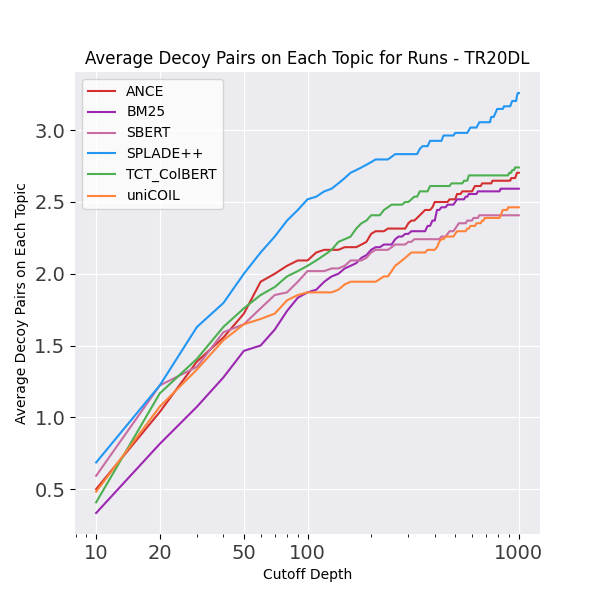}
    \caption*{}
\end{subfigure}
\begin{subfigure}[t]{0.32\textwidth}
    \centering
    \includegraphics[width=5cm]{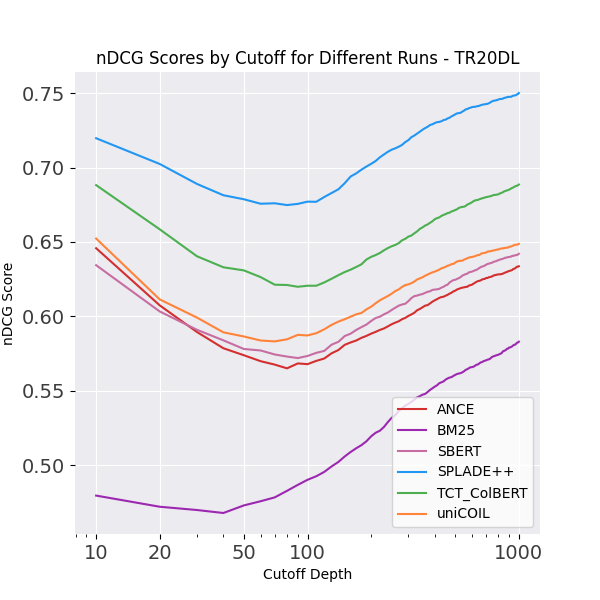}
    \caption*{}
\end{subfigure}
\begin{subfigure}[t]{0.32\textwidth}
    \centering
    \includegraphics[width=5cm]{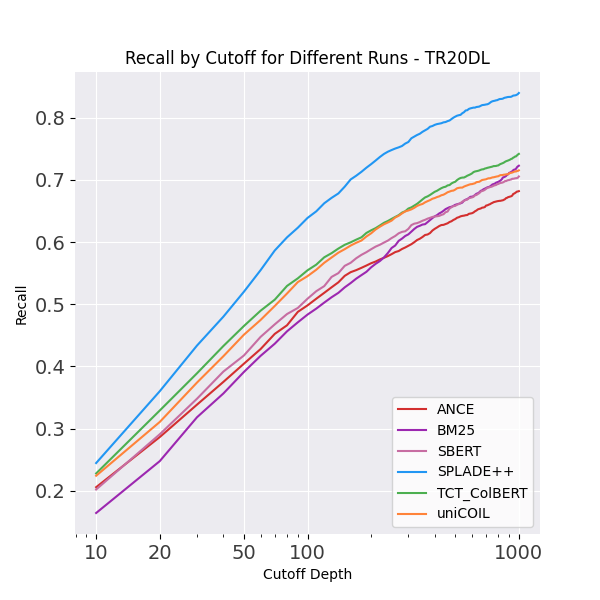}
    \caption*{}
\end{subfigure}

\caption{The number of average decoy pairs on each topic~(left), the score of nDCG~(middle), and the score of recall~(right) for BM25, SPLADE++, uniCOIL, ANCE, TCT-ColBERT, and SBERT across the TREC 19 DL collection~(top) and TREC 20 DL collection~(bottom).}
\label{fig:metrics}
\end{figure*}

\begin{figure}[t]
    \centering
    \includegraphics[width=.75\linewidth]{
    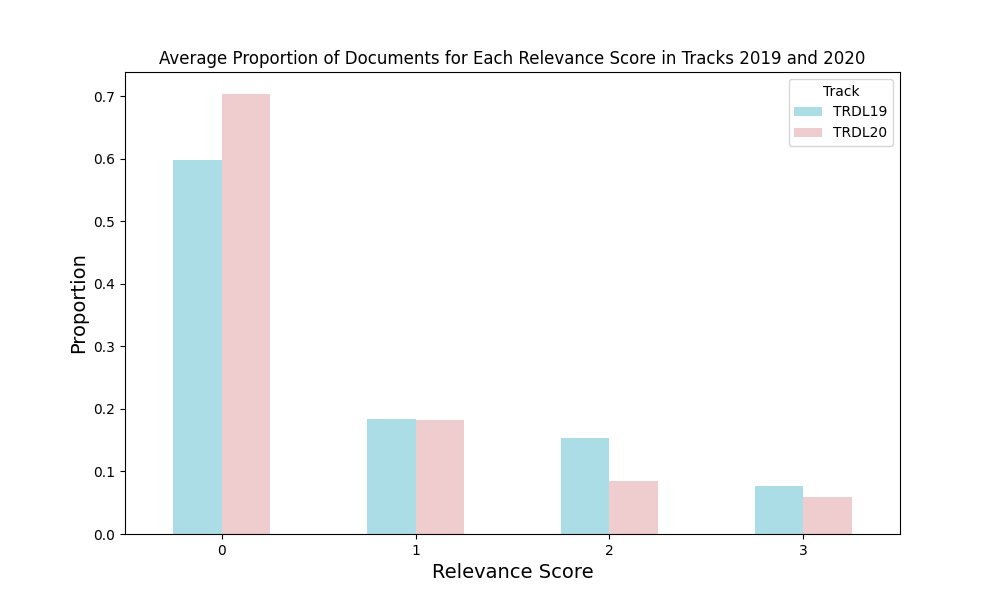
    }
    \caption{\label{fig:docs_per_score} The proportion of documents for each relevance score in TREC 19 DL and TREC 20 DL.}
\end{figure}

Figure~\ref{fig:metrics} shows the average numbers of the decoy pairs on each topic, the nDCG scores, and the recall scores for 6 retrievers at different depths in the TREC 19 DL and TREC 20 DL tasks. From Figure~\ref{fig:metrics} it can be firstly observed that, similar to recall, the number of decoy pairs increase as the cutoff becomes large. This observation is intuitive. Since our definition of decoy pairs involves relevance~(refer to Section~\ref{subsec:exp3setting}), similar to recall, the number of decoy pairs increases as more documents are retrieved. Based on the previous definition of decoy pairs, when the cutoff depth increases, existing decoy pairs remain unchanged; At the same time, newly retrieved documents may become either target documents or decoy documents, thus increasing the number of decoy pairs. 

Another observed result is that, compared to recall and nDCG, the system ranking measured by the number of decoy pairs is volatile and changes with the variation in cutoff depth. From Figure~\ref{fig:metrics}, we can see that in TREC 19 DL, except for the swapping of ranks between ANCE and uniCOIL, the ranking of other systems remained almost unchanged. Similarly, in TREC 20 DL, apart from a rank exchange between ANCE and SBERT, the rankings of the other systems also showed little to no change. The system rankings measured by recall are also similar to those measured by nDCG. After excluding BM25, the relative rankings of the five systems in both tracks showed almost no change. This observation reflects the complexity of the decoy effect mechanism. According to our definition, the identification of a decoy pair depends not only on the relative quality between the document pair but also on factors like their similarity and ranking. 

As previously mentioned, the factors influencing the number of decoy pairs are complex. Therefore, relying solely on the number of decoy pairs as a metric may not effectively reflect the vulnerability of retrieval systems to decoy results and lead to biased judgments. Consider the example of SPLADE++. As shown in Figure~\ref{fig:metrics}, on TREC 20 DL, it returned the most decoy pairs. However, as Figure~\ref{fig:docs_per_score} shows, if we investigate the distribution of document proportions for different relevance scores on TREC 19 DL and TREC 20 DL, we find that TREC 20 DL has nearly $90\%$ of its documents with relevance scores of 0 or 1, with only a few documents having a relevance score of 2 or 3 (highly relevant). Therefore, the reason for this result might be because SPLADE++ returned more highly relevant documents than other systems, thereby leading to more decoy pairs. Consider another example of BM25, the number of decoy pairs returned by BM25 is the least in two tracks when the cutoff is less than 100. Nonetheless, this outcome may be attributed to the fact that, in comparison with other systems, BM25 retrieves a smaller quantity of (highly) relevant documents, as illustrated in Figure~\ref{fig:metrics}. 

Consider a more extreme example, where a system returns only irrelevant documents to the topic. According to our previous definition, the number of decoy pairs returned by this system is undoubtedly 0. Nevertheless, such a system is fundamentally incapable of providing users with any helpful documents to address their information needs and there fore should definitely not be preferred in information retrieval system evaluation practice like TREC. In other words, discussing the vulnerability of an information retrieval system to the decoy effect is meaningless without considering its effectiveness.




\subsection{A Heuristic Metric to Measure the Vulnerability to the Decoy Effect}
\label{ch:7_4}

\begin{figure}[t]
    \centering
    \includegraphics[width=0.85\linewidth]{
    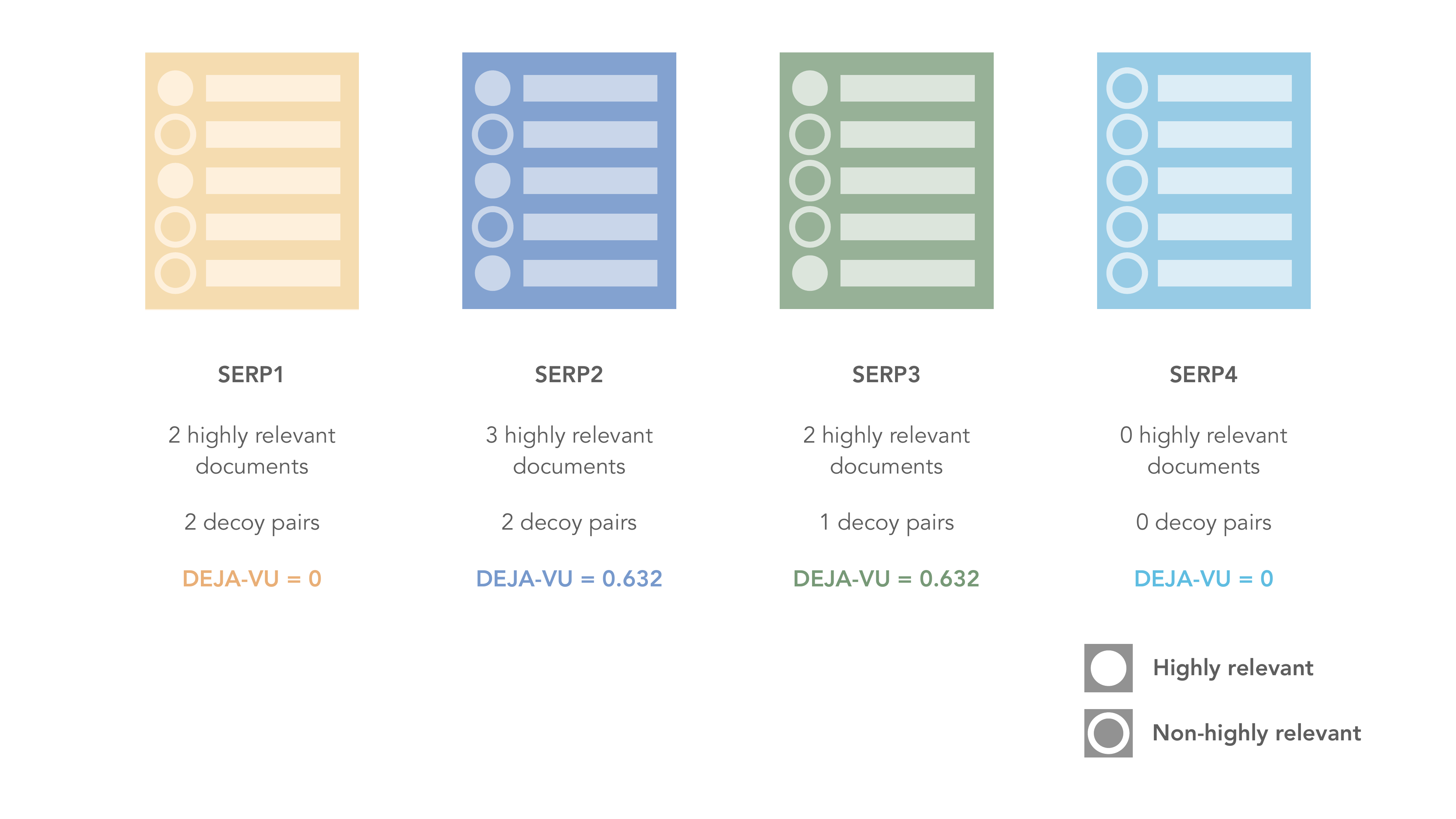
    }
    \caption{Example of DEJA-VU (more precisely, DEJAVU@5) scores. In the example, SERP 1 contains 2 highly relevant documents and 2 pairs of decoy documents, resulting in a DEJA-VU score of 0; SERP 2 contains 3 highly relevant documents and 2 pairs of decoy documents, resulting in a DEJA-VU score of 0.632; SERP 3 contains 2 highly relevant documents and 1 pair of decoy documents, resulting in a DEJA-VU score of 0.632; SERP 4 contains no highly relevant documents, therefore, no decoy pairs exist, resulting in a DEJA-VU score of 0. Solid circles represent highly relevant documents, while hollow circles represent non-highly relevant documents.}
    \label{fig:dejavu_example}
\end{figure}

Inspired by the above discussion, we hope to design a metric that rewards systems which return more highly relevant results while yielding fewer decoy pairs; and penalizes systems that return fewer highly relevant results while producing more decoy pairs. In this subsection, we propose a heuristic metric, namely \textbf{DEcoy Judgement and Assessment VUlnerability (DEJA-VU)}, to measure a system's vulnerability to the decoy effect while trying to balance the relevance scores.

The calculation for DEJA-VU when the cutoff is $k$~(DEJA-VU@$k$) is as follows:
\begin{equation}
\label{eq:dejavu}
    \text{DEJA-VU@}k = 1 - \exp(d - r)
\end{equation}

Where $d$ is the number of decoy pairs at the cutoff $k$ and $r$ is the number of highly relevant documents~(in this experiment is the documents whose relevance score is 2 or 3) at the cutoff $k$. DEJA-VU possesses the following properties:

\begin{itemize}
    \item A system with a higher DEJA-VU score shall be preferred. When the number of highly relevant documents returned (\textit{i.e.,} $r$) is the same, systems that return fewer decoy pairs (\textit{i.e.,} $d$) have higher DEJA-VU scores. Conversely, when the number of returned decoy pairs is the same, systems that return more highly relevant documents have higher DEJA-VU scores.
    
    \item The range of DEJA-VU score is within $\left[0, 1\right)$. According to the conditions in Section~\ref{subsec:exp3setting}, a highly relevant document is paired with only one decoy, hence $d \leq r$ always holds. When $d = r$, the DEJA-VU score is 0. This includes the case of $r = 0$ (\textit{i.e.,} when a system fails to return any highly relevant documents). Therefore, DEJA-VU penalizes systems that can only return very few or no highly relevant documents. According to the properties of the exponential function, $\exp(d - r)$ is always greater than 0. Therefore, the DEJA-VU score is always less than 1.
\end{itemize}

Figure~\ref{fig:dejavu_example} gives an example of the computation of DEJA-VU. In the example, SERP 1 contains 2 highly relevant documents and 2 pairs of decoy documents, resulting in a DEJA-VU score of $1 - \exp(2 - 2) = 0$. SERP 2 contains 3 highly relevant documents and 2 pairs of decoy documents, resulting in a DEJA-VU score of $1 - \exp(2 - 3) \approx 0.632$; SERP 3 contains 2 highly relevant documents and 1 pair of decoy documents, resulting in a DEJA-VU score of $1 - \exp(1 - 2) \approx 0.632$; SERP 4 contains no highly relevant documents, therefore, no decoy pairs exist, resulting in a DEJA-VU score of $1 - \exp(0 - 0) = 0$. It should be noted that, unlike most offline evaluation metrics mentioned earlier, the DEJA-VU score does not focus on how users accumulate gains from relevant documents during interaction with search pages. Instead, the DEJA-VU score considers the system's ability to return highly relevant documents while also accounting for the presence of decoy pairs in the results. Therefore, for some SERP pairs, DEJA-VU preferences may differ from metrics based on user utility models (\textit{e.g.,} DCG~\citep{Jarvelin-2002}, RBP~\citep{Moffat-2008}, ERR~\citep{Chapelle-2009}). For example, in the above example, metrics like DCG based on user utility models may assign a higher score to SERP 1 compared to SERP 4 as SERP 1 returned more highly relevant documents. However, according to the scores provided by DEJA-VU, both are 0 as DEJA-VU penalizes SERP 1 for returning a high proportion of decoy pairs. In the evaluation practice of search systems, it may be worth considering integrating DEJA-VU scores with existing user utility-based metrics in some way. We leave this issue for future research.

The rationale for the hyperparameter settings in DEJA-VU, such as the criteria for measuring similarity, the threshold for similarity, methods for measuring document quality, and the use of an exponential function in the score calculation process, is as follows: Due to the limited research in the IR field on the decoy effect in search interactions, we have arbitrarily defined these hyperparameters based on our understanding of the traditional definition of the decoy effect. Regarding the measurement of similarity, in this study, we used the relatively simple cosine similarity, but other more sophisticated measures based on embeddings, such as semantic similarity, could also be explored. The threshold for cosine similarity can also reference the work of ~\citet{Eickhoff2018}, where it is stipulated that the similarity between the target and the decoy should be greater than or equal to 0.7. The methods for measuring document quality have been discussed in Section~\ref{ch:data_processing}, where, in addition to relevance, other criteria can be selected based on different circumstances. Since the dataset used in this study directly provides annotations for relevance, we use relevance scores as the measure of document quality. In cases where relevance scores are unavailable, one can refer to the method used by~\citet{Eickhoff2018}, selecting alternatives such as BM25 scores as a substitute for relevance scores. The use of the exponential function is to facilitate smoothly controlling the range of DEJA-VU scores between 0 and 1. The advantage of this approach is that it allows for linear combination with evaluation metrics such as nDCG, enabling a comprehensive assessment of an information retrieval system's robustness to the decoy effect and retrieval effectiveness. We discussed this in Section~\ref{ch:7_5}.

It should also be noted that DEJA-VU is merely a heuristic metric, primarily designed to balance the relationship between the number of highly relevant documents and the number of decoy pairs when computing the system score for ranking IR systems, rather than providing an entirely accurate measure of a system's vulnerability to the decoy effect. We leave the research question of how to construct a precise measure of the vulnerability of IR systems to the decoy effect for future studies. 

\begin{figure*}[htbp]
\centering
\begin{subfigure}[t]{0.48\textwidth}
    \centering
    \includegraphics[width= 6cm]{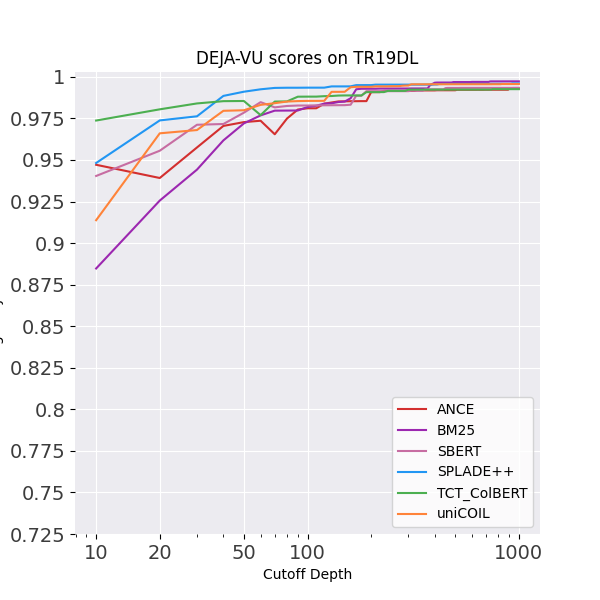}
    \caption*{}
\end{subfigure}
\begin{subfigure}[t]{0.48\textwidth}
    \centering
    \includegraphics[width=6cm]{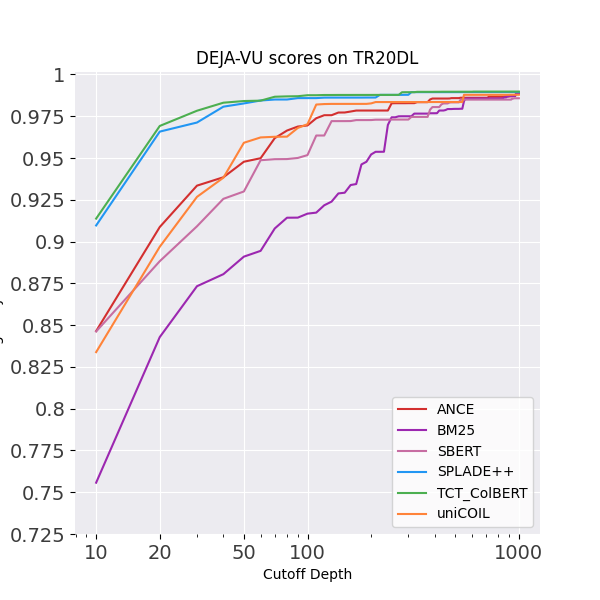}
    \caption*{}
\end{subfigure}
\caption{Average DEJA-VU scores of the retrievers on TREC 19 DL~(left) and TREC 20 DL~(right) respectively.}
\label{fig:dejavu}
\end{figure*}

Figure~\ref{fig:dejavu} displays the DEJA-VU scores of retrievers on the TREC 19 DL track and TREC 20 DL track. It can be observed that as the cutoff becomes very large, the DEJA-VU scores of different systems become very close, which is determined by the nature of the exponential function. From here, it can be seen that when the cutoff becomes very large, DEJA-VU cannot effectively differentiate between various systems. We will leave the research question of designing a ``better'' decoy vulnerability evaluation metric for future studies.

Considering that in real-life scenarios, most users do not go beyond the first 10 or 20 results when browsing a SERP~\cite{wicaksono2018}, discussing DEJA-VU@$10$ or DEJA-VU@$20$ is more meaningful. Table~\ref{tab:metric_19} and Table~\ref{tab:metric_20} urther present the DEJA-VU, nDCG, and recall scores for various retrievers on TRDL19 and TRDL20, with cutoffs set at 10 and 20, respectively. From Table~\ref{tab:metric_19} and Table~\ref{tab:metric_20}, it can be observed that regardless of whether the cutoff is 10 or 20, TCT-ColBERT consistently has the highest DEJA-VU score among retrievers. Following it closely is SPLADE++. On the TREC 20 DL track, SPLADE++ and TCT-ColBERT have only a marginal difference in their rankings. BM25, on the other hand, ranks last in both tracks. 

Additionally, from Table~\ref{tab:metric_19} and Table~\ref{tab:metric_20}, we observe that when the cutoff is set at 10 or 20, the system rankings provided by DEJA-VU scores and those provided by nDCG scores exhibit a certain degree of similarity. In Figure~\ref{fig:dejavu}, TCT-ColBERT and SPLADE++ form the best-performing cluster, while COIL, SBERT, and ANCE constitute the intermediate cluster, with BM25 performing the worst. This conclusion aligns to some extent with the findings obtained from nDCG scores in Figure~\ref{fig:metrics}. Issues regarding the correlation between DEJA-VU scores and commonly used offline evaluation metrics (e.g., DCG~\cite{Jarvelin-2002}, nDCG~\cite{Jarvelin-2002}, RBP~\cite{Moffat-2008}, ERR~\cite{Chapelle-2009}), as well as the consistency of the system rankings provided, will be left for future research. A series of meta-evaluations around DEJA-VU, such as whether its ranking of systems aligns with the preferences of real users and to what extent it can statistically discriminate between system pairs~\citep{sakai2006}, will also need to be addressed in future research.

\begin{table}[ht]
\centering
\begin{tabular}{l|ccc|ccc}
\hline
\multirow{2}{*}{Run (Retriever)} & \multicolumn{3}{c|}{Cutoff = 10} & \multicolumn{3}{c}{Cutoff = 20} \\ \cline{2-7} 
                     & DEJA-VU   & nDCG    & Recall   & DEJA-VU   & nDCG    & Recall   \\ \hline
ANCE        & 0.947      & 0.645    & 0.145     & 0.939      & 0.616    & 0.240     \\
BM25                & 0.885      & 0.506    & 0.129     & 0.926     & 0.491    & 0.201     \\
SBERT & 0.940     & 0.693   & 0.148     & 0.956     & 0.656    & 0.232     \\
SPLADE++      & \underline{0.948}      & \textbf{0.731}   & \textbf{0.172}    & \underline{0.974}      & \textbf{0.717}    & \textbf{0.289}     \\
TCT-ColBERT   &\textbf{0.974}     & \underline{0.720}    & \underline{0.167}    & \textbf{0.981}      & \underline{0.680}    & \underline{0.261}     \\
uniCOIL       & 0.914      & 0.643    & 0.162     & 0.966      & 0.614    & 0.247     \\ \hline

\end{tabular}
\caption{Scores of DEJA-VU, nDCG, and Recall for the six runs on TRDL 19 at cutoffs of 10 and 20 respectively. Bold indicates the highest score for a run, while underscore denotes the second highest score.}
\label{tab:metric_19}
\end{table}
\begin{table}[ht]
\centering
\begin{tabular}{l|ccc|ccc}
\hline
\multirow{2}{*}{Run (Retriever)} & \multicolumn{3}{c|}{Cutoff = 10} & \multicolumn{3}{c}{Cutoff = 20} \\ \cline{2-7} 
                     & DEJA-VU   & nDCG    & Recall   & DEJA-VU   & nDCG    & Recall   \\ \hline
ANCE        & 0.846     & 0.646    & 0.206     & 0.909      & 0.607    & 0.287     \\
BM25                & 0.756      & 0.480    & 0.164     & 0.843     & 0.472   & 0.248     \\
SBERT & 0.846     & 0.634   & 0.202     & 0.888     & 0.603    & 0.291     \\
SPLADE++      & \underline{0.910}      & \textbf{0.720}   & \textbf{0.245}    & \underline{0.966}      & \textbf{0.702}    & \textbf{0.360}     \\
TCT-ColBERT   &\textbf{0.914}     & \underline{0.688}    & \underline{0.228}    & \textbf{0.969}      & \underline{0.659}    & \underline{0.329}     \\
uniCOIL       & 0.834      & 0.652    & 0.224     & 0.897     & 0.611    & 0.311     \\ \hline

\end{tabular}
\caption{Scores of DEJA-VU, nDCG, and Recall for the six runs on TRDL 20 at cutoffs of 10 and 20 respectively. Bold indicates the highest score for a run, while underscore denotes the second highest score.}
\label{tab:metric_20}
\end{table}

\subsection{Combining DEJA-VU with Effectiveness-Oriented Evaluation Metrics}
\label{ch:7_5}
In the previous subsection, we computed DEJA-VU scores for six runs across two collections, TRDL19 and TRDL20 passage tracks, and compared them with their nDCG and Recall scores. As previously mentioned, DEJA-VU focuses on measuring a system's vulnerability to the decoy effect (with higher DEJA-VU scores indicating lower vulnerability to the decoy effect). However, in practice, the effectiveness of a retrieval system has consistently been regarded as a crucial evaluation dimension. Therefore, this subsection presents a simple and intuitive framework that linearly combines DEJA-VU scores with commonly used effectiveness-oriented metrics such as nDCG~\cite{Jarvelin-2002}, RBP\cite{Moffat-2008}, and ERR~\cite{Chapelle-2009}). This approach reflects the trade-off between low vulnerability to the decoy effect and high effectiveness in the metric scores.

\begin{table}[htbp]
\centering
\begin{tabular}{lcccccc}
\hline
\textbf{Metric} & \textbf{ANCE} & \textbf{BM25} & \textbf{SBERT} & \textbf{SPLADE++} & \textbf{TCT-ColBERT} & \textbf{UniCOIL} \\
\hline
\multicolumn{7}{c}{\textbf{TRDL19}} \\
\hline
DEJAVU@10 & 0.947 & 0.885 & 0.940 & \underline{0.948} & \textbf{0.974} & 0.914 \\
DEJAVU@20 & 0.939 & 0.926 & 0.956 & \underline{0.974} & \textbf{0.974} & 0.966 \\
nDCG@10 & 0.645 & 0.506 & 0.693 & \textbf{0.731} & \underline{0.720} & 0.643 \\
nDCG@20 & 0.616 & 0.491 & 0.656 & \textbf{0.717} & \underline{0.680} & 0.614 \\
RBP($\phi=0.8$)@10& 0.434 & 0.339 & 0.478 & \textbf{0.492} & \underline{0.480} & 0.426 \\
RBP($\phi=0.8$)@20 & 0.521 & 0.406 & 0.565 & \textbf{0.591} & \underline{0.576} & 0.509 \\
ERR@10 & 0.636 & 0.517 & \textbf{0.731} & \underline{0.701} & 0.697 & 0.624 \\
ERR@20 & 0.639 & 0.520 & \textbf{0.733} & \underline{0.703} & 0.699 & 0.627 \\
LCw/nDCG@10 & 0.796 & 0.695 & 0.816 & \textbf{0.840} & \underline{0.847} & 0.779 \\
LCw/nDCG@20 & 0.778 & 0.709 & 0.806 & \underline{0.846} & \textbf{0.831} & 0.790 \\
LCw/RBP($\phi=0.8$)@10 &  0.691 & 0.612 & 0.709 & \underline{0.720} & \textbf{0.727} & 0.670 \\
LCw/RBP($\phi=0.8$)@20 & 0.730 & 0.666 & 0.760 & \textbf{0.782} &  \underline{0.778} & 0.738 \\
LCw/ERR@10 &0.792 & 0.701 & \underline{0.836} & 0.825 & \textbf{0.848} & 0.769 \\
LCw/ERR@20 & 0.789 & 0.723 & \textbf{0.859} & 0.847 & \underline{0.850} & 0.797 \\

\hline
\multicolumn{7}{c}{\textbf{TRDL20}} \\
\hline
DEJAVU@10 & 0.846 & 0.756 & 0.846 & \underline{0.910} & \textbf{0.914} & 0.834 \\
DEJAVU@20 & 0.909 & 0.843 & 0.888 & \underline{0.966} & \textbf{0.969} & 0.897 \\
nDCG@10 & 0.646 & 0.480 & 0.634 & \textbf{0.720} & \underline{0.688} & 0.652 \\
nDCG@20 & 0.607 & 0.472 & 0.603 & \textbf{0.702} & \underline{0.659} & 0.611 \\
RBP($\phi=0.8$)@10& 0.120 & 0.087 & 0.117 & \textbf{0.130} & \underline{0.125} & 0.120 \\
RBP($\phi=0.8$)@20 & 0.140 & 0.102 & 0.137 & \textbf{0.154} &\underline{0.147} & 0.141\\
ERR@10 & 0.181 & 0.149 & 0.183 & \underline{0.189} & \textbf{0.191} & 0.181 \\
ERR@20 & 0.181 & 0.151 & 0.184 & \underline{0.190} & \textbf{0.192} & 0.181 \\
LCw/nDCG@10 & 0.746 & 0.618 & 0.740 & \textbf{0.815} & \underline{0.801} & 0.743 \\
LCw/nDCG@20 & 0.758 & 0.658 & 0.746 & \underline{0.834} & \textbf{0.814} & 0.754 \\
LCw/RBP($\phi=0.8$)@10  & 0.483 & 0.421 & 0.482 & \textbf{0.520} &  \underline{0.519} & 0.477 \\
LCw/RBP($\phi=0.8$)@20  & 0.522 & 0.473 & 0.513 & \textbf{0.560} & \underline{0.558} & 0.519 \\
LCw/ERR@10 & 0.514 & 0.452 & 0.515 & \underline{0.549} & \textbf{0.552} & 0.481 \\
LCw/ERR@20  & 0.516 & 0.497 & 0.536 & \underline{0.548} & \textbf{0.561} & 0.504 \\
\hline
\end{tabular}
\caption{The scores of six runs—ANCE, BM25, SBERT, SPLADE++, TCT-ColBERT, and UniCOIL—on different metrics for the TRDL19 and TRDL20 collections. Bold indicates the
highest score for a run, while underscore denotes the second highest score.}
\label{tab:linear}
\end{table}

In this subsection, we selected two commonly used effectiveness-oriented evaluation metrics, RBP and ERR, in addition to nDCG. 

The calculation for RBP score is as follows~\citep{Moffat-2008}:
\begin{equation}
    \text{RBP} = (1 - \phi)  \sum_{i=1}^{d} r_i  \phi^{i-1},
\end{equation}

where $\phi$ is a parameter, $r_i$ represents the gain of the document at position $i$, and $d$ is the cutoff depth. The user model for RBP assumes that after examining the document at position $i-1$, the user will continue to examine the document at position $i$ with a fixed probability of $\phi$, or will leave with a probability of $(1-\phi)$. The larger the value of $\phi$, the more patient the user is, increasing the likelihood of examining results at later positions~\citep{moffat2013}. In our experiment, we set $\phi$ to 0.8.

The calculation for ERR score is as follows~\citep{Chapelle-2009}:
\begin{equation}
    \text{ERR} = \sum_{i=1}^{d} \frac{1}{i} \prod_{j=1}^{i-1} (1 - R_j) R_i,
\end{equation}

where $R$ is a mapping from relevance grades to ``probability of relevance'', and can be calculated from the following equation~\citep{Chapelle-2009}:
\begin{equation}
    \label{eq:err_map}
    \mathcal{R}(g) = \frac{2^g - 1}{2^{g_{\text{max}}}}, \quad g \in \{0, \dots, g_{\text{max}}\}. 
\end{equation}
In the above equation, $g$ represents the relevance grade of the document at the current position, and $g_{\text{max}}$ represents the maximum relevance grade. The user model for ERR assumes that the user has a probability of $R_i$ of stopping at the document at position $i$, or a probability of $(1-R_i)$ of continuing to browse the search results~\citep{Chapelle-2009}. 

In our experiment, when calculating RBP, we linearly normalized the original relevance scores from TRDL19 and TRDL20 (which have four levels) to obtain $r_i$. That is, $r_i \in \{0, 1/3, 2/3, 1\}$. When calculating ERR, we mapped the original relevance scores $g$ from TRDL19 and TRDL20 to $R_i \in \{0, 1/8, 3/8, 7/8\}$ according to Equation~\ref{eq:err_map}.

Since the nDCG, RBP, ERR, and DEJA-VU scores $\mathcal{M}$ all satisfy $\mathcal{M} \in [0, 1)$\footnote{For the properties of the nDCG, RBP, and ERR scores, please refer to the original papers. For the properties of the DEJA-VU score, please refer to~\ref{ch:7_4}.}, we can simply combine their scores using a linear weighted method:
\begin{equation}
\text{Linear Combination (LC)} = \alpha M_{\text{DEJA-VU}} + (1-\alpha) M_{\text{nDCG|RBP|ERR}},
\end{equation}
where $ M_{\text{DEJA-VU}}$ is the score of DEJA-VU, and $M_{\text{nDCG|RBP|ERR}}$ can be the score of nDCG, RBP or ERR; The value of $\alpha$ ranges from $\left[0,1\right]$. If $\alpha$ is 0, the score calculated by the metric is the nDCG, RBP, or ERR score. If $\alpha$ is 1, the score calculated by the metric is the DEJA-VU score. $\alpha$ represents the evaluator's perceived relative importance of a system's robustness to the decoy effect. In our experiment, we set $\alpha$ to 0.5.

Table~\ref{tab:linear} presents the scores calculated using different metrics for the six runs on the TRDL19 and TRDL20 collections. Overall, whether measured by the DEJA-VU score for robustness to the decoy effect or by nDCG, RBP, and ERR for effectiveness, TCT-ColBERT consistently performs as the best or second-best run; in most cases, SPLADE++ also ranks as the best or second-best run. A few exceptions include the ERR score on TRDL19, where SBERT is ranked as the best run. Specifically, on both collections, RBP, like nDCG, ranks SPLADE++ as the best run, while DEJA-VU ranks TCT-ColBERT as the best run. ERR provides slightly different conclusions; it ranks SBERT as the best run on TRDL19, while on TRDL20, it ranks TCT-ColBERT as the best run. As a linear combination of DEJA-VU and effectiveness-oriented metrics, LC generally does not produce conclusions that significantly differ from the individual metrics mentioned above. However, an interesting point is that on TRDL19, ERR ranks SBERT as the best, while DEJA-VU ranks TCT-ColBERT as the best. Both ERR and DEJA-VU rank SPLADE++ as the second best. When ERR and DEJA-VU are linearly combined, TCT-ColBERT ranks first at cutoff = $10$, with SBERT in second place; at cutoff = 
 $20$, SBERT ranks first, with TCT-ColBERT in second place, while SPLADE++ consistently ranks third.

In summary, in this section, we first analyzed the relationship between the number of decoy pairs in the results returned by various retrievers and the number of highly relevant documents. We found that the vulnerability of text retrieval systems to the decoy effect cannot be simply measured by the number of decoy pairs. We then introduced a heuristic metric, namely DEJA-VU, to assess the system's performance in terms of having a lower vulnerability to the decoy effect while achieving higher effectiveness. We also introduced a framework that linearly combines DEJA-VU scores with the scores of three metrics commonly used in offline evaluation practices: nDCG, RBP, and ERR. Based on the rankings provided by DEJA-VU scores, we can answer the \textbf{RQ4}: In general, when the cutoff is small, TCT-ColBERT and SPLADE++ can achieve higher effectiveness while having a lower vulnerability to the decoy effect.
\section{Conclusion and Discussion}
\label{ch:8}

Contrary to the implicit assumptions underlying various formal models of information seeking, it is posited that users exhibit bounded rationality, and typically do not base their search decisions on precise estimations of search gains and costs~\citep{Azzopardi2021, liu2023behavioral}. Despite the increasing attention paid to data biases and algorithm biases in the the field of computing~\citep{Kordzadeh2022, lalor2024, chenjiawei2023}, only a few studies have focused on the issue of properly recognizing and effectively mitigating the negative effects of biases on human decision-making processes. Within the realm of cognitive biases influencing decision-making, the decoy effect reflects how users alter their in-situ preferences and judgments on presented options. Although prior research has examined the impact of the decoy effect on relevance judgments under the crowdsourcing scenario~\citep{Eickhoff2018}, how decoy results presented on SERPs influence users' interaction behaviors, and how this effect interacts with search contextual factors such as users' prior topical knowledge, the difficulty of the search task, remains unclear.

To address this gap and advance towards the vision of human-centered information retrieval, in this study, we proposed 4 research questions~(see Section~\ref{ch:rq}) and seek to comprehend:~(1)~How the \textit{decoy effect} at the document level impacts users’ interaction behaviors on SERPs, such as clicks, dwell time, and usefulness perceptions~(\textbf{RQ1});~(2)~How the decoy effect influences user behavior differently across varying levels of task difficulty and the user's knowledge scale~(\textbf{RQ2 and RQ3});~(3)~How to assess the vulnerability of different text retrieval models to the decoy effect~(\textbf{RQ4}). To answer the above research questions, we conducted data analysis and experiments on several publicly available user behavior datasets and test collections.  

The results of this study empirically confirm the insights from behavioral economic theories in a novel context, and also shed light on innovative ways for developing computational models to explain boundedly rational user search behaviors and assess user-centric search systems. As far as we know, we are the first to addresses how the decoy effect influences user interactions on SERPs. This work extends the endeavors of the IR community in exploring how cognitive biases impact user behaviors in document examining and relevance judgment, providing evidence from the perspective of the decoy effect. This study also contributes to human-centric evaluation practice  in information retrieval by introducing a novel metric for assessing systems’ vulnerability to the decoy effect.

\subsection{Main Findings and Implications}
The following are the main findings of this study, which address our research questions. 

\subsubsection{RQ1: Influence of a Decoy on Users' Interaction Behavior.} From the result of the first experiment, it can be observed that, when a decoy is present, in comparison to when it is absent, there is a increase in the likehood of a document being clicked and its perceived usefulness, given conditions such as the ranking of the document, the type of the task and the individual characteristics unchanged. 

Going beyond document assessment experiments that iosolated search factors~(\textit{e.g.,}~\citep{Eickhoff2018,Scholer2013}), this study identified the impact of decoy options on users' click behavior and perceived usefulness by analysing user interaction logs collected under  laboratory-based environment. This study provides empirical insights into the debiasing of human biases in IR algorithms and developing user-centered IR systems. The finding of this study suggests that, when designing algorithms, it is important not only to consider document relevance and quality but also to take into account the influence of cognitive biases, such as the decoy effect, that users may experience. This implies that algorithms may need to incorporate the understanding of boundedly rational users' behavior patterns in order to help users achieve the globally optimal search outcome~\citep{liu2023}.

\subsubsection{RQ2 and RQ3: The Variance of the Impact of Decoy Effect Across Task Difficulties and Users' Prior Knowledge Levels.} The result of the second experiment indicates that the extent to which the decoy effect distorts user-document interaction behavior is influenced by both task difficulty and the user's level of prior knowledge. More specifically, when the search task is more difficult, users are less likely to click on documents having a decoy compared to situations where the search task is less difficult. They also spend less time on documents having a decoy and assign lower usefulness scores for such documents. When users have a lower level of prior knowledge about the search topic, they tend to assign higher usefulness scores to the documents who has a decoy. 

Previous research has shown associations between the difficulty of search tasks and the level of users' domain knowledge with certain behavioral signals exhibited by users during the search process~\citep{liu2010b, Liu2010, mao2016, cole2015}. However, only a limited number of studies have examined how cognitive biases interact with varying task difficulties and users with different levels of domain knowledge, thus influencing search behavior. 
This study addresses this gap by empirically demonstrating how the influence of the decoy effect on user search behavior varies across different levels of task difficulty and user knowledge levels. 
The finding in this study expands the understanding of the IR community regarding how tasks, domain knowledge, and cognitive biases influence users' behavior during the search process. It also provides a new perspective for the design of task-oriented information retrieval systems. For instance, in a multi-query search session, the system can adaptively adjust algorithms and page layouts based on different task states and user types predicted from various online signals~\citep{cole2015, wang2023}, thus mitigating the impact of the decoy effect.

\subsubsection{RQ4: Measuring IR System Vulnerability to Decoy Effect.} The result of the third experiment indicates that, the vulnerability of text retrieval systems to the decoy effect cannot be simply measured by the number of decoy pairs. When evaluating the vulnerability of text retrieval models to the decoy effect, it is essential to consider various factors, including effectiveness, in a comprehensive manner. Based on the above points, we have developed a metric for measuring the vulnerability to the decoy effect. According to the metric scores, when the cutoff is small, TCT-ColBERT and SPLADE++ can generally achieve higher effectiveness while having a lower vulnerability to the decoy effect.

Traditional evaluation methods for IR systems are based on the assumption of globally rational users, which may not fully capture the complexity of the boundedly rational users' behavior under the influence of cognitive biases. Therefore, there is a need to develop new evaluation metrics to more accurately measure system performance and ensure they reflect users' actual experiences and preferences. Although some recent efforts have incorporated cognitive biases into the calculation of IR system evaluation metrics~\cite{chen2022, chen2023, zhang2020}, they have not considered the decoy effect. The introduction of DEJA-VU expands the work on IR system evaluation by considering the vulerability of IR systems to the decoy effect. By integrating the DEJA-VU metric with other evaluation metrics, researchers can fine-tune and improve search algorithms to enhance the effectiveness of search systems while mitigating the negative impact of the decoy effect on user decisions. Alternatively, they can design personalized reranking algorithms capable of adapting to changes in user behavior.

\subsection{Limitations and Future Work}
In this study, we sought to address the challenge of how the decoy effect influences user interactions on SERPs and how to measure the vulnerability of IR systems to the decoy effect. Nevertheless, our study is still in a preliminary stage toward the vision of human-bias-aware IR modeling, with numerous aspects awaiting further exploration.

In Section~\ref{ch:4}, when processing data, many hyper-parameters were arbitrarily determined by the authors, based on the definition of the decoy effect from other fields and its potential mechanisms in the context of interactive information retrieval. That is because this study examines the decoy effect in the context of user interactions with SERPs, and there are very few papers highly relevant to our work, and many of the hyper-parameters in the experimental setup of this study could not reference prior work. Future research could explore different window length, methods for measuring document similarity, and document similarity threshold when extracting decoy pairs.

In Section~\ref{ch:5}, 
we used an OLS regression model to investigate the impact of the presence of a decoy on users' document click behavior, document viewing duration, and document usefulness ratings. But there are still some variables related to other cognitive biases that have not been included in the model, such as the number of irrelevant items per page/per batch of 5. Future experiments could consider incorporating these factors into the model when considering additional cognitive bias factors. 

In Section~\ref{ch:6}, we discussed the relationship between varying levels of topic difficulty, different degrees of user prior knowledge, and the extent of the decoy effect. Since the THUIR2016 and THU-KDD datasets did not collect users' perceived difficulty of the topics and users' knowledge on the topics, we only used the THUIR2018 dataset for the experiment. On the other hand, previous research has shown that cognitive abilities, such as working memory capacity and associative memory ability, influence users' search behavior~\cite{choi2023,brennan2014}. Due to dataset limitations, these factors could not be examined in this study. Future research could investigate whether users with different cognitive abilities experience varying degrees of the decoy effect on tasks of differing complexity, particularly examining how this effect manifests in search behavior-related signals such as session length, SERP clicks, and SERP dwell time.

Regarding the DEJA-VU metric proposed in Section~\ref{ch:7}, further meta-evaluation~(\textit{e.g.,}~\citep{chen2017, sakai2020}) is required to understand whether the scores it provides align with users' actual preferences and to what extent it can statistically significantly differentiate retrieval systems in reproducible offline evaluation~\cite{liu2022toward}. Moreover, although there are existing studies~(\textit{e.g.,}~\citep{chen2022, chen2023,zhang2020}, \textit{etc.}) that have integrated cognitive biases into the score computation of evaluation metrics, there is still a need for a broader theoretical framework to integrate them together. 

Since effectiveness is considered an important dimension in offline information system evaluation, we propose a linear combination method in Section~\ref{ch:7_4} to achieve a trade-off between effectiveness and robustness to the decoy effect in the evaluation. However, there are other ways to combine with effectiveness, such as utilizing the C/W/L/A framework~\citep{moffat2022batch}. In this context, the concept of DEJA-VU can be viewed as an \textit{aggregation function} within the C/W/L/A framework, which can be weighted and combined with other aggregation functions (\textit{e.g.,} Expected Rate of Gain) and different user browsing models to derive various evaluation metrics. The advantage of this approach is that it decouples the user browsing model from the user utility accumulation model (aggregation function), allowing for flexible combinations of different user browsing models and user utility accumulation models according to practical needs. 
These aspects require further research for exploration in the future.

\subsection{Open Questions}
In this paper, our work primarily focuses on the decoy effect in search scenarios. But there are still many open questions worth discussing regarding cognitive biases in interactive search and other information behaviors.

\subsubsection{The Impact of Cognitive Biases: Positive or Negative?}
While cognitive biases are generally believed to lead to poorer decisions, this is not always the case~\citep{Azzopardi2021}. For example, the influence of the decoy effect depends on the nature of the information presented in the SERP. If the target document contains incorrect information or malicious rumors, the utility for users affected by the decoy effect may be harmed, which is problematic. In political or legal topics, the decoy effect may interact with confirmation bias, leading users to focus on information that aligns with their existing beliefs, thereby trapping them in an echo chamber, which is also undesirable. However, the effects of the decoy effect may not be entirely negative. A possible scenario is that providers might use the decoy effect to control the position of their products in the results, thereby increasing their sales. If a business provides high-quality products offline, then in this case, the decoy effect can be beneficial for users. 

Although the decoy effect can also lead to positive outcomes, we advocate for reducing cognitive biases to enable users to make more rational decisions. This constitutes the motivation behind our proposal of DEJA-VU.  

\subsubsection{Causal Inference in Identifing and Mitigating Cognitive Biases.}

As mentioned earlier, in the experiment of this study, we found through regression analysis that the presence of a decoy document can influence users' behavior when browsing the target document, specifically showing a positive correlation between the presence of a decoy document and clickthrough likelihood, browsing dwell time, and usefulness score. However, whether such correlations represent causal relationships remains unknown. Although causal bias correction has received attention in some literature related to search and recommendation systems, these work have mainly focused on biases in the data, such as popularity bias~\citep{wei2021, wang2021}, exposure bias~\citep{liu2021miti,xu2023}, \textit{etc}. Researchers still have a limited understanding of how cognitive biases of users affect data through explicit or implicit feedback, further amplify system bias, and thus lead to the loop of bias.

\subsubsection{Cognitive Biases in Generative Artificial Intelligence}
In recent years, generative Artificial Intelligence~(AI) such as Large Language Models (LLMs)~\citep{gpt3, touvron2023llama}  have garnered widespread attention fromhe computing community. Many efforts in the fields of search and recommendation have attempted to incorporate LLMs into traditional pipelines (such as using LLMs as content encoder~\citep{ONCE, harte2023}, directly using LLMs to return search or recommendation results~\citep{sun2023chatgpt,ma2023zeroshot,qin2023}, \textit{etc.}) or to build new pipelines using LLMs~(\textit{e.g.,} using LLMs for generative retrieval or recommendation~\citep{li2023large,ji2023genrec}). Recent studies have found that LLMs can be influenced by context when generating outputs, leading to biases resembling human cognitive biases~(\textit{e.g.,} the recency effect~\citep{zhao2021calibrate}, the framing effect~\citep{jones2022capturing}, the bandwargon effect~\citep{koo2023benchmarking}, and the priming effect~\citep{chen2024ai}). By using techniques like prompt engineering to trigger LLMs to generate biased results, could this lead to suboptimal search and recommendation outcomes, thus reducing overall user utility? Would biases in the generation process of LLMs interact with users' cognitive biases in online information seeking behaviors, resulting in a loop of biases? These are all questions that need to be addressed.  Considering that individuals from diverse backgrounds may have different triggers for cognitive biases, it is also worthwhile to investigate how to personalize the outputs of LLMs in search and recommendation ranking based on user profiles, in order to reduce the negative impact of cognitive biases on individual decision-making. 

\bibliographystyle{ACM-Reference-Format}
\bibliography{cas-refs}

\clearpage
\appendix
\section{Details of the User Study Datasets}
\label{ch:appendix_dataset}
This section provides details on the three publicly available datasets used in Experiment1 and Experiment2, including the topics of the search tasks, the process of collecting user feedback, and the instructions given to external annotators. The information can be found in the
original papers~\citep{Liu2018www,mao2016,Liu2019}.

\subsection{THUIR2016}
\subsubsection{Task Description}
\label{ch:mao_task_description}
In the study by~\citet{mao2016}, participants are asked to complete 9 tasks. These tasks were based on the TREC Session Track topics and were modified to ensure that the tasks were clearly understood by all participants and were complex enough to involve multiple queries. The task descriptions were provided in Chinese. Table~\ref{tab:task_kdd} reports the descriptions of these nine tasks.
\subsubsection{Self-rating Scale from Users}
\begin{figure}[t]
    \centering
    \includegraphics[width=.75\linewidth]{
    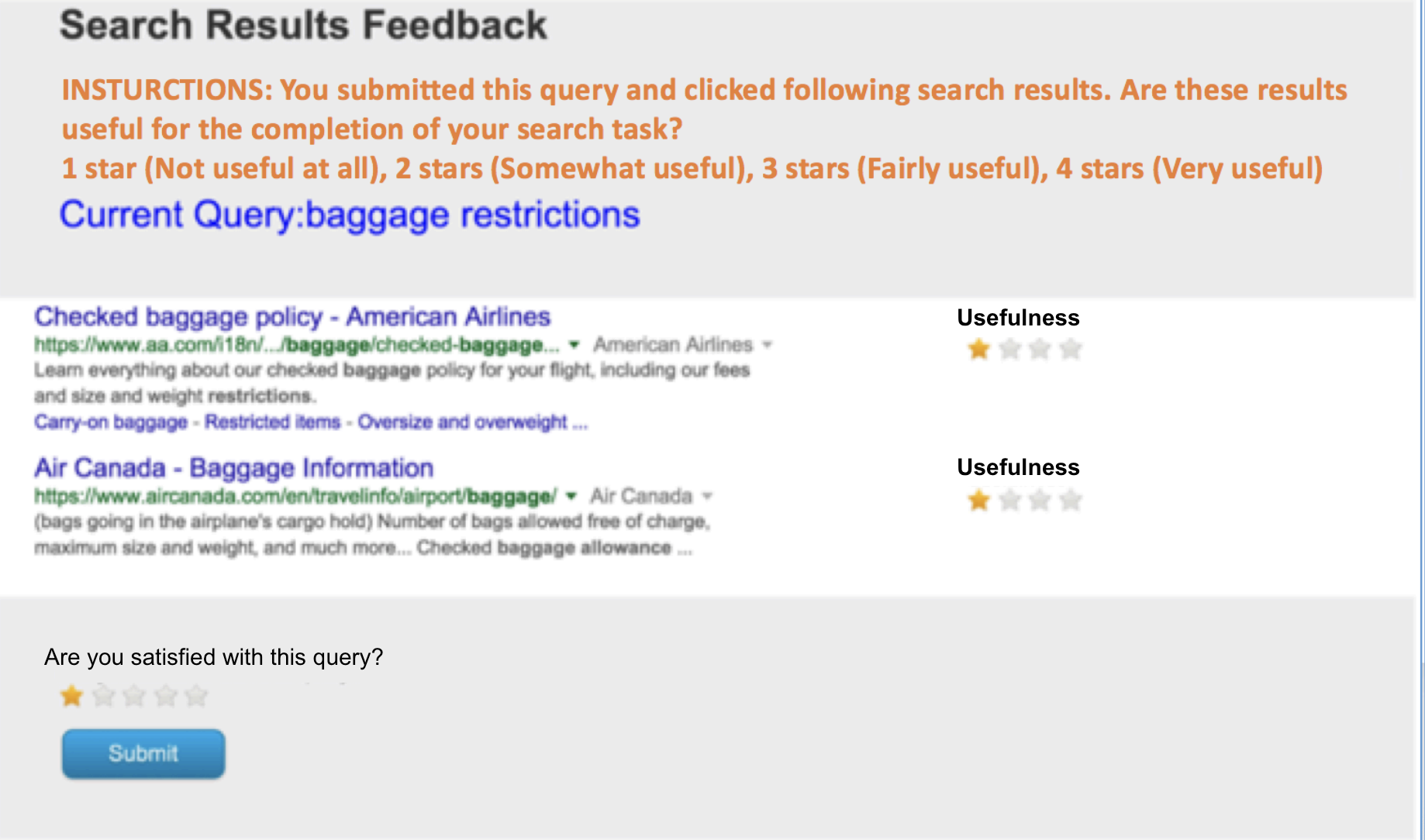
    }
    \caption{\label{fig:usefulness_mao}The interface for reporting document usefulness and query-level SERP satisfaction used in the study by~\citet{mao2016}.
}
\end{figure}

Participants need to self-report the usefulness of each document they encounter on a scale from 1 (not at all useful) to 4 (very useful). They also provide a satisfaction rating for each query on a scale from 1 (very dissatisfied) to 5 (very satisfied), and an overall satisfaction rating for the entire search session, also on a scale from 1 to 5. Table~\ref{tab:user_anno_kdd} presents the variables of user self-ratings along with their definitions. The interface used by \citet{mao2016} to collect user feedback on document usefulness and query-level SERP satisfaction is shown in Figure~\ref{fig:usefulness_mao}.

\subsubsection{The Interface for Annotation}
\begin{figure}[t]
    \centering
    \includegraphics[width=.75\linewidth]{
    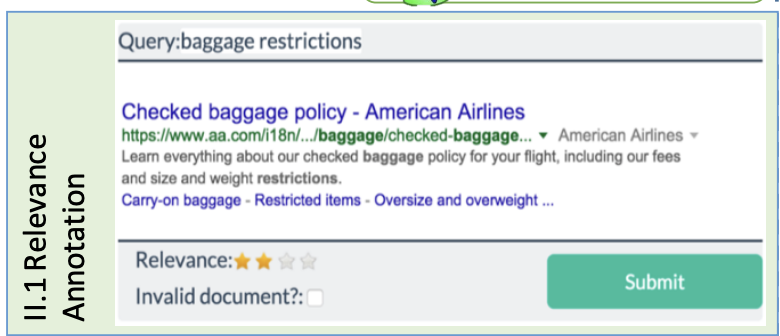
    }
    \caption{\label{fig:relevance_mao}The interface for annotating document relevance used in the study by~\citet{mao2016}.
}
\end{figure}

9 undergraduate students were asked to complete the relevance annotation task. They were asked to rate the relevance of the documents on a scale from 0 to 4, with 0 being the lowest and 4 being the highest. Figure~\ref{fig:relevance_mao} shows the interface they used. 

\subsection{THU-KDD}
\subsubsection{Task Description}
In the study by~\citet{Liu2019}, the selected tasks are exactly the same as those chosen by \citet{mao2016}. For information on these tasks, please refer to Section~\ref{ch:mao_task_description} and Table~\ref{tab:task_kdd}.

\subsubsection{Self-rating Scale from Users}
In the study by \citet{Liu2019}, users were similarly asked to rate the usefulness of documents on a 4-point Likert scale, from 0 (not at all useful) to 3 (very useful). They also provided a satisfaction rating for each query on a scale from 1 (very dissatisfied) to 5 (very satisfied), as well as an overall satisfaction rating for the entire search session, also on a scale from 1 to 5. Table~\ref{tab:user_anno_kdd} presents the variables of user self-ratings along with their definitions.

\subsubsection{Instructions Given to Annotators}
\begin{figure}[t]
    \centering
    \includegraphics[width=.75\linewidth]{
    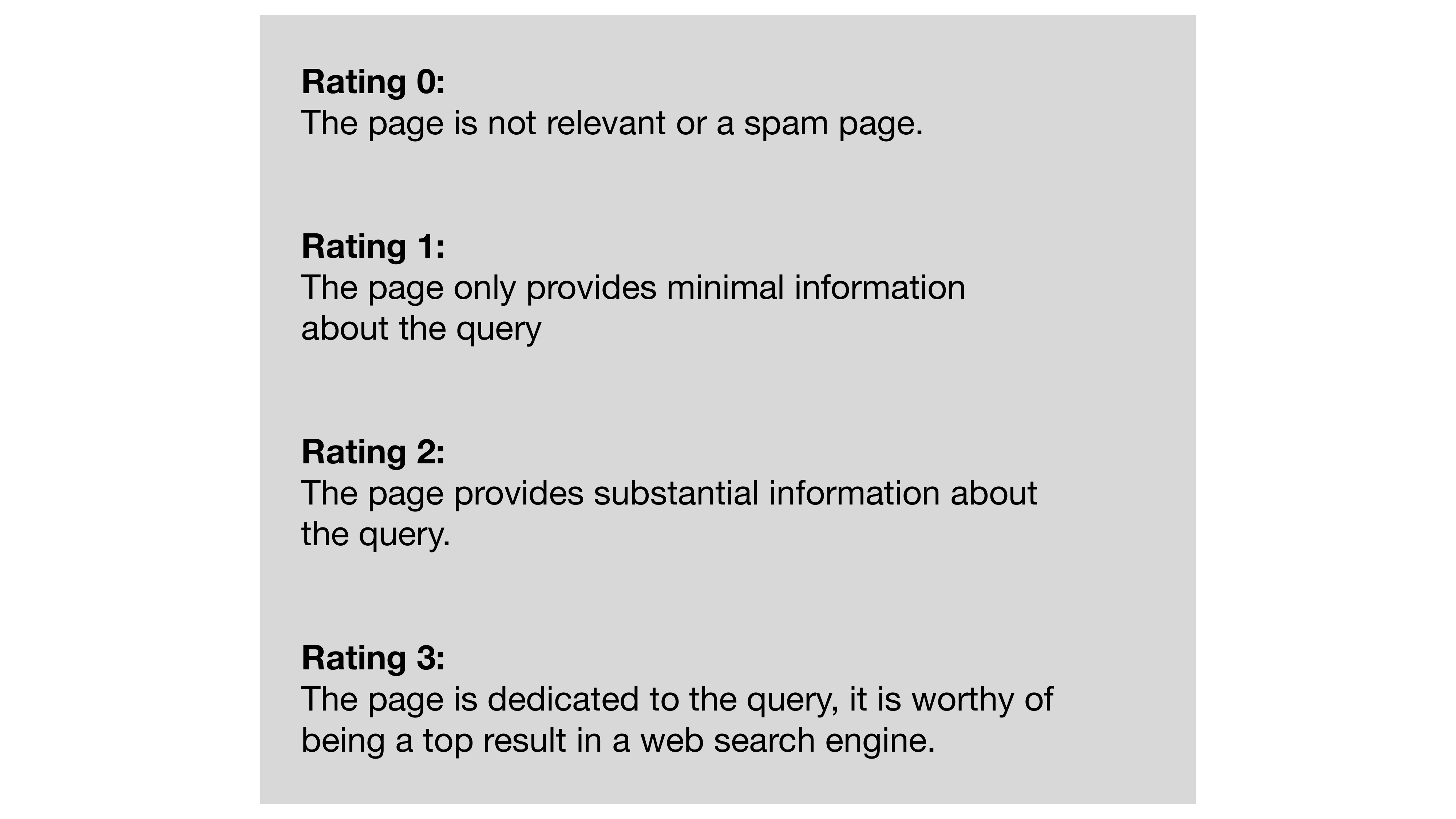
    }
    \caption{\label{fig:kdd_anno}The Instruction to external annotators used in the study by~\citet{Liu2019}}
\end{figure}

External assessors evaluated the relevance of documents through a crowdsourcing platform. Each assessor was given a query-document pair and was asked to assign a relevance score based on the rating criteria shown in Figure~\ref{fig:kdd_anno}.

\subsection{THUIR2018}
\subsubsection{Task Description}
In the study by \citet{Liu2018www}, each participant was required to complete 6 tasks from three domains: Environment, Medicine, and Politics. The tasks were designed by experts and aimed to be complex enough to require more than just a few simple search interactions. The task descriptions were provided in Chinese. Table~\ref{tab:task_www} presents the descriptions of the six search tasks used in the THUIR2018 dataset. 
\subsubsection{Self-rating Scale from Users}
Participants reported their perceived difficulty, prior knowledge, and interest in the topic, through a 5-point Likert scale (1: not at all, 2: slightly, 3: somewhat, 4: moderately, 5: very), before starting a task. After completing the task, they were asked to rate their satisfaction and the perceived usefulness of each result (1: not at all, 2: somewhat, 3: fairly, 4: very). Table~\ref{tab:user_anno_www} presents the variables of user self-ratings along with their definitions.

\subsubsection{Instructions Given to Annotators}
\begin{figure}[t]
    \centering
    \includegraphics[width=.99\linewidth]{
    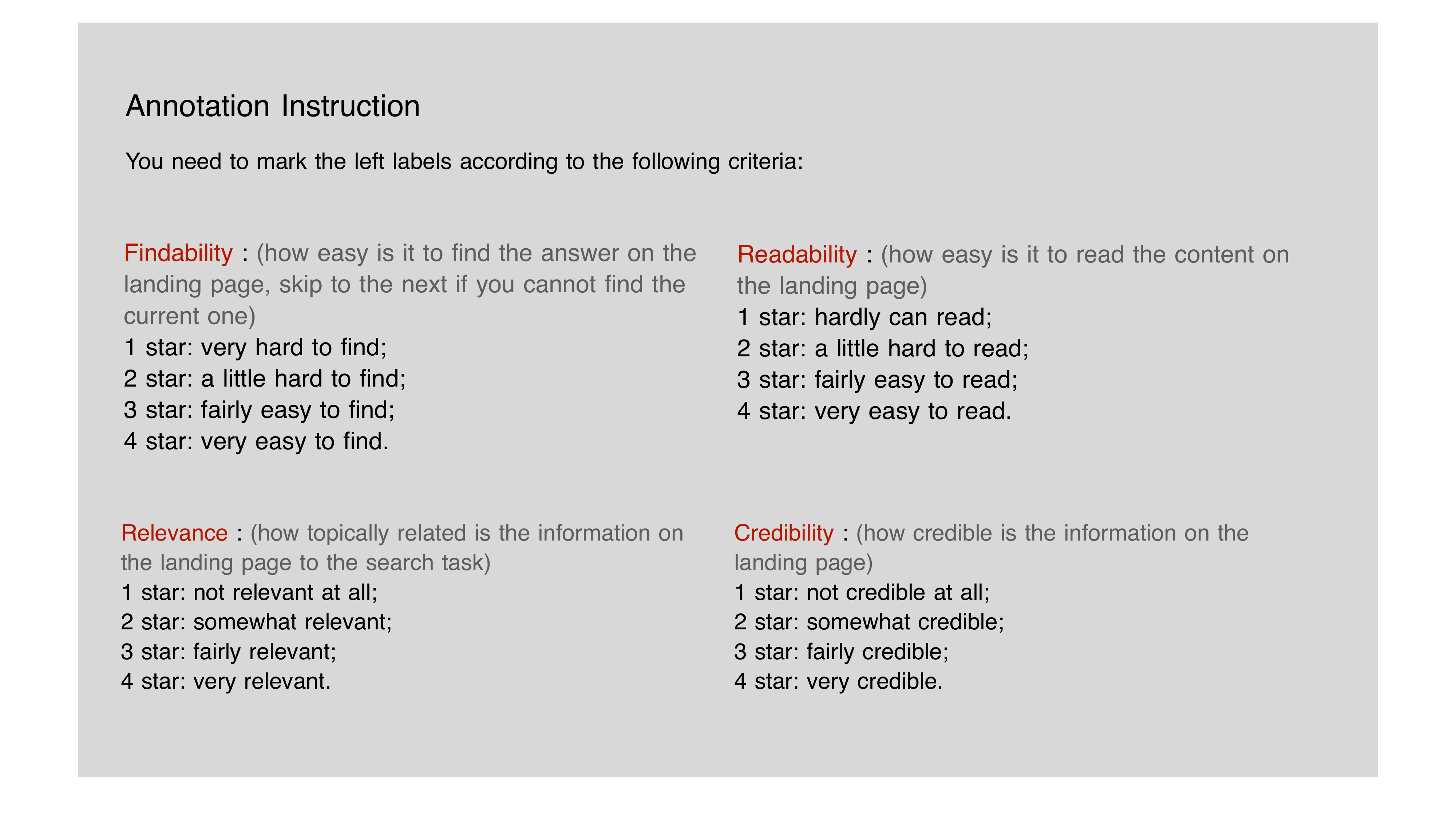
    }
    \caption{\label{fig:www_anno}The Instruction to external annotators used in the study by~\citet{Liu2018www}}
\end{figure}

30 graduate and undergraduate students were hired for the annotation. Annotators are required to read the instructional guide before the annotation. Figure~\ref{fig:www_anno} shows how \citet{Liu2018www} instructed external annotators to perform data annotation.

\begin{table}[h]
\centering
\begin{tabular}{|c|p{12cm}|}
\hline
\textbf{Task No.} & \textbf{Task Description} \\ \hline
1 & Xiaoming's department has recently hired ten new employees and needs to conduct new employee training. Please find a suitable icebreaker game for the training that is lively and effective, and describe in one sentence what preparations are needed for this game. \\ \hline
2 & Xiaoming has a week off in May and wants to explore travel options around Beijing. Please consider the costs (including transportation, accommodation, and entertainment) and recommend the best travel plan. Describe the advantages and disadvantages of this plan in three sentences. \\ \hline
3 & Xiaoming wants to buy a fixed-gear bicycle near Tsinghua University. Please find possible purchase channels, precautions, and price ranges. \\ \hline
4 & Xiaoming is preparing to fly to the US for studying and might need to carry a lot of luggage. Please research the regulations for carry-on luggage weight and prohibited items on international flights, and state three precautions. \\ \hline
7 & A friend of Xiaoming wants to quit smoking, so he wants to know the benefits of quitting, side effects, and effective ways to quit smoking. \\ \hline
8 & Xiaoming wants to know how to travel to Huangshan during the Qingming Festival. Please find the most suitable transportation and route, and state one reason for your choice. \\ \hline
10 & Xiaoming wants to know which mobile phone brands are the most popular right now and select one that he would most like to buy. \\ \hline
11 & Xiaoming wants to buy a white T-shirt on Taobao. Please suggest search keywords and explain why you chose them. \\ \hline
12 & Xiaoming's family wants to know how to choose healthy vegetables. Please find relevant information and provide three suggestions. \\ \hline
\end{tabular}
\caption{Search tasks in the user study of~\citet{mao2016} and ~\citet{Liu2019}, with their descritions.}
\label{tab:task_kdd}
\end{table}
\begin{table}[h]
\centering
\begin{tabular}{|l|l|l|}
\hline
\textbf{Measure}   & \textbf{Type}  & \textbf{Description} \\ \hline
usefulness        & 0(low)\textasciitilde 3(high) & User’s usefulness feedback on a document \\ \hline
serp\_satisfaction      & 1(low)\textasciitilde 5(high) & User’s satisfaction feedback on a query-level SERP \\ \hline
session\_satisfaction      & 1(low)\textasciitilde 5(high) & User’s satisfaction feedback on the whold search session \\ \hline
\end{tabular}
\caption{(A part of) User self-rating variables and their definitions in the study by~\citet{mao2016} and~\citet{Liu2019}.}
\label{tab:user_anno_kdd}
\end{table}

\begin{table}[h]
\centering
\begin{tabular}{|c|p{10cm}|>{\centering\arraybackslash}m{3cm}|}
\hline
\textbf{Task No.} & \textbf{Task Description} & \textbf{Domain} \\ \hline
1 & What are the characteristics of pollution particulate matter in China? Your answer should cover its compositions, its time-varying patterns, and its geographical characteristics. & Environment \\ \hline
2 & Why ultraviolet disinfection cannot completely supplant chlorination when disinfecting drinking water? And what are the advantages and disadvantages of them? & Environment \\ \hline
3 & What are the most commonly-used methods for cancer treatment in clinics? & Medicine \\ \hline
4 & What are the potential applications of 3D printing for "Precision Medicine"? & Medicine \\ \hline
5 & Political scientists have noted that the trend of political polarization during the US presidential election is increasingly evident. What are the reasons behind it? & Politics \\ \hline
6 & In order to achieve their own interests, what kind of strategies do the US interest groups often take? & Politics \\ \hline
\end{tabular}
\caption{Search tasks in the user study by~\citet{Liu2018www}, with their descritions.}
\label{tab:task_www}
\end{table}
\begin{table}[h]
\centering
\begin{tabular}{|l|l|l|}
\hline
\textbf{Measure}   & \textbf{Type}  & \textbf{Description} \\ \hline
pre\_difficulty   & 1(low)\textasciitilde 5(high)  & User perceived task difficulty \\ \hline
pre\_knowledge    & 1(low)\textasciitilde 5(high)  & User’s prior knowledge about a task \\ \hline
pre\_interest     & 1(low)\textasciitilde 5(high)  & User’s interest about a task \\ \hline
usefulness        & 1(low)\textasciitilde 4(high) & User’s usefulness feedback on a document \\ \hline
satisfaction      & 1(low)\textasciitilde 5(high) & User’s satisfaction feedback on a search session \\ \hline
\end{tabular}
\caption{User self-rating variables and their definitions. in the study by~\citet{Liu2019}..}
\label{tab:user_anno_www}
\end{table}

\clearpage
\section{Complete Regression Results of Experiment 1}
\label{ch:appendix_regression1}
This section presents the complete regression results on the THUIR2016, THU-KDD, and THUIR2018 datasets in Experiment 1 Table~\ref{tab:full_reg_click_thuir2} shows the results of logistic regression on the dependent variable \texttt{is\_clicked} in the THUIR2016 dataset. Table~\ref{tab:full_reg_time_thuir2} shows the results of OLS regression on the dependent variable \texttt{duration} in the THUIR2016 dataset. Table~\ref{tab:full_reg_usefulness_thuir2} shows the results of OLS regression on the dependent variable \texttt{usefulness} in the THUIR2016 dataset. Table~\ref{tab:full_reg_click_kdd} shows the results of logistic regression on the dependent variable \texttt{is\_clicked} in the THU-KDD dataset. Table~\ref{tab:full_reg_time_kdd} shows the results of OLS regression on the dependent variable \texttt{duration} in the THU-KDD dataset. Table~\ref{} shows the results of OLS regression on the dependent variable \texttt{usefulness} in the THU-KDD dataset. Table~\ref{tab:full_reg_click_www} shows the results of logistic regression on the dependent variable \texttt{is\_clicked} in the THUIR2018 dataset. Table~\ref{tab:full_reg_time_www} shows the results of OLS regression on the dependent variable \texttt{duration} in the THUIR2018 dataset. Table~\ref{tab:full_reg_usefulness_www} shows the results of OLS regression on the dependent variable \texttt{usefulness} in the THUIR2018 dataset.

\begin{table}[htbp]
\centering
\scriptsize
\begin{tabular}{lllllll}
\toprule
 \texttt{Variable} & Coefficient & Standard Error & z-Value & p-Value & CI Lower & CI Upper \\
\midrule
\texttt{const} &      0.3676 &          0.648 &   0.568 &    0.57 &   -0.902 &    1.637 \\
\texttt{has\_decoy} &      0.8791 &           0.22 &   3.989 &       0 &    0.447 &    1.311 \\
\texttt{rank\_2} &     -0.3345 &          0.386 &  -0.867 &   0.386 &   -1.091 &    0.422 \\
\texttt{rank\_3} &     -1.3361 &           0.43 &  -3.108 &   0.002 &   -2.179 &   -0.494 \\
\texttt{rank\_4} &     -0.7929 &          0.403 &   -1.97 &   0.049 &   -1.582 &   -0.004 \\
\texttt{rank\_5} &     -1.1756 &          0.437 &  -2.691 &   0.007 &   -2.032 &   -0.319 \\
\texttt{rank\_6} &      -1.913 &          0.466 &  -4.101 &       0 &   -2.827 &   -0.999 \\
\texttt{rank\_7} &      -2.175 &          0.452 &  -4.812 &       0 &   -3.061 &   -1.289 \\
\texttt{rank\_8} &     -1.8464 &          0.462 &  -3.997 &       0 &   -2.752 &   -0.941 \\
\texttt{rank\_9} &      -1.577 &          0.525 &  -3.004 &   0.003 &   -2.606 &   -0.548 \\
\texttt{rank\_10} &     -1.6736 &          0.449 &  -3.725 &       0 &   -2.554 &   -0.793 \\
\texttt{task\_id\_2} &      0.1104 &          0.323 &   0.342 &   0.732 &   -0.523 &    0.744 \\
\texttt{task\_id\_3} &      0.4123 &          0.294 &   1.403 &   0.161 &   -0.164 &    0.988 \\
\texttt{task\_id\_4} &       1.173 &          0.374 &   3.137 &   0.002 &     0.44 &    1.906 \\
\texttt{task\_id\_5} &     -1.2566 &           0.44 &  -2.854 &   0.004 &    -2.12 &   -0.394 \\
\texttt{task\_id\_6} &      0.6954 &          0.457 &   1.521 &   0.128 &   -0.201 &    1.591 \\
\texttt{student\_2} &     -0.7307 &          0.694 &  -1.053 &   0.292 &    -2.09 &    0.629 \\
\texttt{student\_3} &     -0.5232 &          0.786 &  -0.665 &   0.506 &   -2.064 &    1.018 \\
\texttt{student\_4} &     -2.3829 &          0.823 &  -2.896 &   0.004 &   -3.996 &    -0.77 \\
\texttt{student\_5} &      -1.071 &           0.84 &  -1.276 &   0.202 &   -2.717 &    0.575 \\
\texttt{student\_6} &     -0.7689 &          0.708 &  -1.086 &   0.277 &   -2.156 &    0.618 \\
\texttt{student\_7} &     -0.5979 &          0.778 &  -0.769 &   0.442 &   -2.122 &    0.926 \\
\texttt{student\_8} &     -1.6778 &           1.18 &  -1.422 &   0.155 &    -3.99 &    0.634 \\
\texttt{student\_9} &     -1.4871 &          0.767 &   -1.94 &   0.052 &    -2.99 &    0.015 \\
\texttt{student\_10} &     -1.4507 &          0.776 &  -1.869 &   0.062 &   -2.972 &    0.071 \\
\texttt{student\_11} &     -0.7464 &          0.741 &  -1.008 &   0.314 &   -2.198 &    0.705 \\
\texttt{student\_12} &     -1.7335 &          0.885 &  -1.959 &    0.05 &   -3.467 &    0.001 \\
\texttt{student\_13} &     -1.6054 &          0.765 &  -2.099 &   0.036 &   -3.104 &   -0.106 \\
\texttt{student\_14} &     -0.9688 &           0.69 &  -1.405 &    0.16 &    -2.32 &    0.383 \\
\texttt{student\_15} &     -1.2802 &          0.711 &  -1.802 &   0.072 &   -2.673 &    0.112 \\
\texttt{student\_16} &     -0.8133 &          0.733 &   -1.11 &   0.267 &    -2.25 &    0.623 \\
\texttt{student\_17} &     -0.6112 &          0.719 &  -0.851 &   0.395 &   -2.019 &    0.797 \\
\texttt{student\_18} &     -0.4251 &          0.665 &  -0.639 &   0.523 &   -1.729 &    0.879 \\
\texttt{student\_19} &      -1.014 &          0.735 &   -1.38 &   0.168 &   -2.455 &    0.427 \\
\texttt{student\_20} &     -0.5814 &          0.746 &   -0.78 &   0.436 &   -2.043 &     0.88 \\
\texttt{student\_21} &     -1.3491 &          0.837 &  -1.611 &   0.107 &    -2.99 &    0.292 \\
\texttt{student\_22} &     -1.2576 &          0.831 &  -1.513 &    0.13 &   -2.887 &    0.372 \\
\texttt{student\_23} &      0.5394 &          0.709 &   0.761 &   0.447 &    -0.85 &    1.929 \\
\texttt{student\_24} &     -1.2317 &          0.749 &  -1.644 &     0.1 &     -2.7 &    0.237 \\
\texttt{student\_25} &     -2.2676 &           0.98 &  -2.313 &   0.021 &   -4.189 &   -0.346 \\
\texttt{student\_26} &     -0.9942 &          0.741 &  -1.341 &    0.18 &   -2.447 &    0.458 \\
\texttt{student\_27} &     -0.3854 &          0.699 &  -0.551 &   0.582 &   -1.756 &    0.985 \\
\texttt{student\_28} &      -1.075 &          0.808 &  -1.331 &   0.183 &   -2.658 &    0.508 \\
\bottomrule
\end{tabular}
\caption{Complete regression results of Experiment 1 for the dependent variable \texttt{is\_clicked} on the THUIR2018 dataset.}
\label{tab:full_reg_click_www}
\end{table}
\begin{table}[htbp]
\centering
\scriptsize
\begin{tabular}{lllllll}
\toprule
 \texttt{Variable} & Coefficient & Standard Error & t-Value & p-Value & CI Lower & CI Upper \\
\midrule
        \texttt{const} &    67.835 &  36.565 &  1.855 &  0.064 &   -3.951 &  139.621 \\
    \texttt{has\_decoy} &   51.5206 &  11.736 &   4.39 &      0 &   28.481 &   74.561 \\
       \texttt{rank\_2} &   -1.5181 &  23.354 & -0.065 &  0.948 &  -47.368 &   44.332 \\
       \texttt{rank\_3} &   -24.065 &  23.872 & -1.008 &  0.314 &  -70.931 &   22.801 \\
       \texttt{rank\_4} &  -17.4902 &  23.596 & -0.741 &  0.459 &  -63.815 &   28.834 \\
       \texttt{rank\_5} &  -20.4777 &  24.582 & -0.833 &  0.405 &  -68.739 &   27.783 \\
       \texttt{rank\_6} &   -56.211 &   23.56 & -2.386 &  0.017 & -102.466 &   -9.956 \\
       \texttt{rank\_7} &  -52.8948 &  22.732 & -2.327 &   0.02 &  -97.523 &   -8.267 \\
       \texttt{rank\_8} &  -44.6302 &   23.56 & -1.894 &  0.059 &  -90.884 &    1.624 \\
       \texttt{rank\_9} &  -50.7611 &  27.376 & -1.854 &  0.064 & -104.507 &    2.985 \\
      \texttt{rank\_10} &  -48.0745 &  23.686 &  -2.03 &  0.043 &  -94.577 &   -1.572 \\
    \texttt{task\_id\_2} &   11.9637 &  16.088 &  0.744 &  0.457 &  -19.622 &   43.549 \\
    \texttt{task\_id\_3} &    5.4719 &  15.382 &  0.356 &  0.722 &  -24.727 &   35.671 \\
    \texttt{task\_id\_4} &   25.5298 &  19.494 &   1.31 &  0.191 &  -12.742 &   63.801 \\
    \texttt{task\_id\_5} &  -30.7869 &  18.534 & -1.661 &  0.097 &  -67.174 &      5.6 \\
    \texttt{task\_id\_6} &  -26.7088 &  24.575 & -1.087 &  0.277 &  -74.956 &   21.538 \\
 \texttt{student\_id\_2} &  -23.6528 &  37.507 & -0.631 &  0.528 &  -97.289 &   49.983 \\
 \texttt{student\_id\_3} &  -12.2459 &  44.454 & -0.275 &  0.783 &  -98.988 &   74.496 \\
 \texttt{student\_id\_4} &  -50.6921 &  42.838 & -1.183 &  0.238 & -135.659 &   34.275 \\
 \texttt{student\_id\_5} &   55.3693 &  43.144 &  1.283 &  0.201 &  -29.392 &  140.131 \\
 \texttt{student\_id\_6} &  -33.9244 &  38.182 & -0.888 &  0.375 & -108.885 &   41.036 \\
 \texttt{student\_id\_7} &  -25.2137 &  43.157 & -0.584 &  0.559 & -109.942 &   59.514 \\
 \texttt{student\_id\_8} &  -24.7322 &  47.047 & -0.526 &  0.599 & -117.097 &   67.632 \\
 \texttt{student\_id\_9} &  -39.7449 &  39.684 & -1.002 &  0.317 & -117.655 &   38.165 \\
\texttt{student\_id\_10} &  -14.9165 &  41.329 & -0.361 &  0.718 &  -96.055 &   66.222 \\
\texttt{student\_id\_11} &     1.049 &  42.298 &  0.025 &   0.98 &  -81.992 &    84.09 \\
\texttt{student\_id\_12} &   -40.299 &  43.728 & -0.922 &  0.357 & -126.148 &   45.551 \\
\texttt{student\_id\_13} &  -37.0122 &  37.626 & -0.984 &  0.326 & -110.881 &   36.856 \\
\texttt{student\_id\_14} &  -33.7491 &  37.077 &  -0.91 &  0.363 &  -106.54 &   39.042 \\
\texttt{student\_id\_15} &   -4.5565 &  37.202 & -0.122 &  0.903 &  -77.593 &    68.48 \\
\texttt{student\_id\_16} &   88.2735 &   42.93 &  2.056 &   0.04 &    3.991 &  172.556 \\
\texttt{student\_id\_17} &   -9.6742 &  39.272 & -0.246 &  0.805 &  -86.774 &   67.426 \\
\texttt{student\_id\_18} &   -2.7298 &  36.683 & -0.074 &  0.941 &  -74.748 &   69.289 \\
\texttt{student\_id\_19} &   48.1401 &  40.665 &  1.184 &  0.237 &  -31.695 &  127.975 \\
\texttt{student\_id\_20} &   34.4275 &  42.272 &  0.814 &  0.416 &  -48.563 &  117.418 \\
\texttt{student\_id\_21} &   13.3694 &  45.195 &  0.296 &  0.767 &  -75.359 &  102.098 \\
\texttt{student\_id\_22} &  -17.6902 &  43.057 & -0.411 &  0.681 & -102.222 &   66.842 \\
\texttt{student\_id\_23} &    3.3337 &  39.818 &  0.084 &  0.933 &   -74.84 &   81.507 \\
\texttt{student\_id\_24} &   25.7126 &    38.6 &  0.666 &  0.506 &  -50.069 &  101.494 \\
\texttt{student\_id\_25} &  -49.7761 &  43.212 & -1.152 &   0.25 & -134.612 &    35.06 \\
\texttt{student\_id\_26} &   -7.4722 &  39.075 & -0.191 &  0.848 &  -84.186 &   69.242 \\
\texttt{student\_id\_27} &   16.1828 &  39.525 &  0.409 &  0.682 &  -61.414 &   93.779 \\
\texttt{student\_id\_28} &  -37.8587 &  42.686 & -0.887 &  0.375 & -121.662 &   45.945 \\
\bottomrule
\end{tabular}
\caption{Complete regression results Experiment 1  for the dependent variable \texttt{duration} on the THUIR2018 dataset.}
\label{tab:full_reg_time_www}
\end{table}
\begin{table}[htbp]
\centering
\scriptsize
\begin{tabular}{lllllll}
\toprule
\texttt{Variable} & Coefficient & Standard Error & t-Value & p-Value & CI Lower & CI Upper \\
\midrule
        \texttt{const} &   0.779 &   0.203 &  3.846 &      0 &  0.381 &  1.177 \\
    \texttt{has\_decoy} &  0.3579 &   0.065 &  5.505 &      0 &   0.23 &  0.486 \\
       \texttt{rank\_2} & -0.2148 &   0.129 &  -1.66 &  0.097 & -0.469 &  0.039 \\
       \texttt{rank\_3} & -0.4331 &   0.132 & -3.275 &  0.001 & -0.693 & -0.173 \\
       \texttt{rank\_4} & -0.4489 &   0.131 & -3.434 &  0.001 & -0.706 & -0.192 \\
       \texttt{rank\_5} & -0.4531 &   0.136 & -3.327 &  0.001 &  -0.72 & -0.186 \\
       \texttt{rank\_6} & -0.5962 &   0.131 & -4.567 &      0 & -0.852 &  -0.34 \\
       \texttt{rank\_7} &   -0.67 &   0.126 &  -5.32 &      0 & -0.917 & -0.423 \\
       \texttt{rank\_8} & -0.6105 &   0.131 & -4.677 &      0 & -0.867 & -0.354 \\
       \texttt{rank\_9} & -0.5754 &   0.152 & -3.794 &      0 & -0.873 & -0.278 \\
      \texttt{rank\_10} & -0.5939 &   0.131 & -4.526 &      0 & -0.852 & -0.336 \\
    \texttt{task\_id\_2} &  0.1573 &   0.089 &  1.765 &  0.078 & -0.018 &  0.332 \\
    \texttt{task\_id\_3} &  0.1389 &   0.085 &  1.629 &  0.104 & -0.028 &  0.306 \\
    \texttt{task\_id\_4} &  0.3764 &   0.108 &  3.485 &  0.001 &  0.164 &  0.588 \\
    \texttt{task\_id\_5} & -0.2161 &   0.103 & -2.104 &  0.036 & -0.418 & -0.014 \\
    \texttt{task\_id\_6} &  0.0573 &   0.136 &   0.42 &  0.674 &  -0.21 &  0.325 \\
 \texttt{student\_id\_2} & -0.0743 &   0.208 & -0.357 &  0.721 & -0.482 &  0.334 \\
 \texttt{student\_id\_3} & -0.1073 &   0.246 & -0.436 &  0.663 & -0.591 &  0.376 \\
 \texttt{student\_id\_4} & -0.4411 &   0.204 & -2.162 &  0.031 & -0.842 & -0.041 \\
 \texttt{student\_id\_5} &  0.2168 &   0.256 &  0.845 &  0.398 & -0.287 &   0.72 \\
 \texttt{student\_id\_6} & -0.1587 &   0.212 &  -0.75 &  0.453 & -0.574 &  0.257 \\
 \texttt{student\_id\_7} &  0.0076 &   0.239 &  0.032 &  0.975 & -0.462 &  0.477 \\
 \texttt{student\_id\_8} & -0.2887 &   0.261 & -1.107 &  0.268 &   -0.8 &  0.223 \\
 \texttt{student\_id\_9} & -0.3127 &    0.22 & -1.422 &  0.155 & -0.744 &  0.119 \\
\texttt{student\_id\_10} & -0.2081 &   0.229 & -0.909 &  0.364 & -0.658 &  0.241 \\
\texttt{student\_id\_11} & -0.2281 &   0.234 & -0.973 &  0.331 & -0.688 &  0.232 \\
\texttt{student\_id\_12} & -0.1915 &   0.242 &  -0.79 &   0.43 & -0.667 &  0.284 \\
\texttt{student\_id\_13} & -0.3407 &   0.208 & -1.635 &  0.103 &  -0.75 &  0.069 \\
\texttt{student\_id\_14} & -0.1637 &   0.205 & -0.797 &  0.426 & -0.567 &   0.24 \\
\texttt{student\_id\_15} & -0.2297 &   0.206 & -1.114 &  0.266 & -0.634 &  0.175 \\
\texttt{student\_id\_16} & -0.0607 &   0.238 & -0.255 &  0.799 & -0.528 &  0.406 \\
\texttt{student\_id\_17} & -0.0389 &   0.218 & -0.179 &  0.858 & -0.466 &  0.388 \\
\texttt{student\_id\_18} & -0.0226 &   0.203 & -0.111 &  0.912 & -0.422 &  0.376 \\
\texttt{student\_id\_19} & -0.2175 &   0.225 & -0.965 &  0.335 &  -0.66 &  0.225 \\
\texttt{student\_id\_20} & -0.0718 &   0.234 & -0.307 &  0.759 & -0.532 &  0.388 \\
\texttt{student\_id\_21} & -0.1422 &    0.25 & -0.568 &   0.57 & -0.634 &  0.349 \\
\texttt{student\_id\_22} &  0.0526 &   0.239 &  0.221 &  0.825 & -0.416 &  0.521 \\
\texttt{student\_id\_23} &  0.0354 &   0.221 &   0.16 &  0.873 & -0.398 &  0.468 \\
\texttt{student\_id\_24} & -0.2288 &   0.214 &  -1.07 &  0.285 & -0.649 &  0.191 \\
\texttt{student\_id\_25} & -0.4601 &   0.239 & -1.922 &  0.055 &  -0.93 &   0.01 \\
\texttt{student\_id\_26} & -0.0217 &   0.216 &   -0.1 &   0.92 & -0.447 &  0.403 \\
\texttt{student\_id\_27} &  0.0007 &   0.219 &  0.003 &  0.997 & -0.429 &  0.431 \\
\texttt{student\_id\_28} & -0.3037 &   0.236 & -1.284 &  0.199 & -0.768 &  0.161 \\
\bottomrule
\end{tabular}
\caption{Complete regression results Experiment 1  for the dependent variable \texttt{usefulness} on the THUIR2018 dataset.}
\label{tab:full_reg_usefulness_www}
\end{table}
\begin{table}[htbp]
\centering
\scriptsize
\begin{tabular}{lllllll}
\toprule
 \texttt{Variable} & Coefficient & Standard Error & z-Value & p-Value & CI Lower & CI Upper \\
\midrule
        \texttt{const} & -1.3902 &   0.422 & -3.296 &  0.001 & -2.217 & -0.564 \\
    \texttt{has\_decoy} &  0.3631 &   0.158 &  2.297 &  0.022 &  0.053 &  0.673 \\
       \texttt{rank\_1} & -0.4594 &   0.214 & -2.146 &  0.032 & -0.879 &  -0.04 \\
       \texttt{rank\_2} & -0.9429 &   0.239 & -3.953 &      0 &  -1.41 & -0.475 \\
       \texttt{rank\_3} & -1.8111 &   0.292 & -6.205 &      0 & -2.383 & -1.239 \\
       \texttt{rank\_4} & -1.8203 &   0.304 & -5.988 &      0 & -2.416 & -1.225 \\
       \texttt{rank\_5} & -1.4252 &   0.286 &  -4.99 &      0 & -1.985 & -0.865 \\
       \texttt{rank\_6} &  -1.238 &   0.345 &  -3.59 &      0 & -1.914 & -0.562 \\
       \texttt{rank\_7} & -1.0311 &   0.322 & -3.199 &  0.001 & -1.663 & -0.399 \\
       \texttt{rank\_8} & -1.6891 &   0.412 & -4.102 &      0 & -2.496 & -0.882 \\
       \texttt{rank\_9} & -0.8941 &   0.296 & -3.019 &  0.003 & -1.475 & -0.314 \\
   \texttt{task\_id\_10} &  0.7163 &   0.336 &  2.132 &  0.033 &  0.058 &  1.375 \\
   \texttt{task\_id\_11} &  0.5444 &   0.272 &      2 &  0.046 &  0.011 &  1.078 \\
   \texttt{task\_id\_12} &  0.6176 &   0.344 &  1.797 &  0.072 & -0.056 &  1.291 \\
    \texttt{task\_id\_2} &  0.2825 &   0.302 &  0.935 &   0.35 &  -0.31 &  0.875 \\
    \texttt{task\_id\_3} & -0.1449 &   0.283 & -0.512 &  0.609 &   -0.7 &   0.41 \\
    \texttt{task\_id\_4} & -0.2481 &   0.383 & -0.648 &  0.517 & -0.998 &  0.502 \\
    \texttt{task\_id\_7} & -0.0685 &   0.332 & -0.206 &  0.836 & -0.719 &  0.582 \\
    \texttt{task\_id\_8} &  1.2105 &   0.327 &  3.704 &      0 &   0.57 &  1.851 \\
 \texttt{student\_id\_2} &  -0.155 &   0.513 & -0.302 &  0.762 &  -1.16 &   0.85 \\
 \texttt{student\_id\_3} &  0.2898 &   0.468 &  0.619 &  0.536 & -0.628 &  1.207 \\
 \texttt{student\_id\_4} & -1.3485 &   0.689 & -1.958 &   0.05 & -2.698 &  0.001 \\
 \texttt{student\_id\_5} &  0.1614 &   0.475 &   0.34 &  0.734 &  -0.77 &  1.093 \\
 \texttt{student\_id\_6} & -0.4368 &   0.559 & -0.782 &  0.434 & -1.532 &  0.659 \\
 \texttt{student\_id\_7} &  0.6382 &   0.463 &  1.378 &  0.168 & -0.269 &  1.546 \\
 \texttt{student\_id\_8} & -0.8768 &   0.577 & -1.519 &  0.129 & -2.008 &  0.254 \\
 \texttt{student\_id\_9} &  1.2037 &   0.423 &  2.845 &  0.004 &  0.374 &  2.033 \\
\texttt{student\_id\_10} &   0.451 &   0.457 &  0.987 &  0.324 & -0.445 &  1.347 \\
\texttt{student\_id\_11} &  -0.163 &   0.464 & -0.351 &  0.725 & -1.072 &  0.746 \\
\texttt{student\_id\_12} &  1.6076 &   0.403 &   3.99 &      0 &  0.818 &  2.397 \\
\texttt{student\_id\_13} &  0.0042 &   0.495 &  0.008 &  0.993 & -0.966 &  0.974 \\
\texttt{student\_id\_14} & -0.9966 &   0.573 & -1.739 &  0.082 &  -2.12 &  0.127 \\
\texttt{student\_id\_15} & -0.0451 &    0.44 & -0.102 &  0.918 & -0.908 &  0.818 \\
\texttt{student\_id\_16} & -1.6169 &   0.572 & -2.828 &  0.005 & -2.738 & -0.496 \\
\texttt{student\_id\_17} &  0.4838 &   0.463 &  1.044 &  0.296 & -0.424 &  1.392 \\
\texttt{student\_id\_18} &  0.0188 &   0.456 &  0.041 &  0.967 & -0.874 &  0.912 \\
\texttt{student\_id\_19} &  -2.253 &   0.795 & -2.833 &  0.005 & -3.812 & -0.694 \\
\texttt{student\_id\_20} & -1.1865 &   0.688 & -1.725 &  0.085 & -2.535 &  0.162 \\
\texttt{student\_id\_21} &  0.1718 &   0.432 &  0.397 &  0.691 & -0.676 &  1.019 \\
\texttt{student\_id\_22} &  0.5019 &   0.489 &  1.026 &  0.305 & -0.457 &   1.46 \\
\texttt{student\_id\_23} &  0.6892 &   0.465 &  1.481 &  0.139 & -0.223 &  1.601 \\
\texttt{student\_id\_24} &  0.0746 &   0.426 &  0.175 &  0.861 &  -0.76 &  0.909 \\
\texttt{student\_id\_25} &   0.346 &   0.461 &  0.751 &  0.453 & -0.557 &  1.249 \\
\bottomrule
\end{tabular}
\caption{Complete regression results of Experiment 1 for the dependent variable \texttt{is\_clicked} on the THUIR2016 dataset.}
\label{tab:full_reg_click_thuir2}
\end{table}
\begin{table}[htbp]
\centering
\scriptsize
\begin{tabular}{lllllll}
\toprule
 \texttt{Variable} & Coefficient & Standard Error & t-Value & p-Value & CI Lower & CI Upper \\
\midrule
  \texttt{const} &  11.5842 &   3.312 &  3.498 &      0 &   5.089 &  18.079 \\
    \texttt{has\_decoy} &   1.9164 &   1.209 &  1.585 &  0.113 &  -0.455 &   4.288 \\
       \texttt{rank\_1} &  -5.0637 &    1.97 & -2.571 &   0.01 &  -8.927 &  -1.201 \\
       \texttt{rank\_2} &   -3.464 &   2.013 & -1.721 &  0.085 &  -7.412 &   0.484 \\
       \texttt{rank\_3} &  -7.9568 &   2.059 & -3.864 &      0 & -11.995 &  -3.918 \\
       \texttt{rank\_4} &  -8.2297 &   2.183 &  -3.77 &      0 &  -12.51 &  -3.949 \\
       \texttt{rank\_5} &  -3.2812 &   2.188 &   -1.5 &  0.134 &  -7.572 &   1.009 \\
       \texttt{rank\_6} &  -9.4029 &   2.625 & -3.582 &      0 & -14.551 &  -4.254 \\
       \texttt{rank\_7} &   -7.243 &   2.587 &   -2.8 &  0.005 & -12.316 &   -2.17 \\
       \texttt{rank\_8} &  -6.6333 &    2.74 & -2.421 &  0.016 & -12.006 &   -1.26 \\
       \texttt{rank\_9} &  -4.0138 &   2.472 & -1.623 &  0.105 &  -8.862 &   0.835 \\
   \texttt{task\_id\_10} &   0.3123 &   2.704 &  0.116 &  0.908 &   -4.99 &   5.615 \\
   \texttt{task\_id\_11} &  -0.9547 &   2.033 & -0.469 &  0.639 &  -4.942 &   3.033 \\
   \texttt{task\_id\_12} &   8.8792 &   2.634 &  3.371 &  0.001 &   3.714 &  14.045 \\
    \texttt{task\_id\_2} &    3.733 &   2.178 &  1.714 &  0.087 &  -0.539 &   8.005 \\
    \texttt{task\_id\_3} &  -3.1521 &   2.023 & -1.558 &  0.119 &  -7.119 &   0.815 \\
    \texttt{task\_id\_4} &  -0.0657 &   2.648 & -0.025 &   0.98 &  -5.258 &   5.127 \\
    \texttt{task\_id\_7} &   1.3948 &   2.317 &  0.602 &  0.547 &   -3.15 &   5.939 \\
    \texttt{task\_id\_8} &   4.5707 &   2.015 &  2.268 &  0.023 &   0.621 &   8.521 \\
    \texttt{task\_id\_9} &   1.1564 &   2.559 &  0.452 &  0.651 &  -3.861 &   6.174 \\
 \texttt{student\_id\_2} &  -3.6587 &   3.843 & -0.952 &  0.341 & -11.195 &   3.877 \\
 \texttt{student\_id\_3} &  -6.6009 &   3.752 & -1.759 &  0.079 &  -13.96 &   0.758 \\
 \texttt{student\_id\_4} &  -7.4009 &   3.873 & -1.911 &  0.056 & -14.997 &   0.195 \\
 \texttt{student\_id\_5} &   2.4163 &   3.728 &  0.648 &  0.517 &  -4.896 &   9.728 \\
 \texttt{student\_id\_6} &  -3.6167 &   4.062 &  -0.89 &  0.373 & -11.583 &   4.349 \\
 \texttt{student\_id\_7} &  -0.5733 &    3.82 &  -0.15 &  0.881 &  -8.065 &   6.918 \\
 \texttt{student\_id\_8} &  -7.8228 &   3.738 & -2.093 &  0.036 & -15.153 &  -0.493 \\
 \texttt{student\_id\_9} &   4.3682 &   3.799 &   1.15 &   0.25 &  -3.082 &  11.819 \\
\texttt{student\_id\_10} &  -2.2482 &   3.831 & -0.587 &  0.557 &  -9.761 &   5.264 \\
\texttt{student\_id\_11} &  -3.5599 &   3.571 & -0.997 &  0.319 & -10.562 &   3.442 \\
\texttt{student\_id\_12} &   2.6864 &   3.554 &  0.756 &   0.45 &  -4.284 &   9.657 \\
\texttt{student\_id\_13} &  -4.6899 &   3.786 & -1.239 &  0.216 & -12.114 &   2.734 \\
\texttt{student\_id\_14} &  -6.4066 &   3.564 & -1.797 &  0.072 & -13.396 &   0.583 \\
\texttt{student\_id\_15} &  -1.0607 &   3.376 & -0.314 &  0.753 &  -7.682 &   5.561 \\
\texttt{student\_id\_16} &  -7.2004 &   3.281 & -2.195 &  0.028 & -13.634 &  -0.767 \\
\texttt{student\_id\_17} &  -1.8364 &   3.708 & -0.495 &   0.62 &  -9.108 &   5.435 \\
\texttt{student\_id\_18} &  -5.4404 &   3.522 & -1.545 &  0.123 & -12.348 &   1.467 \\
\texttt{student\_id\_19} &  -7.8868 &    3.46 &  -2.28 &  0.023 & -14.672 &  -1.102 \\
\texttt{student\_id\_20} &  -6.6542 &   3.966 & -1.678 &  0.093 & -14.431 &   1.123 \\
\texttt{student\_id\_21} &  -2.1011 &   3.428 & -0.613 &   0.54 &  -8.825 &   4.623 \\
\texttt{student\_id\_22} &   2.6174 &   4.137 &  0.633 &  0.527 &  -5.497 &  10.731 \\
\texttt{student\_id\_23} &    1.785 &   3.945 &  0.453 &  0.651 &  -5.951 &   9.521 \\
\texttt{student\_id\_24} &  -1.9769 &   3.317 & -0.596 &  0.551 &  -8.482 &   4.528 \\
\texttt{student\_id\_25} &  -5.2637 &   3.785 & -1.391 &  0.164 & -12.686 &   2.158 \\
\bottomrule
\end{tabular}
\caption{Complete regression results of Experiment 1 for the dependent variable \texttt{duration} on the THUIR2016 dataset.}
\label{tab:full_reg_time_thuir2}
\end{table}
\begin{table}[htbp]
\centering
\scriptsize
\begin{tabular}{lllllll}
\toprule
\texttt{Variable} & Coefficient & Standard Error & t-Value & p-Value & CI Lower & CI Upper \\
\midrule
   \texttt{const} &  0.5795 &    0.126 &  4.606 &  0.000 &   0.333 &   0.826 \\
    \texttt{has\_decoy} &  0.1355 &    0.046 &  2.948 &  0.003 &   0.045 &   0.226 \\
       \texttt{rank\_1} & -0.1996 &    0.075 & -2.668 &  0.008 &  -0.346 &  -0.053 \\
       \texttt{rank\_2} & -0.3049 &    0.076 & -3.987 &  0.000 &  -0.455 &  -0.155 \\
       \texttt{rank\_3} & -0.5336 &    0.078 & -6.822 &  0.000 &  -0.687 &  -0.380 \\
       \texttt{rank\_4} & -0.5202 &    0.083 & -6.275 &  0.000 &  -0.683 &  -0.358 \\
       \texttt{rank\_5} & -0.3684 &    0.083 & -4.433 &  0.000 &  -0.531 &  -0.205 \\
       \texttt{rank\_6} & -0.4099 &    0.100 & -4.110 &  0.000 &  -0.605 &  -0.214 \\
       \texttt{rank\_7} & -0.3455 &    0.098 & -3.516 &  0.000 &  -0.538 &  -0.153 \\
       \texttt{rank\_8} & -0.4864 &    0.104 & -4.673 &  0.000 &  -0.690 &  -0.282 \\
       \texttt{rank\_9} & -0.2362 &    0.094 & -2.515 &  0.012 &  -0.420 &  -0.052 \\
   \texttt{task\_id\_10} &  0.2642 &    0.103 &  2.573 &  0.010 &   0.063 &   0.466 \\
   \texttt{task\_id\_11} &  0.0378 &    0.077 &  0.490 &  0.624 &  -0.114 &   0.189 \\
   \texttt{task\_id\_12} &  0.1077 &    0.100 &  1.076 &  0.282 &  -0.089 &   0.304 \\
    \texttt{task\_id\_2} &  0.0292 &    0.083 &  0.353 &  0.724 &  -0.133 &   0.191 \\
    \texttt{task\_id\_3} & -0.1196 &    0.077 & -1.556 &  0.120 &  -0.270 &   0.031 \\
    \texttt{task\_id\_4} & -0.1092 &    0.101 & -1.086 &  0.278 &  -0.306 &   0.088 \\
    \texttt{task\_id\_7} &  0.0323 &    0.088 &  0.367 &  0.713 &  -0.140 &   0.205 \\
    \texttt{task\_id\_8} &  0.0108 &    0.110 &  0.098 &  0.922 &  -0.205 &   0.227 \\
 \texttt{student\_id\_2} &  0.0851 &    0.146 &  0.585 &  0.559 &  -0.200 &   0.370 \\
 \texttt{student\_id\_3} & -0.1386 &    0.136 & -1.022 &  0.307 &  -0.405 &   0.127 \\
 \texttt{student\_id\_4} &  0.7134 &    0.135 &  5.284 &  0.000 &   0.449 &   0.978 \\
 \texttt{student\_id\_5} & -0.1294 &    0.144 & -0.900 &  0.368 &  -0.411 &   0.153 \\
 \texttt{student\_id\_6} & -0.1974 &    0.135 & -1.458 &  0.145 &  -0.463 &   0.068 \\
 \texttt{student\_id\_7} & -0.0117 &    0.128 & -0.091 &  0.927 &  -0.263 &   0.240 \\
 \texttt{student\_id\_8} & -0.2733 &    0.125 & -2.193 &  0.028 &  -0.518 &  -0.029 \\
 \texttt{student\_id\_9} &  0.2555 &    0.141 &  1.814 &  0.070 &  -0.021 &   0.532 \\
\texttt{student\_id\_10} & -0.0508 &    0.134 & -0.380 &  0.704 &  -0.313 &   0.212 \\
\texttt{student\_id\_11} & -0.3020 &    0.131 & -2.298 &  0.022 &  -0.560 &  -0.044 \\
\texttt{student\_id\_12} & -0.0285 &    0.146 & -0.196 &  0.845 &  -0.315 &   0.258 \\
\texttt{student\_id\_13} & -0.2484 &    0.151 & -1.649 &  0.099 &  -0.544 &   0.047 \\
\texttt{student\_id\_14} & -0.0198 &    0.130 & -0.152 &  0.879 &  -0.275 &   0.236 \\
\texttt{student\_id\_15} &  0.2073 &    0.157 &  1.319 &  0.187 &  -0.101 &   0.515 \\
\texttt{student\_id\_16} &  0.3659 &    0.150 &  2.442 &  0.015 &   0.072 &   0.660 \\
\texttt{student\_id\_17} & -0.0300 &    0.126 & -0.238 &  0.812 &  -0.277 &   0.217 \\
\texttt{student\_id\_18} &  0.1061 &    0.144 &  0.738 &  0.461 &  -0.176 &   0.388 \\
\texttt{student\_id\_19} &  0.0141 &    0.143 &  0.099 &  0.921 &  -0.265 &   0.294 \\
\texttt{student\_id\_20} & -0.2487 &    0.147 & -1.691 &  0.091 &  -0.537 &   0.040 \\
\texttt{student\_id\_21} & -0.0637 &    0.142 & -0.450 &  0.653 &  -0.341 &   0.214 \\
\texttt{student\_id\_22} & -0.1725 &    0.154 & -1.118 &  0.264 &  -0.475 &   0.130 \\
\texttt{student\_id\_23} &  0.2111 &    0.145 &  1.455 &  0.146 &  -0.073 &   0.496 \\
\texttt{student\_id\_24} & -0.1700 &    0.142 & -1.197 &  0.231 &  -0.448 &   0.108 \\
\texttt{student\_id\_25} &  0.4176 &    0.144 &  2.894 &  0.004 &   0.135 &   0.701 \\
\bottomrule
\end{tabular}
\caption{Complete regression results Experiment 1  for the dependent variable \texttt{usefulness} on the THUIR2016 dataset.}
\label{tab:full_reg_usefulness_thuir2}
\end{table}
\begin{table}[htbp]
\centering
\scriptsize
\begin{tabular}{lllllll}
\toprule
 \texttt{Variable} & Coefficient & Standard Error & z-Value & p-Value & CI Lower & CI Upper \\
\midrule
        \texttt{const} &  0.5982 &   0.377 &  1.588 &  0.112 & -0.140 &  1.336 \\
    \texttt{has\_decoy} &  0.2172 &   0.105 &  2.062 &  0.039 &  0.011 &  0.424 \\
       \texttt{rank\_1} & -0.7115 &   0.135 & -5.255 &  0.000 & -0.977 & -0.446 \\
       \texttt{rank\_2} & -1.1671 &   0.155 & -7.553 &  0.000 & -1.470 & -0.864 \\
       \texttt{rank\_3} & -1.5748 &   0.170 & -9.271 &  0.000 & -1.908 & -1.242 \\
       \texttt{rank\_4} & -1.5123 &   0.162 & -9.328 &  0.000 & -1.830 & -1.195 \\
       \texttt{rank\_5} & -1.8044 &   0.177 & -10.206 &  0.000 & -2.151 & -1.458 \\
       \texttt{rank\_6} & -2.3518 &   0.235 & -9.996 &  0.000 & -2.813 & -1.891 \\
       \texttt{rank\_7} & -2.0788 &   0.214 & -9.722 &  0.000 & -2.498 & -1.660 \\
       \texttt{rank\_8} & -2.1094 &   0.237 & -8.895 &  0.000 & -2.574 & -1.645 \\
       \texttt{rank\_9} & -3.9716 &   0.514 & -7.720 &  0.000 & -4.980 & -2.963 \\
   \texttt{task\_id\_10} & -0.2717 &   0.226 & -1.203 &  0.229 & -0.714 &  0.171 \\
   \texttt{task\_id\_11} & -0.3483 &   0.238 & -1.466 &  0.143 & -0.814 &  0.117 \\
   \texttt{task\_id\_12} & -0.5420 &   0.240 & -2.263 &  0.024 & -1.011 & -0.073 \\
    \texttt{task\_id\_2} & -0.2550 &   0.229 & -1.113 &  0.266 & -0.704 &  0.194 \\
    \texttt{task\_id\_3} & -0.1727 &   0.219 & -0.790 &  0.429 & -0.601 &  0.256 \\
    \texttt{task\_id\_4} & -0.0623 &   0.228 & -0.273 &  0.785 & -0.509 &  0.384 \\
    \texttt{task\_id\_7} & -0.3642 &   0.234 & -1.560 &  0.119 & -0.822 &  0.093 \\
    \texttt{task\_id\_8} & -0.1673 &   0.233 & -0.718 &  0.473 & -0.624 &  0.289 \\
\texttt{student\_id\_2} & -0.8214 &   0.570 & -1.442 &  0.149 & -1.938 &  0.295 \\
\texttt{student\_id\_3} &  0.2732 &   0.424 &  0.645 &  0.519 & -0.557 &  1.103 \\
\texttt{student\_id\_4} & -0.7808 &   0.860 & -0.908 &  0.364 & -2.466 &  0.905 \\
\texttt{student\_id\_5} & -0.3962 &   0.559 & -0.709 &  0.479 & -1.492 &  0.700 \\
\texttt{student\_id\_6} & -0.8525 &   0.409 & -2.086 &  0.037 & -1.653 & -0.052 \\
\texttt{student\_id\_7} &  0.3024 &   0.407 &  0.743 &  0.458 & -0.495 &  1.100 \\
\texttt{student\_id\_8} & -1.1595 &   0.657 & -1.765 &  0.078 & -2.447 &  0.128 \\
\texttt{student\_id\_9} & -1.1246 &   0.512 & -2.198 &  0.028 & -2.127 & -0.122 \\
\texttt{student\_id\_10} & -0.5343 &   0.510 & -1.048 &  0.295 & -1.534 &  0.465 \\
\texttt{student\_id\_11} & -1.1104 &   0.440 & -2.524 &  0.012 & -1.973 & -0.248 \\
\texttt{student\_id\_12} & -0.2163 &   0.452 & -0.479 &  0.632 & -1.102 &  0.670 \\
\texttt{student\_id\_13} & -0.6042 &   0.447 & -1.351 &  0.177 & -1.480 &  0.272 \\
\texttt{student\_id\_14} & -0.1328 &   0.419 & -0.317 &  0.751 & -0.954 &  0.688 \\
\texttt{student\_id\_15} & -0.4631 &   0.440 & -1.052 &  0.293 & -1.326 &  0.400 \\
\texttt{student\_id\_16} & -0.4803 &   0.462 & -1.040 &  0.298 & -1.385 &  0.425 \\
\texttt{student\_id\_17} &  0.1523 &   0.467 &  0.326 &  0.744 & -0.763 &  1.067 \\
\texttt{student\_id\_18} & -0.6466 &   0.434 & -1.489 &  0.136 & -1.498 &  0.204 \\
\texttt{student\_id\_19} & -0.7664 &   0.417 & -1.838 &  0.066 & -1.583 & -0.051 \\
\texttt{student\_id\_20} & -0.1351 &   0.544 & -0.249 &  0.804 & -1.201 &  0.930 \\
\texttt{student\_id\_21} & -0.1178 &   0.446 & -0.264 &  0.792 & -0.992 &  0.756 \\
\texttt{student\_id\_22} & -0.4189 &   0.421 & -0.995 &  0.319 & -1.244 &  0.406 \\
\texttt{student\_id\_23} & -0.6313 &   0.430 & -1.468 &  0.142 & -1.474 &  0.211 \\
\texttt{student\_id\_24} &  0.3608 &   0.453 &  0.797 &  0.426 & -0.527 &  1.248 \\
\texttt{student\_id\_25} & -0.6202 &   0.436 & -1.422 &  0.155 & -1.475 &  0.235 \\
\texttt{student\_id\_26} & -0.1058 &   0.479 & -0.221 &  0.825 & -1.044 &  0.832 \\
\texttt{student\_id\_27} &  0.0908 &   0.407 &  0.223 &  0.823 & -0.707 &  0.888 \\
\texttt{student\_id\_28} & -0.4374 &   0.414 & -1.056 &  0.291 & -1.250 &  0.375 \\
\texttt{student\_id\_29} & -0.7691 &   0.447 & -1.720 &  0.085 & -1.646 &  0.107 \\
\texttt{student\_id\_30} & -0.2069 &   0.483 & -0.428 &  0.669 & -1.154 &  0.741 \\
\texttt{student\_id\_31} & -1.0851 &   0.510 & -2.129 &  0.033 & -2.084 & -0.086 \\
\texttt{student\_id\_32} & -0.3567 &   0.431 & -0.828 &  0.407 & -1.201 &  0.487 \\
\texttt{student\_id\_33} & -0.7434 &   0.462 & -1.609 &  0.108 & -1.649 &  0.162 \\
\texttt{student\_id\_34} & -0.9385 &   0.462 & -2.033 &  0.042 & -1.843 & -0.034 \\
\texttt{student\_id\_35} & -0.7976 &   0.469 & -1.699 &  0.089 & -1.718 &  0.122 \\
\texttt{student\_id\_36} & -0.7576 &   0.488 & -1.554 &  0.120 & -1.713 &  0.198 \\
\texttt{student\_id\_37} & -0.7586 &   0.436 & -1.741 &  0.082 & -1.612 &  0.095 \\
\texttt{student\_id\_38} &  0.4308 &   0.481 &  0.897 &  0.370 & -0.511 &  1.373 \\
\texttt{student\_id\_39} & -0.6976 &   0.441 & -1.581 &  0.114 & -1.563 &  0.167 \\
\texttt{student\_id\_40} &  0.0777 &   0.423 &  0.184 &  0.854 & -0.752 &  0.907 \\
\texttt{student\_id\_41} & -0.8009 &   0.402 & -1.991 &  0.046 & -1.589 & -0.013 \\
\texttt{student\_id\_42} & -1.2704 &   0.503 & -2.526 &  0.012 & -2.256 & -0.285 \\
\texttt{student\_id\_43} & -1.2256 &   0.555 & -2.208 &  0.027 & -2.313 & -0.138 \\
\texttt{student\_id\_44} & -0.4811 &   0.418 & -1.152 &  0.249 & -1.300 &  0.337 \\
\texttt{student\_id\_45} & -0.3647 &   0.390 & -0.934 &  0.350 & -1.129 &  0.400 \\
\texttt{student\_id\_46} & -0.3134 &   0.443 & -0.707 &  0.479 & -1.182 &  0.555 \\
\texttt{student\_id\_47} & -0.6779 &   0.456 & -1.485 &  0.137 & -1.573 &  0.217 \\
\texttt{student\_id\_48} &  0.0170 &   0.440 &  0.039 &  0.969 & -0.845 &  0.879 \\
\texttt{student\_id\_49} & -0.0711 &   0.401 & -0.178 &  0.859 & -0.856 &  0.714 \\
\texttt{student\_id\_50} & -0.9144 &   0.591 & -1.546 &  0.122 & -2.073 &  0.245 \\
\bottomrule
\end{tabular}
\caption{Complete regression results of Experiment 1 for the dependent variable \texttt{is\_clicked} on the THU-KDD dataset.}
\label{tab:full_reg_click_kdd}
\end{table}
\begin{table}[htbp]
\centering
\scriptsize
\begin{tabular}{lllllll}
\toprule
 \texttt{Variable} & Coefficient & Standard Error & t-Value & p-Value & CI Lower & CI Upper \\
\midrule
    \texttt{const} &      14.6956 &      4.805 &        3.058 &    0.002 &       5.274 &      24.117 \\
    \texttt{has\_decoy} &       1.9132 &      1.277 &        1.499 &    0.134 &      -0.590 &       4.416 \\
       \texttt{rank\_1} &      -4.2052 &      1.985 &       -2.118 &    0.034 &      -8.098 &      -0.313 \\
       \texttt{rank\_2} &      -6.5017 &      2.120 &       -3.067 &    0.002 &     -10.658 &      -2.346 \\
       \texttt{rank\_3} &      -2.5839 &      2.160 &       -1.196 &    0.232 &      -6.819 &       1.652 \\
       \texttt{rank\_4} &      -7.8692 &      2.105 &       -3.739 &    0.000 &     -11.995 &      -3.743 \\
       \texttt{rank\_5} &      -9.0114 &      2.128 &       -4.234 &    0.000 &     -13.184 &      -4.839 \\
       \texttt{rank\_6} &     -10.7383 &      2.331 &       -4.607 &    0.000 &     -15.308 &      -6.169 \\
       \texttt{rank\_7} &      -9.7773 &      2.308 &       -4.236 &    0.000 &     -14.302 &      -5.252 \\
       \texttt{rank\_8} &      -9.0515 &      2.483 &       -3.645 &    0.000 &     -13.921 &      -4.182 \\
       \texttt{rank\_9} &     -12.9871 &      2.523 &       -5.148 &    0.000 &     -17.933 &      -8.041 \\
   \texttt{task\_id\_10} &      -2.1766 &      2.765 &       -0.787 &    0.431 &      -7.597 &       3.244 \\
   \texttt{task\_id\_11} &      -0.7174 &      2.861 &       -0.251 &    0.802 &      -6.326 &       4.892 \\
   \texttt{task\_id\_12} &      -2.8597 &      2.866 &       -0.998 &    0.318 &      -8.478 &       2.759 \\
    \texttt{task\_id\_2} &      -1.8943 &      2.827 &       -0.670 &    0.503 &      -7.437 &       3.648 \\
    \texttt{task\_id\_3} &       3.6144 &      2.704 &        1.336 &    0.181 &      -1.688 &       8.917 \\
    \texttt{task\_id\_4} &       1.1273 &      2.820 &        0.400 &    0.689 &      -4.401 &       6.656 \\
    \texttt{task\_id\_7} &      -1.0875 &      2.820 &       -0.386 &    0.700 &      -6.616 &       4.441 \\
    \texttt{task\_id\_8} &      -1.9504 &      2.886 &       -0.676 &    0.499 &      -7.608 &       3.708 \\
 \texttt{student\_id\_2} &      -5.2057 &      6.624 &       -0.786 &    0.432 &     -18.193 &       7.782 \\
 \texttt{student\_id\_3} &       5.2426 &      5.576 &        0.940 &    0.347 &      -5.690 &      16.175 \\
 \texttt{student\_id\_4} &      -0.9130 &      9.369 &       -0.097 &    0.922 &     -19.282 &      17.456 \\
 \texttt{student\_id\_5} &      -3.1495 &      7.038 &       -0.448 &    0.655 &     -16.948 &      10.649 \\
 \texttt{student\_id\_6} &      -3.2744 &      4.893 &       -0.669 &    0.503 &     -12.867 &       6.318 \\
 \texttt{student\_id\_7} &       5.8010 &      5.366 &        1.081 &    0.280 &      -4.720 &      16.322 \\
 \texttt{student\_id\_8} &      -0.4306 &      7.319 &       -0.059 &    0.953 &     -14.780 &      13.919 \\
 \texttt{student\_id\_9} &      -0.8358 &      5.730 &       -0.146 &    0.884 &     -12.070 &      10.399 \\
\texttt{student\_id\_10} &      -4.2130 &      6.284 &       -0.670 &    0.503 &     -16.534 &       8.108 \\
\texttt{student\_id\_11} &      -3.9263 &      5.028 &       -0.781 &    0.435 &     -13.785 &       5.933 \\
\texttt{student\_id\_12} &      -1.9897 &      5.682 &       -0.350 &    0.726 &     -13.129 &       9.150 \\
\texttt{student\_id\_13} &       5.9915 &      5.423 &        1.105 &    0.269 &      -4.641 &      16.624 \\
\texttt{student\_id\_14} &       7.4104 &      5.355 &        1.384 &    0.166 &      -3.089 &      17.909 \\
\texttt{student\_id\_15} &       6.6235 &      5.475 &        1.210 &    0.226 &      -4.110 &      17.357 \\
\texttt{student\_id\_16} &      -3.2462 &      5.689 &       -0.571 &    0.568 &     -14.400 &       7.907 \\
\texttt{student\_id\_17} &       1.7870 &      6.174 &        0.289 &    0.772 &     -10.318 &      13.892 \\
\texttt{student\_id\_18} &      -4.5010 &      5.212 &       -0.864 &    0.388 &     -14.720 &       5.718 \\
\texttt{student\_id\_19} &      -5.0407 &      5.048 &       -0.998 &    0.318 &     -14.939 &       4.857 \\
\texttt{student\_id\_20} &      -3.6142 &      6.831 &       -0.529 &    0.597 &     -17.007 &       9.779 \\
\texttt{student\_id\_21} &      33.8699 &      5.705 &        5.937 &    0.000 &      22.684 &      45.056 \\
\texttt{student\_id\_22} &      -3.8870 &      5.312 &       -0.732 &    0.464 &     -14.302 &       6.528 \\
\texttt{student\_id\_23} &      -3.9145 &      5.221 &       -0.750 &    0.453 &     -14.151 &       6.322 \\
\texttt{student\_id\_24} &      23.6064 &      6.007 &        3.930 &    0.000 &      11.828 &      35.384 \\
\texttt{student\_id\_25} &      -4.3527 &      5.306 &       -0.820 &    0.412 &     -14.755 &       6.050 \\
\texttt{student\_id\_26} &       2.9870 &      6.140 &        0.486 &    0.627 &      -9.052 &      15.026 \\
\texttt{student\_id\_27} &      -4.4778 &      5.258 &       -0.852 &    0.394 &     -14.787 &       5.831 \\
\texttt{student\_id\_28} &      -2.3534 &      5.145 &       -0.457 &    0.647 &     -12.442 &       7.735 \\
\texttt{student\_id\_29} &      -6.1016 &      5.379 &       -1.134 &    0.257 &     -16.648 &       4.445 \\
\texttt{student\_id\_30} &      -3.1811 &      6.075 &       -0.524 &    0.601 &     -15.093 &       8.731 \\
\texttt{student\_id\_31} &       0.3445 &      5.736 &        0.060 &    0.952 &     -10.902 &      11.591 \\
\texttt{student\_id\_32} &      -4.0883 &      5.388 &       -0.759 &    0.448 &     -14.653 &       6.476 \\
\texttt{student\_id\_33} &      -6.3012 &      5.444 &       -1.158 &    0.247 &     -16.974 &       4.372 \\
\texttt{student\_id\_34} &      -6.6902 &      5.457 &       -1.226 &    0.220 &     -17.390 &       4.009 \\
\texttt{student\_id\_35} &      -1.5817 &      5.602 &       -0.282 &    0.778 &     -12.565 &       9.401 \\
\texttt{student\_id\_36} &       2.7925 &      5.518 &        0.506 &    0.613 &      -8.027 &      13.611 \\
\texttt{student\_id\_37} &      -3.8948 &      5.293 &       -0.736 &    0.462 &     -14.272 &       6.482 \\
\texttt{student\_id\_38} &      -5.3146 &      6.595 &       -0.806 &    0.420 &     -18.246 &       7.617 \\
\texttt{student\_id\_39} &      -3.2159 &      5.487 &       -0.586 &    0.558 &     -13.974 &       7.542 \\
\texttt{student\_id\_40} &      -0.9360 &      5.527 &       -0.169 &    0.866 &     -11.773 &       9.901 \\
\texttt{student\_id\_41} &      -3.0055 &      4.874 &       -0.617 &    0.538 &     -12.562 &       6.551 \\
\texttt{student\_id\_42} &      -6.2148 &      5.567 &       -1.116 &    0.264 &     -17.130 &       4.701 \\
\texttt{student\_id\_43} &      -4.9234 &      5.969 &       -0.825 &    0.410 &     -16.626 &       6.780 \\
\texttt{student\_id\_44} &      -3.4944 &      5.222 &       -0.669 &    0.503 &     -13.733 &       6.745 \\
\texttt{student\_id\_45} &       0.5996 &      4.925 &        0.122 &    0.903 &      -9.057 &      10.256 \\
\texttt{student\_id\_46} &      -3.0053 &      5.559 &       -0.541 &    0.589 &     -13.904 &       7.893 \\
\texttt{student\_id\_47} &      -5.4970 &      5.511 &       -0.997 &    0.319 &     -16.302 &       5.309 \\
\texttt{student\_id\_48} &      -2.4336 &      5.700 &       -0.427 &    0.669 &     -13.608 &       8.741 \\
\texttt{student\_id\_49} &      -2.3961 &      5.085 &       -0.471 &    0.638 &     -12.367 &       7.574 \\
\texttt{student\_id\_50} &      -6.4358 &      6.526 &       -0.986 &    0.324 &     -19.230 &       6.359 \\
\bottomrule
\end{tabular}
\caption{Complete regression results Experiment 1  for the dependent variable \texttt{duration} on the THU-KDD dataset.}
\label{tab:full_reg_time_kdd}
\end{table}
\begin{table}[htbp]
\centering
\scriptsize
\begin{tabular}{lllllll}
\toprule
\texttt{Variable} & Coefficient & Standard Error & t-Value & p-Value & CI Lower & CI Upper \\
\midrule
 \texttt{const} & 2.2088 & 0.196 & 11.245 & 0 & 1.824 & 2.594 \\
\texttt{has\_decoy} & 0.1559 & 0.052 & 2.986 & 0.003 & 0.054 & 0.258 \\
\texttt{rank\_1} & -0.5271 & 0.081 & -6.495 & 0 & -0.686 & -0.368 \\
\texttt{rank\_2} & -0.8244 & 0.087 & -9.514 & 0 & -0.994 & -0.655 \\
\texttt{rank\_3} & -1.1594 & 0.088 & -13.129 & 0 & -1.333 & -0.986 \\
\texttt{rank\_4} & -0.9703 & 0.086 & -11.278 & 0 & -1.139 & -0.802 \\
\texttt{rank\_5} & -1.2743 & 0.087 & -14.647 & 0 & -1.445 & -1.104 \\
\texttt{rank\_6} & -1.3442 & 0.095 & -14.108 & 0 & -1.531 & -1.157 \\
\texttt{rank\_7} & -1.3014 & 0.094 & -13.794 & 0 & -1.486 & -1.116 \\
\texttt{rank\_8} & -1.3266 & 0.102 & -13.068 & 0 & -1.526 & -1.128 \\
\texttt{rank\_9} & -1.5749 & 0.103 & -15.272 & 0 & -1.777 & -1.373 \\
\texttt{task\_id\_10} & -0.3423 & 0.113 & -3.028 & 0.002 & -0.564 & -0.121 \\
\texttt{task\_id\_11} & -0.1781 & 0.117 & -1.523 & 0.128 & -0.407 & 0.051 \\
\texttt{task\_id\_12} & -0.2404 & 0.117 & -2.052 & 0.04 & -0.47 & -0.011 \\
\texttt{task\_id\_2} & -0.2394 & 0.116 & -2.071 & 0.038 & -0.466 & -0.013 \\
\texttt{task\_id\_3} & -0.4078 & 0.111 & -3.688 & 0 & -0.625 & -0.191 \\
\texttt{task\_id\_4} & -0.0711 & 0.115 & -0.617 & 0.538 & -0.297 & 0.155 \\
\texttt{task\_id\_7} & -0.3997 & 0.115 & -3.468 & 0.001 & -0.626 & -0.174 \\
\texttt{task\_id\_8} & -0.1578 & 0.118 & -1.338 & 0.181 & -0.389 & 0.073 \\
\texttt{student\_id\_2} & -0.5731 & 0.271 & -2.116 & 0.034 & -1.104 & -0.042 \\
\texttt{student\_id\_3} & -0.1151 & 0.228 & -0.505 & 0.613 & -0.562 & 0.332 \\
\texttt{student\_id\_4} & 0.1296 & 0.383 & 0.338 & 0.735 & -0.621 & 0.881 \\
\texttt{student\_id\_5} & -0.4893 & 0.288 & -1.701 & 0.089 & -1.053 & 0.075 \\
\texttt{student\_id\_6} & -0.3563 & 0.200 & -1.781 & 0.075 & -0.748 & 0.036 \\
\texttt{student\_id\_7} & -0.1877 & 0.219 & -0.856 & 0.392 & -0.618 & 0.242 \\
\texttt{student\_id\_8} & -0.8291 & 0.299 & -2.771 & 0.006 & -1.416 & -0.242 \\
\texttt{student\_id\_9} & -0.3292 & 0.234 & -1.405 & 0.160 & -0.788 & 0.130 \\
\texttt{student\_id\_10} & -0.6290 & 0.257 & -2.449 & 0.014 & -1.134 & -0.124 \\
\texttt{student\_id\_11} & -0.2913 & 0.206 & -1.417 & 0.157 & -0.694 & 0.112 \\
\texttt{student\_id\_12} & -0.2483 & 0.232 & -1.069 & 0.285 & -0.704 & 0.207 \\
\texttt{student\_id\_13} & -0.6135 & 0.222 & -2.767 & 0.006 & -1.048 & -0.179 \\
\texttt{student\_id\_14} & -0.1977 & 0.219 & -0.903 & 0.367 & -0.627 & 0.232 \\
\texttt{student\_id\_15} & -0.2476 & 0.224 & -1.106 & 0.269 & -0.686 & 0.191 \\
\texttt{student\_id\_16} & -0.3262 & 0.233 & -1.403 & 0.161 & -0.782 & 0.130 \\
\texttt{student\_id\_17} & -0.2254 & 0.252 & -0.893 & 0.372 & -0.720 & 0.269 \\
\texttt{student\_id\_18} & -0.5053 & 0.213 & -2.372 & 0.018 & -0.923 & -0.088 \\
\texttt{student\_id\_19} & -0.6719 & 0.206 & -3.256 & 0.001 & -1.077 & -0.267 \\
\texttt{student\_id\_20} & -0.1887 & 0.279 & -0.676 & 0.499 & -0.736 & 0.359 \\
\texttt{student\_id\_21} & -0.1976 & 0.233 & -0.847 & 0.397 & -0.655 & 0.238 \\
\texttt{student\_id\_22} & -0.0526 & 0.217 & -0.242 & 0.809 & -0.478 & 0.373 \\
\texttt{student\_id\_23} & -0.1804 & 0.213 & -0.845 & 0.398 & -0.599 & 0.238 \\
\texttt{student\_id\_24} & 0.1348 & 0.246 & 0.549 & 0.583 & -0.347 & 0.616 \\
\texttt{student\_id\_25} & 0.0112 & 0.217 & 0.052 & 0.959 & -0.414 & 0.436 \\
\texttt{student\_id\_26} & -0.3446 & 0.251 & -1.373 & 0.170 & -0.837 & 0.148 \\
\texttt{student\_id\_27} & 0.0535 & 0.215 & 0.249 & 0.803 & -0.368 & 0.475 \\
\texttt{student\_id\_28} & -0.3127 & 0.210 & -1.487 & 0.137 & -0.725 & 0.100 \\
\texttt{student\_id\_29} & -0.3582 & 0.220 & -1.629 & 0.103 & -0.789 & 0.073 \\
\texttt{student\_id\_30} & -0.2375 & 0.248 & -0.956 & 0.339 & -0.724 & 0.249 \\
\texttt{student\_id\_31} & -0.4528 & 0.234 & -1.931 & 0.054 & -0.913 & 0.007 \\
\texttt{student\_id\_32} & -0.0599 & 0.220 & -0.272 & 0.786 & -0.492 & 0.372 \\
\texttt{student\_id\_33} & -0.4830 & 0.223 & -2.171 & 0.030 & -0.919 & -0.047 \\
\texttt{student\_id\_34} & -0.4745 & 0.223 & -2.127 & 0.033 & -0.912 & -0.037 \\
\texttt{student\_id\_35} & -0.5683 & 0.229 & -2.482 & 0.013 & -1.017 & -0.119 \\
\texttt{student\_id\_36} & -0.4044 & 0.226 & -1.793 & 0.073 & -0.847 & 0.038 \\
\texttt{student\_id\_37} & -0.3952 & 0.216 & -1.826 & 0.068 & -0.819 & 0.029 \\
\texttt{student\_id\_38} & -0.2506 & 0.270 & -0.930 & 0.353 & -0.779 & 0.278 \\
\texttt{student\_id\_39} & -0.6242 & 0.224 & -2.783 & 0.005 & -1.064 & -0.184 \\
\texttt{student\_id\_40} & 0.3221 & 0.226 & 1.425 & 0.154 & -0.121 & 0.765 \\
\texttt{student\_id\_41} & -0.2878 & 0.199 & -1.445 & 0.149 & -0.679 & 0.103 \\
\texttt{student\_id\_42} & -0.7060 & 0.228 & -3.102 & 0.002 & -1.152 & -0.260 \\
\texttt{student\_id\_43} & -0.6425 & 0.244 & -2.633 & 0.008 & -1.121 & -0.164 \\
\texttt{student\_id\_44} & -0.3255 & 0.213 & -1.525 & 0.127 & -0.744 & 0.093 \\
\texttt{student\_id\_45} & 0.3962 & 0.201 & 1.968 & 0.049 & 0.001 & 0.791 \\
\texttt{student\_id\_46} & -0.0187 & 0.227 & -0.082 & 0.934 & -0.464 & 0.427 \\
\texttt{student\_id\_47} & -0.3802 & 0.225 & -1.688 & 0.092 & -0.822 & 0.062 \\
\texttt{student\_id\_48} & 0.1291 & 0.233 & 0.554 & 0.580 & -0.328 & 0.586 \\
\texttt{student\_id\_49} & -0.1471 & 0.208 & -0.708 & 0.479 & -0.555 & 0.260 \\
\texttt{student\_id\_50} & -0.5608 & 0.267 & -2.102 & 0.036 & -1.084 & -0.038 \\
\bottomrule
\end{tabular}
\caption{Complete regression results Experiment 1  for the dependent variable \texttt{usefulness} on the THU-KDD dataset.}
\label{tab:full_reg_usefulness_kdd}
\end{table}
\clearpage
\section{Complete Regression Results of Experiment 2}
\label{ch:appendix_regression2}
This section presents the complete regression results on the THUIR2018 dataset in Experiment 2. Table~\ref{tab:full_reg_click_interaction} shows the results of logistic regression on the dependent variable \texttt{is\_clicked} in the THUIR2018 dataset. Table~\ref{tab:full_reg_time_interaction} shows the results of OLS regression on the dependent variable \texttt{duration} in the THUIR2018 dataset. Table~\ref{tab:full_reg_usefulness_interaction} shows the results of OLS regression on the dependent variable \texttt{usefulness} in the THUIR2018 dataset.

\begin{table}[htbp]
\centering
\scriptsize
\begin{tabular}{lllllll}
\toprule
 \texttt{Variable} & Coefficient & Standard Error & z-Value & p-Value & CI Lower & CI Upper \\
\midrule
              \texttt{const} &  0.5581 &    0.654 &  0.853 &  0.393 &  -0.724 &   1.840 \\
          \texttt{has\_decoy} &  1.0641 &    0.320 &  3.327 &  0.001 &   0.437 &   1.691 \\
    \texttt{high\_load\_decoy} & -1.1559 &    0.461 & -2.505 &  0.012 &  -2.060 &  -0.252 \\
          \texttt{is\_low\_kn} & -0.2210 &    0.304 & -0.727 &  0.468 &  -0.817 &   0.375 \\
\texttt{low\_knowledge\_decoy} &  0.6983 &    0.450 &  1.553 &  0.121 &  -0.183 &   1.580 \\
             \texttt{rank\_2} & -0.1829 &    0.392 & -0.466 &  0.641 &  -0.951 &   0.586 \\
             \texttt{rank\_3} & -1.1667 &    0.438 & -2.661 &  0.008 &  -2.026 &  -0.307 \\
             \texttt{rank\_4} & -0.7021 &    0.407 & -1.723 &  0.085 &  -1.501 &   0.097 \\
             \texttt{rank\_5} & -1.0709 &    0.442 & -2.421 &  0.015 &  -1.938 &  -0.204 \\
             \texttt{rank\_6} & -1.8589 &    0.474 & -3.920 &  0.000 &  -2.788 &  -0.930 \\
             \texttt{rank\_7} & -2.0816 &    0.461 & -4.519 &  0.000 &  -2.984 &  -1.179 \\
             \texttt{rank\_8} & -1.7427 &    0.465 & -3.746 &  0.000 &  -2.655 &  -0.831 \\
             \texttt{rank\_9} & -1.4939 &    0.532 & -2.810 &  0.005 &  -2.536 &  -0.452 \\
            \texttt{rank\_10} & -1.4864 &    0.456 & -3.260 &  0.001 &  -2.380 &  -0.593 \\
          \texttt{task\_id\_2} & -0.2184 &    0.361 & -0.605 &  0.545 &  -0.926 &   0.489 \\
          \texttt{task\_id\_3} &  0.0209 &    0.342 &  0.061 &  0.951 &  -0.649 &   0.691 \\
          \texttt{task\_id\_4} &  0.8726 &    0.400 &  2.183 &  0.029 &   0.089 &   1.656 \\
          \texttt{task\_id\_5} & -1.1283 &    0.470 & -2.403 &  0.016 &  -2.049 &  -0.208 \\
          \texttt{task\_id\_6} &  0.6943 &    0.467 &  1.487 &  0.137 &  -0.221 &   1.609 \\
       \texttt{student\_id\_2} & -0.8000 &    0.697 & -1.147 &  0.251 &  -2.167 &   0.567 \\
       \texttt{student\_id\_3} & -0.6629 &    0.801 & -0.828 &  0.408 &  -2.233 &   0.907 \\
       \texttt{student\_id\_4} & -2.5434 &    0.834 & -3.048 &  0.002 &  -4.179 &  -0.908 \\
       \texttt{student\_id\_5} & -1.1777 &    0.866 & -1.360 &  0.174 &  -2.875 &   0.519 \\
       \texttt{student\_id\_6} & -0.8018 &    0.711 & -1.127 &  0.260 &  -2.196 &   0.593 \\
       \texttt{student\_id\_7} & -0.6661 &    0.786 & -0.847 &  0.397 &  -2.207 &   0.875 \\
       \texttt{student\_id\_8} & -1.6833 &    1.180 & -1.426 &  0.154 &  -3.997 &   0.630 \\
       \texttt{student\_id\_9} & -1.5757 &    0.781 & -2.018 &  0.044 &  -3.107 &  -0.045 \\
      \texttt{student\_id\_10} & -1.4453 &    0.789 & -1.831 &  0.067 &  -2.992 &   0.102 \\
      \texttt{student\_id\_11} & -0.7468 &    0.747 & -1.000 &  0.317 &  -2.211 &   0.717 \\
      \texttt{student\_id\_12} & -1.6931 &    0.892 & -1.898 &  0.058 &  -3.442 &   0.056 \\
      \texttt{student\_id\_13} & -1.5241 &    0.777 & -1.961 &  0.050 &  -3.047 &  -0.001 \\
      \texttt{student\_id\_14} & -0.9180 &    0.695 & -1.322 &  0.186 &  -2.279 &   0.443 \\
      \texttt{student\_id\_15} & -1.2895 &    0.716 & -1.802 &  0.072 &  -2.692 &   0.113 \\
      \texttt{student\_id\_16} & -0.8160 &    0.772 & -1.057 &  0.290 &  -2.328 &   0.696 \\
      \texttt{student\_id\_17} & -0.5387 &    0.723 & -0.745 &  0.456 &  -1.956 &   0.879 \\
      \texttt{student\_id\_18} & -0.3919 &    0.669 & -0.586 &  0.558 &  -1.704 &   0.920 \\
      \texttt{student\_id\_19} & -1.0245 &    0.738 & -1.389 &  0.165 &  -2.470 &   0.421 \\
      \texttt{student\_id\_20} & -0.6952 &    0.749 & -0.928 &  0.353 &  -2.163 &   0.773 \\
      \texttt{student\_id\_21} & -1.1763 &    0.835 & -1.409 &  0.159 &  -2.812 &   0.459 \\
      \texttt{student\_id\_22} & -1.3546 &    0.833 & -1.625 &  0.104 &  -2.988 &   0.279 \\
      \texttt{student\_id\_23} &  0.5424 &    0.722 &  0.751 &  0.452 &  -0.873 &   1.958 \\
      \texttt{student\_id\_24} & -1.1166 &    0.768 & -1.454 &  0.146 &  -2.622 &   0.389 \\
      \texttt{student\_id\_25} & -2.0834 &    0.992 & -2.100 &  0.036 &  -4.028 &  -0.139 \\
      \texttt{student\_id\_26} & -0.9079 &    0.745 & -1.218 &  0.223 &  -2.369 &   0.553 \\
      \texttt{student\_id\_27} & -0.3934 &    0.708 & -0.556 &  0.578 &  -1.780 &   0.994 \\
      \texttt{student\_id\_28} & -1.1212 &    0.843 & -1.330 &  0.184 &  -2.774 &   0.530 \\
\bottomrule
\end{tabular}
\caption{Complete regression results of Experiment 2 for the dependent variable \texttt{is\_clicked} on the THUIR2018 dataset.}
\label{tab:full_reg_click_interaction}
\end{table}
\begin{table}[htbp]
\centering
\scriptsize
\begin{tabular}{lllllll}
\toprule
 \texttt{Variable} & Coefficient & Standard Error & t-Value & p-Value & CI Lower & CI Upper \\
\midrule
              \texttt{const} &  74.6608 &   36.272 &  2.058 &  0.040 &    3.450 & 145.872 \\
          \texttt{has\_decoy} &  78.8909 &   17.657 &  4.468 &  0.000 &   44.226 & 113.555 \\
    \texttt{high\_load\_decoy} & -84.8992 &   23.084 & -3.678 &  0.000 & -130.219 & -39.580 \\
          \texttt{is\_low\_kn} &   7.2933 &   14.247 &  0.512 &  0.609 &  -20.677 &  35.264 \\
\texttt{low\_knowledge\_decoy} &  31.5345 &   23.290 &  1.354 &  0.176 &  -14.190 &  77.259 \\
             \texttt{rank\_2} &   8.1503 &   23.286 &  0.350 &  0.726 &  -37.567 &  53.867 \\
             \texttt{rank\_3} & -10.1791 &   23.946 & -0.425 &  0.671 &  -57.191 &  36.833 \\
             \texttt{rank\_4} &  -8.4571 &   23.495 & -0.360 &  0.719 &  -54.584 &  37.670 \\
             \texttt{rank\_5} & -13.4217 &   24.467 & -0.549 &  0.583 &  -61.456 &  34.613 \\
             \texttt{rank\_6} & -49.2204 &   23.434 & -2.100 &  0.036 &  -95.228 &  -3.213 \\
             \texttt{rank\_7} & -44.6424 &   22.673 & -1.969 &  0.049 &  -89.156 &  -0.129 \\
             \texttt{rank\_8} & -37.1315 &   23.440 & -1.584 &  0.114 &  -83.150 &   8.887 \\
             \texttt{rank\_9} & -44.9385 &   27.195 & -1.652 &  0.099 &  -98.329 &   8.452 \\
            \texttt{rank\_10} & -39.7140 &   23.729 & -1.674 &  0.095 &  -86.300 &   6.872 \\
          \texttt{task\_id\_2} & -12.1143 &   17.305 & -0.700 &  0.484 &  -46.089 &  21.861 \\
          \texttt{task\_id\_3} & -22.7458 &   17.129 & -1.328 &  0.185 &  -56.375 &  10.883 \\
          \texttt{task\_id\_4} &   2.4951 &   20.334 &  0.123 &  0.902 &  -37.426 &  42.416 \\
          \texttt{task\_id\_5} & -33.0691 &   19.779 & -1.672 &  0.095 &  -71.900 &   5.761 \\
          \texttt{task\_id\_6} & -34.9866 &   25.007 & -1.399 &  0.162 &  -84.082 &  14.108 \\
       \texttt{student\_id\_2} & -27.5829 &   37.197 & -0.742 &  0.459 & -100.611 &  45.445 \\
       \texttt{student\_id\_3} & -22.7295 &   44.144 & -0.515 &  0.607 & -109.396 &  63.937 \\
       \texttt{student\_id\_4} & -61.7338 &   36.552 & -1.689 &  0.092 & -133.494 &  10.027 \\
       \texttt{student\_id\_5} & 110.3034 &   46.280 &  2.383 &  0.017 &   19.444 & 201.163 \\
       \texttt{student\_id\_6} & -34.5152 &   37.844 & -0.912 &  0.362 & -108.813 &  39.783 \\
       \texttt{student\_id\_7} & -22.2099 &   42.881 & -0.518 &  0.605 & -106.396 &  61.976 \\
       \texttt{student\_id\_8} & -21.4848 &   46.667 & -0.460 &  0.645 & -113.105 &  70.135 \\
       \texttt{student\_id\_9} & -44.8027 &   39.376 & -1.138 &  0.256 & -122.109 &  32.503 \\
      \texttt{student\_id\_10} & -19.4754 &   41.239 & -0.472 &  0.637 & -100.437 &  61.487 \\
      \texttt{student\_id\_11} &  -0.7468 &   41.987 & -0.018 &  0.986 &  -83.178 &  81.685 \\
      \texttt{student\_id\_12} & -38.5503 &   43.433 & -0.888 &  0.375 & -123.820 &  46.719 \\
      \texttt{student\_id\_13} & -37.6782 &   37.634 & -1.001 &  0.317 & -111.564 &  36.208 \\
      \texttt{student\_id\_14} & -29.4337 &   36.834 & -0.799 &  0.424 & -101.748 &  42.880 \\
      \texttt{student\_id\_15} &  -3.2992 &   36.901 & -0.089 &  0.929 &  -75.745 &  69.147 \\
      \texttt{student\_id\_16} &  77.2389 &   43.441 &  1.778 &  0.076 &   -8.047 & 162.525 \\
      \texttt{student\_id\_17} &  -9.9520 &   39.035 & -0.255 &  0.799 &  -86.588 &  66.684 \\
      \texttt{student\_id\_18} &   5.1994 &   36.622 &  0.142 &  0.887 &  -66.699 &  77.097 \\
      \texttt{student\_id\_19} &  47.7269 &   40.295 &  1.184 &  0.237 &  -31.382 & 126.836 \\
      \texttt{student\_id\_20} &  30.4401 &   41.963 &  0.725 &  0.468 &  -51.944 & 112.824 \\
      \texttt{student\_id\_21} &  15.9694 &   44.974 &  0.355 &  0.723 &  -72.326 & 104.265 \\
      \texttt{student\_id\_22} & -24.6618 &   42.730 & -0.577 &  0.564 & -108.551 &  59.228 \\
      \texttt{student\_id\_23} &  -3.7334 &   39.828 & -0.094 &  0.925 &  -81.926 &  74.459 \\
      \texttt{student\_id\_24} &  23.6226 &   38.853 &  0.608 &  0.543 &  -52.655 &  99.901 \\
      \texttt{student\_id\_25} & -45.7345 &   43.523 & -1.051 &  0.294 & -131.181 &  39.712 \\
      \texttt{student\_id\_26} &  -3.3204 &   38.807 & -0.086 &  0.932 &  -79.509 &  72.868 \\
      \texttt{student\_id\_27} &  15.2719 &   39.189 &  0.390 &  0.697 &  -61.667 &  92.210 \\
      \texttt{student\_id\_28} & -43.9809 &   42.740 & -1.029 &  0.304 & -127.891 &  39.930 \\
\bottomrule
\end{tabular}
\caption{Complete regression results of Experiment 2 for the dependent variable \texttt{duration} on the THUIR2018 dataset.}
\label{tab:full_reg_time_interaction}
\end{table}
\begin{table}[htbp]
\centering
\scriptsize
\begin{tabular}{lllllll}
\toprule
 \texttt{Variable} & Coefficient & Standard Error & t-Value & p-Value & CI Lower & CI Upper \\
\midrule
              \texttt{const} &  0.821100 &    0.200 &  4.101 &  0.000 &   0.428 &   1.214 \\
          \texttt{has\_decoy} &  0.494400 &    0.097 &  5.072 &  0.000 &   0.303 &   0.686 \\
    \texttt{high\_load\_decoy} & -0.519300 &    0.127 & -4.075 &  0.000 &  -0.769 &  -0.269 \\
          \texttt{is\_low\_kn} &  0.011000 &    0.079 &  0.140 &  0.888 &  -0.143 &   0.165 \\
\texttt{low\_knowledge\_decoy} &  0.275000 &    0.129 &  2.139 &  0.033 &   0.023 &   0.527 \\
             \texttt{rank\_2} & -0.152300 &    0.129 & -1.185 &  0.237 &  -0.405 &   0.100 \\
             \texttt{rank\_3} & -0.347300 &    0.132 & -2.627 &  0.009 &  -0.607 &  -0.088 \\
             \texttt{rank\_4} & -0.394100 &    0.130 & -3.039 &  0.002 &  -0.649 &  -0.139 \\
             \texttt{rank\_5} & -0.406100 &    0.135 & -3.007 &  0.003 &  -0.671 &  -0.141 \\
             \texttt{rank\_6} & -0.548300 &    0.129 & -4.238 &  0.000 &  -0.802 &  -0.294 \\
             \texttt{rank\_7} & -0.615400 &    0.125 & -4.917 &  0.000 &  -0.861 &  -0.370 \\
             \texttt{rank\_8} & -0.560500 &    0.129 & -4.331 &  0.000 &  -0.815 &  -0.306 \\
             \texttt{rank\_9} & -0.534900 &    0.150 & -3.563 &  0.000 &  -0.830 &  -0.240 \\
            \texttt{rank\_10} & -0.533100 &    0.131 & -4.070 &  0.000 &  -0.790 &  -0.276 \\
          \texttt{task\_id\_2} &  0.014400 &    0.096 &  0.151 &  0.880 &  -0.173 &   0.202 \\
          \texttt{task\_id\_3} & -0.028300 &    0.095 & -0.299 &  0.765 &  -0.214 &   0.157 \\
          \texttt{task\_id\_4} &  0.239100 &    0.112 &  2.130 &  0.034 &   0.019 &   0.459 \\
          \texttt{task\_id\_5} & -0.226300 &    0.109 & -2.073 &  0.039 &  -0.441 &  -0.012 \\
          \texttt{task\_id\_6} &  0.012400 &    0.138 &  0.089 &  0.929 &  -0.259 &   0.283 \\
       \texttt{student\_id\_2} & -0.097300 &    0.205 & -0.474 &  0.636 &  -0.500 &   0.306 \\
       \texttt{student\_id\_3} & -0.168200 &    0.244 & -0.690 &  0.490 &  -0.647 &   0.310 \\
       \texttt{student\_id\_4} & -0.485000 &    0.202 & -2.403 &  0.016 &  -0.881 &  -0.089 \\
       \texttt{student\_id\_5} &  0.160300 &    0.255 &  0.627 &  0.531 &  -0.341 &   0.662 \\
       \texttt{student\_id\_6} & -0.158800 &    0.209 & -0.760 &  0.447 &  -0.569 &   0.251 \\
       \texttt{student\_id\_7} &  0.030900 &    0.237 &  0.130 &  0.896 &  -0.434 &   0.496 \\
       \texttt{student\_id\_8} & -0.275300 &    0.258 & -1.069 &  0.286 &  -0.781 &   0.230 \\
       \texttt{student\_id\_9} & -0.340500 &    0.217 & -1.566 &  0.118 &  -0.767 &   0.086 \\
      \texttt{student\_id\_10} & -0.227400 &    0.228 & -0.999 &  0.318 &  -0.674 &   0.220 \\
      \texttt{student\_id\_11} & -0.230600 &    0.232 & -0.995 &  0.320 &  -0.686 &   0.224 \\
      \texttt{student\_id\_12} & -0.171500 &    0.240 & -0.715 &  0.475 &  -0.642 &   0.299 \\
      \texttt{student\_id\_13} & -0.339800 &    0.208 & -1.636 &  0.102 &  -0.748 &   0.068 \\
      \texttt{student\_id\_14} & -0.128000 &    0.203 & -0.629 &  0.529 &  -0.527 &   0.271 \\
      \texttt{student\_id\_15} & -0.218000 &    0.204 & -1.070 &  0.285 &  -0.618 &   0.182 \\
      \texttt{student\_id\_16} & -0.118600 &    0.240 & -0.494 &  0.621 &  -0.589 &   0.352 \\
      \texttt{student\_id\_17} & -0.030600 &    0.215 & -0.142 &  0.887 &  -0.454 &   0.392 \\
      \texttt{student\_id\_18} &  0.033700 &    0.202 &  0.166 &  0.868 &  -0.363 &   0.431 \\
      \texttt{student\_id\_19} & -0.218100 &    0.222 & -0.981 &  0.327 &  -0.655 &   0.219 \\
      \texttt{student\_id\_20} & -0.096800 &    0.232 & -0.418 &  0.676 &  -0.552 &   0.358 \\
      \texttt{student\_id\_21} & -0.114500 &    0.248 & -0.461 &  0.645 &  -0.602 &   0.373 \\
      \texttt{student\_id\_22} &  0.013200 &    0.236 &  0.056 &  0.955 &  -0.450 &   0.476 \\
      \texttt{student\_id\_23} &  0.001700 &    0.220 &  0.008 &  0.994 &  -0.430 &   0.433 \\
      \texttt{student\_id\_24} & -0.226300 &    0.214 & -1.055 &  0.292 &  -0.647 &   0.195 \\
      \texttt{student\_id\_25} & -0.414600 &    0.240 & -1.726 &  0.085 &  -0.886 &   0.057 \\
      \texttt{student\_id\_26} &  0.013200 &    0.214 &  0.061 &  0.951 &  -0.407 &   0.434 \\
      \texttt{student\_id\_27} & -0.000029 &    0.216 &  0.000 &  1.000 &  -0.425 &   0.425 \\
      \texttt{student\_id\_28} & -0.329800 &    0.236 & -1.398 &  0.163 &  -0.793 &   0.133 \\
\bottomrule
\end{tabular}
\caption{Complete regression results Experiment 2 for the dependent variable \texttt{usefulness} on the THUIR2018 dataset.}
\label{tab:full_reg_usefulness_interaction}
\end{table}

\end{document}